%% file: DRAFT_V10.tex
\documentclass[11pt]{article}
\usepackage{natbib}
\usepackage{amsmath}
\usepackage{amssymb}
\usepackage{pifont}
\usepackage[margin=0.8in, footskip=0.25in]{geometry}
\usepackage{amsthm}
\usepackage[capposition=top]{floatrow}
\usepackage{graphicx}
\usepackage{booktabs}
\usepackage[bf]{titlesec}
\usepackage[nottoc]{tocbibind}
\usepackage{setspace}
\usepackage{lscape}
\usepackage{tikz}
\usepackage{tkz-euclide,subfigure}
\usepackage{multicol}
\usepackage{colortbl}
\usepackage{enumerate}
\usepackage{tikz}
\usepackage[flushleft]{threeparttable}
\usepackage{longtable}
\usepackage{xtab}
\usepackage{tabulary}
\newcolumntype{K}[1]{>{\centering\arraybackslash}m{#1}}
\usepackage{threeparttable}
\usepackage{multirow}
\usepackage{fnpct}
\usepackage{pdflscape}
\usepackage{titletoc}
\usepackage[hidelinks,bookmarks=false]{hyperref}
\hypersetup{
	colorlinks,
	linkcolor={blue!80!black},
	citecolor={blue!80!black},
	urlcolor={blue!80!black},
	pdfstartview={}
}
\usepackage{booktabs}
\usepackage{caption}
\definecolor{dgray}{gray}{0.35}

\DeclareMathOperator{\sign}{sgn}

\usepackage{epigraph}

\usepackage{etoolbox}

\makeatletter
\patchcmd{\epigraph}{\@epitext{#1}}{\itshape\@epitext{#1}}{}{}
\makeatother

\usepackage{multibib}

\newcites{app}{Appendix Bibliography}

\title{Minimum Wages and Optimal Redistribution\thanks{Email: \href{damianvergara@berkeley.edu}{damianvergara@berkeley.edu}. First version: September, 2021. This version: \today.  I am especially grateful to Danny Yagan, Pat Kline, and Emmanuel Saez, for continuous guidance and encouragement throghout this project. I also thank Alan Auerbach, Nano Barahona, Sydnee Caldwell, David Card, Patricio Dom\'inguez, Cecile Gaubert, Andr\'es Gonz\'alez, Attila Lindner, Crist\'obal Otero, Pablo Mu\~noz, Michael Reich, Marco Rojas, Jesse Rothstein, Harry Wheeler, Gabriel Zucman, my discussants Benjamin Glass and Thomas Winberry, and seminar participants at the Online Public Finance Workshop for Graduate Students, the NTA 114th Annual Conference on Taxation, the XIII RIDGE Forum Workshop on Public Economics, UC Berkeley, and Universidad Adolfo Ib\'a\~nez for very helpful discussions and suggestions. Early versions of this project benefited from discussions with Jakob Brounstein, Sree Kancherla, Maximiliano Lauletta, Michael Love, Billy Morrison, and M\'onica Saucedo. I acknowledge financial support from the Center of Equitable Growth at UC Berkeley. This paper previously circulated under the title ``When do Minimum Wages Increase Social Welfare? A Sufficient Statistics Analysis with Taxes and Transfers''. Usual disclaimers apply.}}%
\date{\href{https://dvergarad.github.io/files/JMP_DV.pdf}{Updated frequently, click here for the latest version}}

\author{Dami\'an Vergara\\ UC Berkeley \\\textit{Job Market Paper}}

\renewenvironment{abstract}
 {\small
  \begin{center}
  \bfseries \abstractname\vspace{-.5em}\vspace{0pt}
  \end{center}
  \list{}{
    \setlength{\leftmargin}{.5cm}%
    \setlength{\rightmargin}{\leftmargin}%
  }%
  \item\relax}
 {\endlist}

\begin{document}

\maketitle

\begin{abstract}
This paper analyzes whether a minimum wage should be used for redistribution on top of taxes and transfers. I characterize optimal redistribution for a government with three policy instruments -- labor income taxes and transfers, corporate income taxes, and a minimum wage -- using an empirically grounded model of the labor market with positive firm profits. A minimum wage can increase social welfare when it increases the average post-tax wages of low-skill labor market participants and when corporate profit incidence is large. When chosen together with taxes, the minimum wage can help the government redistribute efficiently to low-skill workers by preventing firms from capturing low-wage income subsidies such as the EITC and from enjoying high profits that cannot be redistributed via corporate taxes due to capital mobility in unaffected industries. Event studies show that the average US state-level minimum wage reform over the last two decades increased average post-tax wages of low-skilled labor market participants and reduced corporate profits in affected industries, namely low-skill labor-intensive services. A sufficient statistics analysis implies that US minimum wages typically remain below their optimum under the current tax and transfer system.
\end{abstract}

\thispagestyle{empty}

\newpage
\pagenumbering{arabic} 

\section{Introduction}

Governments use income taxes and transfers to redistribute to low-income individuals. However, those taxes and transfers can be distortionary, yielding an equity-efficiency tradeoff \citep{mirrlees1971exploration,piketty2013optimal}. This paper asks whether a minimum wage can relax that tradeoff and enable more efficient redistribution than income taxes and transfers alone.

Economists have long debated this question. \cite{mill1884principles} suggested that a minimum wage was the simplest way to redistribute profits to raise the incomes of low earners but 
\cite{stigler1946economics} articulated the argument that a minimum wage is inefficient relative to income-based taxes and transfers given its effects on employment. More recently, attempts to formally address this question have provided mixed answers by using frameworks that fail to incorporate empirically relevant channels through which minimum wages can perform redistribution \citep{hungerbuhler2009optimality,lee2012optimal,cahuc2014optimal,lavecchia2020minimum}. Empirically, a growing literature has found that minimum wages increase incomes of low earners with limited reductions in employment (\citealp{lee1999wage}; \citealp{autor2016contribution}; \citealp{cengiz2019effect}; \citealp{dube2019minimum}; \citealp{fortin2021labor}; \citealp{manning2021elusive}), possibly accompanied by reduced corporate profits (\citealp{draca2011minimum}; \citealp{harasztosi2019pays}; \citealp{drucker2019pays}). Yet, even if minimum wages redistribute from high-earning capitalists to low-earning workers, it remains unsettled whether such redistribution is preferred over analogous redistribution via corporate income taxes and income-based transfers alone, or to what extent there exist interactions between the minimum wage and the tax system that help governments to redistribute more efficiently when using all instruments together.

This paper proposes a novel theoretical framework to analyze the redistributive role of the minimum wage when taxes and transfers are also available to the policymaker. I characterize optimal redistribution for a government with three policy instruments: labor income taxes and transfers, corporate income taxes, and a minimum wage. The analysis illustrates the channels through which the minimum wage affects the income distribution, explicitly describing its tradeoffs and interactions with the tax system. Results are expressed as a function of reduced-form ``sufficient statistics'' that can be estimated from appropriate data. The sufficient statistics feature, which is illustrated in an empirical exercise using publicly available US data, provides a direct connection between theory and evidence in the optimal policy analysis.

The model of the labor market uses directed search and two-sided heterogeneity to allow for three potentially relevant features regarding the use of a minimum wage: the possibility of limited employment effects, wage and employment spillovers to non-minimum wage jobs, and positive firm profits. A population of workers with heterogeneous skills and costs of participating in the labor market decides whether to enter the labor market and which sort of job to seek. A corresponding population of capitalists with heterogeneous productivity decides whether to create firms, how many vacancies to post, and attaches a wage to those vacancies. In the model, minimum wage changes affect workers' application strategies which, in turn, affect the posting behavior of firms. These behavioral responses can lead to limited employment effects and spillovers to non-minimum wage jobs. The model features positive profits in equilibrium and reproduces other empirically relevant characteristics of labor markets such as wage dispersion for similar workers \citep{card2018firms} and finite firm-specific labor supply elasticities \citep{sokolova2021monopsony}. Importantly, allocations are constrained efficient in directed search models \citep{moen1997competitive,wright2019directed} so the analysis restricts the attention to the redistributive role of the minimum wage rather than its efficiency rationales usually discussed in the related literature.

I use this model of the labor market to characterize optimal policy when a utilitarian social planner chooses the minimum wage, the labor income tax system, and the corporate tax rate to maximize social welfare taking as given social preferences for redistribution. Considering first a case with no taxes and transfers, the minimum wage affects the welfare of active low- and high-skill workers through its effects on equilibrium wages and employment probabilities, and the welfare of capitalists through its effects on profits. This implies that the minimum wage can affect both the relative welfare within labor income earners and between labor and capital income earners. When income taxes and transfers are present, the optimal minimum wage depends additionally on fiscal externalities from both sides of the market. On the workers' side, changes in wages and employment induce a change in income tax collection and transfer spending. On the firms' side, the change in profits affects corporate tax revenue. The optimal minimum wage increases when the corporate tax rate is low, both because the revenue loss is smaller and the welfare gains from redistributing from capitalists to workers are higher. 

When the planner simultaneously chooses the tax system and the minimum wage, binding minimum wages can be desirable because they can make tax-based redistribution more efficient. To illustrate why, suppose that the optimal tax schedule with no minimum wage resembles an EITC to redistribute to low-wage workers. If firms internalize the effects of the EITC, they will react by lowering pre-tax wages \citep{rothstein2010eitc}. However, the minimum wage prevents firms from decreasing wages, thereby increasing the efficacy of the EITC. In other words, the minimum wage allows the transfer to low-wage workers to be partially paid by firm profits. When corporate taxes distort pre-tax profits, they cannot fully correct this incidence distortion, suggesting that combining minimum wages and corporate taxes is possibly optimal for taxing profits. The fiscal benefit of the minimum wage is accompanied by a vacancy posting distortion that can generate disemployment effects. This distortion is possibly small at low levels but growing as the minimum wage departs from the market wage, generating a tradeoff for the planner. 

Given this intuition, I show that if the effect on vacancies is negligible when the minimum wage is set at the market level, then having a binding minimum wage is desirable if the planner still wants to redistribute profits from the affected firms to other individuals in the economy after having used the corporate tax rate. This condition is met when the corporate tax is distorting enough not to distribute profits to workers at the socially desired level. I also show that, under the optimal minimum wage, the marginal tax rate on employed low-skill workers is negative if the social planner values redistribution toward them. This conclusion extends \cite{lee2012optimal} result on the complementarity between the EITC and the minimum wage using a framework with imperfect labor markets and positive firms profits.

I then study in greater depth the interaction of the minimum wage and the corporate tax rate under international tax competition. Corporate taxes may optimally be low because capital in manufacturing can flow to lower-tax countries \citep{devereux2008countries,devereux2021taxing}. A minimum wage with incidence on manufacturing could similarly cause capital to flow to lower-tax countries. However, if the minimum wage predominantly affects non-tradable services industries, then changes in the minimum wage generate little capital distortions. I show that the minimum wage can be desirable as a kind of industry-specific corporate tax if it affects profits in immobile service firms while leaving corporate taxes low for unaffected mobile firms. Formally, the welfare benefits of increasing the minimum wage are larger when the capital in non-affected firms is more mobile because the optimal corporate tax decreases, making a stronger case for a binding minimum wage that does not encourage undesirable capital flows.

To close the theoretical discussion, I present a suggestive numerical exercise to illustrate the intuitions developed throughout the policy analysis. I calibrate a restricted version of the model to study how the optimal minimum wage changes under different tax systems. Optimal minimum wages are higher when the EITC is larger and when the corporate tax rate is lower, but social welfare is larger when all instruments are used together. Then, a general finding of the analysis is that social planners should not make the tax system and the minimum wage compete for who is the most efficient redistributive policy. Instead, social planners can benefit from using all instruments together. Optimal redistribution possibly consists of a binding minimum wage, a non-trivial corporate tax rate, and a targeted EITC. 

One feature of the policy analysis is that some results can be written as a function of sufficient statistics, meaning that reduced-form causal effects can be used to assess whether minimum wages are too high or too low under a given tax and transfer system. Changes in capitalists' welfare after minimum wage changes coincide with profit effects. Changes in workers' welfare after minimum wage changes are summarized by the change in the expected utility of participating in the labor market which, under the assumptions of the model, equals the change in the average post-tax wage of labor market participants including the unemployed. This sufficient statistic, whose sign is theoretically ambiguous, aggregates all wage, employment, and participation responses that can affect workers' utility in a single elasticity.  

Given this attribute, the final part of the paper provides empirical estimates of these sufficient statistics by exploiting US state-level variation in minimum wages. I follow \cite{cengiz2019effect,cengiz2021ml} and estimate stacked event studies, where events are defined as state-level hourly minimum wage increases of at least \$0.25 (in 2016 dollars) in states where at least 2\% of the pre-event year working population earned less than the new minimum wage and where treated states did not experience other relevant minimum wage increases in the pre-event window. I identify 50 events in the period 1997-2019 for which the outcomes of interest are observed through an eight-year balanced window. 

The data consists of yearly state-level aggregates of different outcomes of interest. To measure workers' outcomes, I follow \cite{cengiz2019effect,cengiz2021ml} and use the individual-level NBER Merged Outgoing Rotation Group of the CPS to compute average pre-tax hourly wages and the Basic CPS monthly files to compute employment and participation rates. I combine these data sources to compute pre-tax versions of the worker-level sufficient statistics at the skill-by-state-by-year level, where low- and high-skill workers are defined by their college attainment. To estimate workers' side fiscal externalities, I use data on income maintenance benefits, medical benefits, and gross federal income taxes taken from the BEA regional accounts. State-level average profits are proxied by the gross operating surplus estimates from the BEA regional accounts and normalized by the average number of private establishments reported in the QCEW data files. Combining both data sources I compute average pre-tax profits per establishment at the industry-by-state-by-year level.

The empirical results imply that minimum wages have increased low-skill workers' welfare with an estimated elasticity of around 0.1. Conversely, all specifications estimate a precise zero elasticity for the high-skill workers' analog. Results for low-skill workers are stable across demographic groups, suggesting that the welfare gains are not concentrated on particular groups of ``winners''. If anything, teens (aged 16-19) and black low-skill workers seem to experience larger welfare gains from minimum wage increases, but no group experiences welfare losses. Consistent with \cite{cengiz2019effect,cengiz2021ml}, decomposing the sufficient statistic of low-skill workers across different margins shows that the entire effect is driven by an increase in the wage conditional on employment. No effect is found on hours, employment, or participation. 

The estimated elasticity of income maintenance benefits to minimum wage changes ranges between -0.31 and -0.39, suggesting sizable fiscal externalities on the workers' side of the market. This result is consistent with \cite{reich2015effects}, who find elasticities of around -0.2 for SNAP expenditures, and \cite{dube2019minimum}, who finds that after-tax income elasticities are one-third smaller than their pre-tax analogs. These fiscal externalities attenuate the welfare gains for low-skill workers. I find no effects on medical transfers and gross federal income tax liabilities, suggesting that most of the worker-level fiscal effects are mediated by targeted transfers based on pre-tax income levels.

When looking at pre-tax profits per establishment, the estimated elasticity is zero when pooling all industries. However, I find a clear negative effect in ``exposed industries'' with large numbers of minimum wage workers (retail trade, low-skill health services, food, accommodation, and social services) with an estimated elasticity of -0.36. No association is found between profits and minimum wage events in manufacturing and non-exposed service industries. The effect on the number of establishments in the exposed services is negligible, suggesting that changes in average pre-tax profits are driven by the intensive margin. The estimated effect of pre-tax profits generates a fiscal externality in terms of corporate tax revenue that is not significantly present in other sources of capital income revenue such as business and dividend income reported in the SOI state-level tables.

To interpret the estimates through the lens of the optimal policy analysis, I plug the empirical estimates into the theoretical results to assess the desirability of small minimum wage increases under different calibration choices. In the absence of preferences for redistribution, increasing the minimum wage is close to being welfare-neutral. However, increasing the minimum wage generates substantial welfare gains when distributional concerns between workers and capitalists are incorporated. Intuitively, results show that minimum wages help low-skill workers, hurt firm owners in exposed industries, and generate fiscal savings in income transfers but fiscal costs in terms of corporate tax revenue. Total gains for low-skill workers are comparable to total losses for capitalists, and the net fiscal effect is approximately zero. However, average post-tax profits per capitalist are between five and six times larger than average per capita post-tax wages of active low-skill workers. As a result, when including social preferences for redistribution, the gains for workers substantially outweigh the losses for capitalists. Hence, results suggest that, under existing tax and transfer systems, the average past increase in state-level minimum wages has increased social welfare, and increasing the minimum wage today would do so as well. 

\paragraph{Related literature and contributions} The main contribution of this paper is to provide a normative analysis of the minimum wage in a framework with taxes and transfers. Previous literature abstracts from firm profits, firm-level heterogeneity, and corporate taxation, and bases the analysis on labor market models that do not explicitly accommodate empirically relevant general equilibrium effects of the minimum wage that can dampen employment impacts. \cite{lee2012optimal} use a competitive supply-demand framework and show that the case for binding minimum wages under optimal taxes depends on labor rationing assumptions. \cite{cahuc2014optimal} contest \cite{lee2012optimal}'s result by arguing that the minimum wage cannot improve welfare on top of an optimal non-linear tax schedule even if the labor demand is modeled as a standard monopsonist. Both analyses abstract from search frictions and firm-level heterogeneity, and do not give a central role to firm profits. \cite{hungerbuhler2009optimality} and \cite{lavecchia2020minimum} consider random search models but also abstract from firm profits, restricting the role of the minimum wage under optimal taxes to solving search and matching inefficiencies. 

A complementary literature studies the welfare consequences of the minimum wage using structural models that abstract from tax-design questions. Some papers also abstract from the distributional dimension and focus on efficiency rationales motivated by labor market imperfections (\citealp{flinn2006minimum}; \citealp{wu2020partially}; \citealp{ahlfeldt2022optimal}; \citealp{drechsel2022macroeconomic}). Two recent papers give an important role to redistribution within the analysis. \cite{bergermw} propose a general equilibrium model of oligopsonistic labor markets and find that welfare improvements from minimum wage increases stem mainly from redistribution because reductions in labor market power can simultaneously generate misallocation as large-productive firms increase their market shares. Consistent with my analysis, they find that the main distributional benefits come from redistributing from capitalists to low-skill workers. \cite{hurst} develop a general equilibrium model to compare the short- and long-run impacts of the minimum wage, finding that the minimum wage encourages capital-labor substitution in the long-run. They argue this generates unintended distributional consequences on low-skill workers that are displaced by capital. Their results favor the tax system but also suggest that a moderate minimum wage can improve the efficacy of the EITC by setting a wage floor for firms. My analysis differs from theirs since I focus on the optimal policy design and its short- and medium-run -- rather than long-run -- consequences. I also explicitly discuss the interactions between the minimum wage and the corporate tax policy.\footnote{\cite{dworczak2020redistribution} indirectly analyzes the redistributive consequences of the minimum wage by studying redistribution through markets and price controls using mechanism design techniques. \cite{loertscherwage} also use mechanism design methods to explore whether minimum wages can reduce involuntary unemployment caused by market power.}

In terms of optimal redistribution, this paper adds to the literature that explores whether the combination of different instruments can improve the efficiency of the tax system. Seminal examples include \cite{diamond1971optimal}, who study the interaction between production and consumption taxes, and \cite{atkinson1976design} and \cite{saez2002desirability}, who study the interaction between income and commodity taxation. Recent papers that study the interaction between a standard income tax system and other policy instruments include \cite{gaubert2020place}, \cite{ferey}, and \cite{ferey2022sufficient}. This paper also adds to the analysis of redistributive policies in labor markets with frictions (\citealp{hungerbuhler2006optimal}; \citealp{stantcheva2014optimal}; \citealp{sleet2017taxation}; \citealp{kroft2020optimal}; \citealp{bagger2021equilibrium}; \citealp{hummel2019unemployment}; \citealp{mousavi2021optimal}; \citealp{craig2022optimal}; \citealp{doligalski2020redistribution}), and to the analysis of redistribution between capital and labor income (\citealp{atesagaoglu2021optimal}; \citealp{eeckhout2021optimal}; \citealp{hummel2020monopsony}). 

Finally, the empirical results add to a large literature that studies the effects of minimum wages on different outcomes. The workers' side results at the skill-level complement the vast literature that studies effects on wages and employment \citep{manning2021elusive}. Results on income maintenance transfers and other fiscal outcomes complement the evidence presented in \cite{reich2015effects} and \cite{dube2019minimum}. Finally, the empirical results on profits are in line with the findings of \cite{draca2011minimum}, \cite{harasztosi2019pays}, and \cite{drucker2019pays} and are, to my knowledge, the first such findings derived using US data. 

\paragraph{Structure of the paper} The rest of the paper is organized as follows. Section \ref{sec2} describes the model of the labor market. Section \ref{sec3} develops the optimal policy analysis. Section \ref{sec4} presents policy applications using a restricted version of the model. Section \ref{sec5} estimates the sufficient statistics. Section \ref{sec6} concludes.

\section{Model of the Labor Market}
\label{sec2}

This section develops a model of the labor market with positive firm profits that can accommodate limited employment effects and spillovers to non-minimum wage jobs after minimum wage increases. 

\subsection{Setup: workers and capitalists}

\paragraph{Overview} The model is static and uses directed search and two-sided heterogeneity. On one side, there is a population of workers that is heterogeneous in two dimensions: skills and costs of participating in the labor market. For simplicity, I assume workers are either low-skill or high-skill. On the other side, there is a population of capitalists with heterogeneous productivities. Labor market interactions are modeled following a directed search approach \citep{moen1997competitive}. Capitalists decide whether to create firms based on expected profits. Conditional on creating a firm, they post wages and vacancies, with all vacancies posted at a given wage forming a \textit{sub-market}.\footnote{The notion of sub-market should not be confounded with the notion of local labor market. Sub-markets only vary with wages and, in principle, all workers are equally able to apply to them. Both concepts could be closer in a more general model with multidimensional firm heterogeneity and heterogeneous application costs.} Labor markets are segmented, meaning that wages and vacancies are skill-specific. Workers observe wages and vacancies and make their labor market participation and application decisions. In equilibrium, there is a continuum of sub-markets indexed by $m$, characterized by skill-specific wages, $w_m^s$, vacancies, $V_m^s$, and applicants, $L_m^s$, with $s\in\{l,h\}$ indexing skill.

\paragraph{Matching technology} There are standard matching frictions within each sub-market. The number of matches within a sub-market is given by the matching function $\mathcal{M}^s(L_m^s, V_m^s)$, with $\mathcal{M}^s$ continuously differentiable, increasing and concave, and possessing constant returns to scale. The matching technology is allowed to be different for low- and high-skill workers \citep{berman1997help,hall2018measuring}.

Under these assumptions, the sub-market skill-specific job-finding rate can be written as
\begin{eqnarray}
p_m^s = \frac{\mathcal{M}^s(L_m^s,V_m^s)}{L_m^s} = \mathcal{M}^s(1,\theta_m^s) = p^s(\theta_m^s),
\end{eqnarray}
with $\partial p^s(\theta_m^s)/\partial \theta_m^s \equiv p_{\theta}^s>0$, where $\theta_m^s=V_m^s/L_m^s$ is the sub-market skill-specific vacancies to applicants ratio, also denoted as \textit{sub-market tightness}. Intuitively, the higher the ratio of vacancies to applicants, the more likely that an applicant will be matched with one of those vacancies. Likewise, the sub-market skill-specific job-filling rate can be written as
\begin{eqnarray}
q_m^s=\frac{\mathcal{M}^s(L_m^s,V_m^s)}{V_m^s}=\mathcal{M}^s\left(\frac{1}{\theta_m^s},1\right) = q^s(\theta_m^s),
\end{eqnarray}
with $\partial q^s(\theta_m^s)/\partial \theta_m^s\equiv q_{\theta}^s<0$. Intuitively, the lower the ratio of vacancies to applicants, the more likely that the firm will be able to fill the vacancy with a worker. Neither workers or firms internalize that their behavior affects equilibrium tightness, so they take $p_m^s$ and $q_m^s$ as given when making their decisions.

\paragraph{Workers} The population of workers is normalized to 1. The exogenous shares of low- and high-skill workers are given by $\alpha_l$ and $\alpha_h$, respectively. Conditional on skill, each worker draws a parameter $c\in \mathcal{C} = [0,C]\subset \mathbb{R}$ that represents the cost of participating in the labor market, which admits different interpretations such as search costs, disutility of (extensive margin) labor supply, or other opportunity costs of working such as home production. Let $f_s$ and $F_s$ be the skill-specific density and cumulative distributions of $c$, respectively, both of which are assumed to be smooth. 

Workers derive utility from the after-tax wage net of labor market participation costs. Since the model abstracts from intensive margin decisions, I refer to wages, incomes, or earnings indistinctly. The utility of not entering the labor market is $u_0 = y_0$, where $y_0$ is a lump-sum transfer paid by the government to non-employed individuals. $u_0$ is the same for all workers, regardless of their $(s,c)$ type. When entering the labor market, workers apply to jobs. Following \cite{moen1997competitive}, I assume that workers can apply to jobs in only one sub-market.\footnote{See \cite{kircher2009efficiency} and \cite{wolthoff2018applications} for models where workers can simultaneously apply to several sub-markets.} Conditional on employment, after-tax wages in sub-market $m$ are given by $y_m^s = w_m^s - T(w_m^s)$, where $T$ is the (possibly non-linear) income tax-schedule, with $T(0)= - y_0$. Then, the expected utility of entering the labor market for a worker of type $(s,c)$ is given by 
\begin{eqnarray}
u_1(s,c) &=& \max_m \left\{p_{m}^sy_{m}^s + (1-p_{m}^s)y_0\right\} - c, \end{eqnarray}
since workers apply to the sub-market that gives them the highest expected after-tax wage internalizing that the application ends in employment with probability $p_m^s$ and unemployment with probability $1 - p_m^s$.

Recall that $p_m^s$ depends on the mass of workers of skill $s$ that apply to jobs in sub-market $m$: given a stock of vacancies, the more workers apply, the smaller the likelihood of being employed. Then, individuals take $p_m^s$ as given but it is endogenously determined by the aggregate application behavior. This implies that, in equilibrium, all markets have the same expected after-tax wage, i.e., $p_{i}^sy_{i}^s + (1-p_{i}^s)y_0 = p_{j}^sy_{j}^s + (1-p_{j}^s)y_0 = \max_m\{p_{m}^sy_{m}^s + (1-p_{m}^s)y_0\}$, for all $i, j$; if not, workers have incentives to change their applications toward markets with higher expected values, pushing downward the job-filling probabilities and restoring the equilibrium. This means that workers face a trade-off between wages and employment probabilities because it is more difficult to get a job in sub-markets that pay higher wages.\footnote{I assume risk neutrality. Incorporating risk aversion does not affect the high-level analysis but affects the empirical approximations of the relevant objects of the optimal policy analysis. I come back to this discussion in the next section.}

In what follows, I define $U^s \equiv \max_m\{p_{m}^sy_{m}^s + (1-p_{m}^s)y_0\}$ so $u_1(s,c) = U^s - c$. The labor market participation decision is given by $l(s,c) = 1\{u_1(s,c)\geq u_0\} = 1\{U^s - y_0\geq c\}$. This implies that $l(s,c)=1$ if $c\leq U^s - y_0$, $l(s,c)=0$ otherwise. Let $L_A^s = \alpha_s\cdot \int l(s,c)dF_s(c)$ denote the mass of active workers of skill $s$, that is, the mass of workers of skill $s$ that enter the labor market. Then, $L_A^s = \alpha_s\cdot F_s(U^s-y_0)$. Inactive workers are given by $L_I = L_I^l+L_I^h = 1 - L_A^l - L_A^h$. Denote by $L^s_m$ the mass of individuals of skill $s$ applying to jobs in sub-market $m$, so $L_A^s = \int L_m^sdm$. I assume away sorting patterns based on $c$, that is, application decisions conditional on participating in the labor market are independent from $c$. 

Note that the expression $U^s = p_{m}^sy_{m}^s + (1-p_{m}^s)y_0$ implies that $\theta_m^s$ can be written as a function of $w_m^s$ and $U^s$, for all $m$ \citep{moen1997competitive}. Formally, $\theta_m^s = \theta_m^s(w_m^s,U^s)$, with $\partial\theta^s_{m}/\partial w_m^s<0$ and $\partial\theta^s_{m}/\partial U^s>0$.\footnote{Since $U^s = p^s(\theta_m^s)\cdot(w_m^s - T(w_m^s)) + (1-p^s(\theta_m^s))\cdot y_0$, then $dU^s = p_{\theta}^s\cdot d\theta_m^s\cdot y_m^s + p_m^s\cdot (1 - T'(w_m^s))\cdot dw_m^s$. Recalling that $p_{\theta}^s>0$ and asssuming $T'(w_m^s)<1$ yields the result.} This result simplifies the analysis below since it implies that, conditional on wages, equilibrium behavior can be summarized by the scalars $U^s$ without needing to characterize the continuous sequence of $\theta_m^s$.

\paragraph{Capitalists} The population of capitalists is normalized to $K$. Each capitalist draws a parameter $\psi\in\Psi = \left[\underline{\psi},\overline{\psi}\right]\subset\mathbb{R}^+$ that represents firm productivity. Let $o$ and $O$ be the density and cumulative distributions of $\psi$, respectively, both of which are assumed to be smooth. 

Capitalists observe $\psi$ and choose whether to create a firm. Firms are price-takers in the output market (with the price normalized to 1). Technology is assumed to depend on $\psi$, low- and high-skill workers, and the (flat) corporate tax rate, $t$, so a firm of productivity $\psi$ that hires $(n^l,n^h)$ workers generates revenue equal to $\phi(\psi,n^l,n^h,t)$, with $\phi$ twice differentiable, $\phi_{\psi}>0$, $\phi_{n^s}>0$ and $\phi_{n^sn^s}\leq0$, $\phi_t \leq 0$, and $\phi_{tn^s}\leq 0$, for $s\in\{l,h\}$. Allowing the revenue function to depend on $t$ accomodates, in a reduced-form fashion, the fact that corporate tax rates can distort pre-tax profits \citep{pat}.\footnote{In Appendix \ref{results}, I propose two different microfoundations of $\phi$ that generate dependence on $t$: a capital allocation problem, where capitalists have a fixed stock of capital that have to allocate between the domestic firm and an international outside option, and an effort allocation problem, where revenue also depends on the owners' effort. The same arguments may lead $\phi$ to also depend on the minimum wage. I use the capital allocation problem in a policy application in Section \ref{sec4}.}

Firms choose skill-specific wages, $w^s$, and vacancies, $v^s$, internalizing that $n^s$ is the result of the matching process. While firms take the job-filling probabilities as given, they internalize that paying higher wages increases the job-filling probabilities. In other words, the wage choice is equivalent to the sub-market choice. I rewrite job-filling probabilities as $\tilde{q}^s(w^s,U^s)=q(\theta^s(w^s,U^s))$, with $\tilde{q}^s_w = q^s_{\theta}\cdot(\partial \theta^s/\partial w^s)>0$ (since firms take $U^s$ as given), so $n^s=\tilde{q}^s(w^s,U^s)\cdot v^s$. Posting $v^s$ vacancies has a cost $\eta^s(v^s)$, with $\eta_v^s>0$ and $\eta_{vv}^s>0$. Then, pre-tax profits are given by revenue net of labor costs:
\begin{eqnarray}
\pi\left(w^l,w^h,v^l,v^h;\psi,t\right) &=& \phi\left(\psi,\tilde{q}^l(w^l,U^l)\cdot v^l,\tilde{q}^h(w^h,U^h)\cdot v^h,t\right)\nonumber\\
&& - \left(w^l\cdot\tilde{q}^l(w^l,U^l)\cdot v^l + \eta^l(v^l)\right) - \left(w^h\cdot\tilde{q}^h(w^h,U^h)\cdot v^h + \eta^h(v^h)\right).
\end{eqnarray} 
Denote the value function by $\Pi(\psi,t) = \max_{w^l,w^h,v^l,v^h}\pi\left(w^l,w^h,v^l,v^h;\psi,t\right)$. Then, after-tax profits are given by $(1-t)\cdot\Pi(\psi,t)$. 

Conditional on $\psi$, firms are homogeneous. Then, the solution to the profit maximizing problem can be characterized by functions $w^s(\psi)$ and $v^s(\psi)$. Appendix \ref{results} derives the first-order conditions and shows that dispersion in productivities leads to dispersion in wages, with wages \textit{marked down} relative to the marginal productivities.\footnote{The effective wage markdown is governed by the wage-dependent job-filling probabilities -- which emulate firm-specific labor supply elasticities -- and the vacancy creation costs.} $m$ indexes sub-markets as well as the productivity levels of capitalists that create firms, so $w_m^s = w^s(\psi_m)$, $v_m^s = v^s(\psi_m)$, and $V_m^s = K\cdot v^s(\psi_m)\cdot o(\psi_m)$. 

Capitalists pay a fixed cost, $\xi$, to create firms, and receive the lump-sum transfer, $y_0$, when remaining inactive, so they create firms when $(1-t)\cdot\Pi(\psi,t)\geq \xi + y_0$. Since profits are increasing in productivity, the entry rule defines a productivity threshold, $\psi^*$, implicitly determined by $(1-t)\cdot\Pi(\psi^*,t)=\xi + y_0$ such that capitalists create firms only if $\psi\geq \psi^*$. Then, the mass of active capitalists is given by $K_A = K\cdot\left(1 - O(\psi^*)\right)$. The mass of inactive capitalists, $K_I$, is given by $K_I = K\cdot O(\psi^*)$, with $K_A + K_I = K$. 

\subsection{Discussion}

Before introducing a minimum wage to the model, I discuss some features and limitations of the proposed framework. This is a non-exhaustive discussion which is continued in Appendix \ref{results}.

\paragraph{Directed search} Directed search models generate efficient outcomes in terms of search and posting behavior \citep{moen1997competitive,wright2019directed}. That is, these models don't exhibit inefficient mixes of applicants and vacancies as can happen in random search models (\citealp{hosios1990efficiency,mangin2021efficiency}). In Appendix \ref{results}, I show that the proposed model maintains this property, which I interpret a feature rather than a design flaw as it fosters a focus on the redistributive role of the minimum wage rather than on its efficiency rationales (e.g., \citealp{burdett1998wage}; \citealp{acemoglu2001good}).\footnote{This result can be thought of as an extension of \cite{moen1997competitive} result to a setting with ex-ante firm-level heterogeneity and positive profits. I show that not only posting is efficient, but also the entry thresholds at the worker- and firm-levels.}

\paragraph{Monopsony power} While search and posting behavior is efficient, the model admits monopsony power through wage-dependent job-filling probabilities that have a similar flavor to the standard monopsony intuition of upward-slopping firm-specific labor supply curves \citep{robinson1933economics,card2018firms} supported by recent empirical evidence \citep{staiger2010there,azar2019estimating,dubemonopsony,sokolova2021monopsony,bassier2020monopsony}. Firms internalize that paying higher wages lead to more applicants, so wages are \textit{marked down} relative to marginal productivities.\footnote{Appendix \ref{results} shows that the standard markdown equation can be derived from the firm's first order conditions.} 

\paragraph{Low-wage labor markets} The equilibrium of the model is consistent with other stylized facts of low-wage labor markets. The model features wage dispersion for similar workers \citep{card2018firms}, wage posting rather than bargaining, which has been found to be more relevant for low-wage jobs \citep{hall2012evidence,caldwell2019outside,lachowska2021workers}, and can rationalize bunching in the wage distribution at the minimum wage \citep{cengiz2019effect}.

\paragraph{Restricted heterogeneity} One limitation of the model is that the dimensions of worker- and firm-level heterogeneity are limited. On the workers' side, the model assumes that all workers of the same skill type get the same expected utility. \cite{hurst} suggest within-skill heterogeneity can mask important distributional effects if there are winners and losers within skill-type after minimum wage changes. Extending the model in this direction would imply that $U^s$ -- which will play an important role in the optimal policy analysis -- can be different for different groups of low- and high-skill workers.\footnote{Formally, consider an additional variable, $\widetilde{s}$ such that $U^{s,\widetilde{s}_1}\neq U^{s,\widetilde{s}_2}$. This could be the case if, for example, workers of type $(s,\widetilde{s}_1)$ can apply to a different pool of firms than workers of type $(s,\widetilde{s}_2)$.} I come back to this discussion in Section \ref{sec5} where I empirically explore for heterogeneities in the estimated welfare changes within skill groups.  

On the firm side, one-dimensional heterogeneity is a convenient simplification. Extending the model to multidimensional heterogeneity, $(\psi,\widetilde{\psi})$, is straightforward so long as workers do not have preferences for these attributes. This could accommodate, for example, variation in factor shares. In this setting, the problem's solution would be given by wage and vacancy functions $w^s(\psi,\widetilde{\psi})$ and $v^s(\psi,\widetilde{\psi})$, and by a set of conditional productivity thresholds, $\psi^*(\widetilde{\psi})$. Such an extension adds little intuition to the general policy analysis while introducing more complicated notation. The above argument requires workers to not have preferences over $\widetilde{\psi}$ beyond its effect on wages and vacancies. Hence, the simple extension to multidimensional heterogeneity does not apply to non-wage amenities.\footnote{For evidence on their importance, see \cite{bonhomme2009pervasive}, \cite{mas2017valuing}, \cite{maestas2018value}, \cite{sorkin2018ranking}, \cite{taber2020estimation}, \cite{jager2021worker}, \cite{le2021gender}, \cite{lindenlaub2021worker}, \cite{sockin2021show}, \cite{lamadon2019imperfect}, and \cite{roussille2022bidding}.} Amenities can affect the policy analysis for two reasons. First, if workers rank firms using a composite index of expected wages and amenities and the latter are not taxed, then the tax system can distort workers' preferences \citep{lamadon2019imperfect}. Second, if amenities are endogenous, minimum wage increases may induce firms to worsen the non-wage attributes of the job \citep{clemens2018minimum,clemens2021firms}. Such effects could attenuate potential welfare gains to workers after minimum wage hikes.

\subsection{Introducing a minimum wage}

I introduce a minimum wage, $\overline{w}$, to explore how the predictions of the model speak to the related empirical literature. I separately explore the effects on workers and capitalists decisions. 

\paragraph{Low-skill workers} In equilibrium, $U^l = p^l(\theta_m^l)\cdot y^l_m + (1-p^l(\theta_m^l))\cdot y_0$, for all sub-markets $m$. Let $i$ be the sub-market constrained by the minimum wage, so $w_i^l = \overline{w}$. Differentiating yields
\begin{eqnarray}
\frac{dU^l}{d\overline{w}} &=& p_{\theta}^l\cdot\frac{d\theta^l_i}{d\overline{w}}\cdot(\overline{w} - T(\overline{w}) - y_0) + p^l(\theta_i^l)\cdot (1-T'(\overline{w})). \label{ls_w}
\end{eqnarray}
Since $p^s(\theta_i^l)>0$, and assuming $T'(\overline{w})<1$, $dU^l/d\overline{w} = d\theta^l_i/d\overline{w} = 0$ is not a feasible solution to equation \eqref{ls_w}. This implies that changes in $\overline{w}$ necessarily affect the equilibrium values of $U^l$, $\theta_i^l$, or both. 

Intuitively, an increase in the minimum wage mechanically makes minimum wage jobs more attractive for low-skill workers. This effect is captured by $p^l(\theta_i^l)\cdot(1-T'(\overline{w}))$: the increase in the attractiveness of this sub-market is the net-of-tax gain conditional on working, $1 - T'(\overline{w})$, times the employment probability, $p^l(\theta_i^l)$. This attracts new applicants toward minimum-wage sub-markets (from other sub-markets and/or from outside the labor force), thus pushing $\theta_i^l$ downwards until the across sub-market equilibrium is restored. This decreases the employment probability in sub-market $i$, whose effect is captured by the change in the employment probability, $p_{\theta}^l\cdot(d\theta^l_i/d\overline{w})$, times after-tax income conditional on employment, $\overline{w} - T(\overline{w}) - y_0$. These two effects capture the standard effects on wages and employment debated in the minimum wage literature. How these effects balance determine the overall impact on expected utility.

This tradeoff captures the essence of the general equilibrium effects of the model: the initial change in applications toward minimum-wage jobs triggers a sequence of reactions that reconfigure labor market outcomes. 
Changes in $\overline{w}$ also affect the equilibrium of unconstrained low-skill sub-markets. To see this, let $j$ be a sub-market that is not constrained by the minimum wage, so $w_j^l>\overline{w}$. Differentiating yields
\begin{eqnarray}
\frac{dU^l}{d\overline{w}} &=& p_{\theta}^l\cdot\frac{d\theta^l_j}{d\overline{w}}\cdot (w_j^l - T(w_j^l) -y_0) + p^l(\theta_j^l)\cdot (1-T'(w_j^l))\cdot \frac{dw_j^l}{d\overline{w}}. \label{ls_w2}
\end{eqnarray}
Equation \eqref{ls_w} suggests that the left-hand-side of equation \eqref{ls_w2} is unlikely to be zero, implying that $\theta^l_j$ or $w_j^l$ or both are possibly affected by changes in the minimum wage. There are two forces that mediate this spillover. First, the change in applicant flows between sub-markets and from in and out of the labor force affects the employment probabilities of all sub-markets until the equilibrium condition of equal expected utilities is restored. This effect is captured by the first term of equation \eqref{ls_w2}. Second, firms can also respond to changes in applicants. The potential wage response is captured in the second term of equation \eqref{ls_w2} and changes in vacancy posting implicitly enter the terms $d\theta^l_m/d\overline{w}$ of equations \eqref{ls_w} and \eqref{ls_w2}.

Changes in $U^l$ also induce changes in labor market participation, since $L_A^l = \alpha_l\cdot F_l(U^l - y_0)$, so $dL_A^l/d\overline{w} = \alpha_l\cdot f_l(U^l - y_0)\cdot \left(dU^l/d\overline{w}\right)$. Then, whenever $dU^l/d\overline{w}>0$, minimum wage hikes increase labor market participation. The behavioral response is scaled by $f_l(U^l)$, which may be negligible. This effect may result in positive impacts on expected utilities with little participation effects at the aggregate level.

\paragraph{High-skill workers} If $\min_m\{w^h_m\}>\overline{w}$, equilibrium effects for high-skill workers take the form of equation \eqref{ls_w2}. Then, the question is whether there are equilibrium forces that rule out solutions of the form $dU^h/d\overline{w} = d\theta^h_i/d\overline{w} =dw_i^h/d\overline{w} = 0$. In this model, effects in high-skill sub-markets are mediated by the production function, since demand for high-skill workers depends on low-skill workers through $\phi$. Then, this model may induce within-firm spillovers explained by a technological force. Changes in low-skill markets affect high-skill posting, thus affecting high-skill workers application decisions. 

\paragraph{Firms} Workers react to changes in the minimum wage by changing their application strategies and extensive margin decisions, thus affecting sub-markets' tightness and the profit maximization problem of the firms. Appendix \ref{results} provides expressions for the effects of minimum wage changes on firms' outcomes, which are analytically complex given the potential non-linearities of the matching, production, and vacancy cost functions. In what follows, I describe the main intuitions behind the analysis.

Firms for which the minimum wage binds optimize low-skill vacancies and high-skill wages and vacancies taking low-skill wages as given. The effect of the minimum wage on low-skill vacancy posting is ambiguous. On one hand, an increase in the minimum wage induces a mechanical increase in labor costs, decreasing the expected value of posting a low-skill vacancy. However, if sub-market tightness decreases given the increase in applicants, job-filling probabilities increase. This effect increases the expected value of posting a low-skill vacancy. Within the minimum wage sub-market, the net effect on vacancies is more likely to be negative the lower the productivity.\footnote{In the model, it is possible to have productivity dispersion across firms that pay the minimum wage. Concretely, all firms whose market low-skill wage is lower than $\overline{w}$ bunch at $\overline{w}$ conditional on entry.} That is, the least productive firms among the constrained group reduce their size after increases in the minimum wage while the most productive firms within this group could have null or positive firm-specific employment effects. 

Firms for which the minimum wage does not bind also react by adapting their posted wages and vacancies to changes in their relevant sub-market tightness. The analytical expression for the wage spillover is difficult to sign and interpret but directly depends on the change in sub-market tightness (see equation \eqref{wage_sp} of Appendix \ref{results}). Since wages and vacancies are positively correlated at the firm and skill level, if wage spillovers are positive, then unconstrained firms also post more vacancies and, therefore, increase their size. Therefore, the model has potential to generate reallocation effects.

Profits are also affected by minimum wage changes. Firms for which the minimum wage binds face a reduction in profits regardless of the employment effect. This in turn leads marginal firms to exit the market after increases in the minimum wage. Firms for which the minimum wage does not bind may also have their profits affected given the change in the equilibrium job-filling probabilities.

\paragraph{Relation to empirical literature} The purposely imposed tractability needed for the optimal policy analysis puts limits on the ability of the model to fully rationalize observed labor market reactions to minimum wage changes.\footnote{For structural models with richer levels of heterogeneity and flexibility, see \cite{haanwinckel2018supply}, \cite{ahlfeldt2022optimal}, \cite{bergermw}, \cite{drechsel2022macroeconomic}, \cite{engbom2018earnings}, and \cite{hurst}.} However, the proposed framework generates predictions consistent with the empirical literature that favor its suitability for the policy analysis.

One systematic finding of the empirical literature is that minimum wage hikes generate positive wage effects with limited -- or \textit{elusive} -- disemployment effects (see \citealp{manning2021elusive} for a recent review). This empirical fact is inconsistent with a perfectly competitive model of the labor market, and is difficult to rationalize with a random search framework since it requires an implausibly large labor force participation response that is at odds with the empirical literature \citep{cengiz2021ml}. The proposed framework can rationalize positive wage effects with limited employment and participation effects through the equilibrium changes in applications. When the minimum wage increases, constrained firms face a mechanical increase in their labor costs. However, job applicants reallocate applications toward these jobs, increasing the expected value of posting vacancies. This effect attenuates the negative shock in labor costs. The reorganization of applications within the mass of active workers can mediate this result when the size of the density at the margin of indifference is low enough to prevent important participation responses.

The empirical literature also finds that minimum wages generate spillovers to non-minimum wage jobs in terms of wages and employment both within and between firms \citep{cengiz2019effect,derenoncourtspil,dustmann2019reallocation,forsythe2022effect,giupponi2018changing}, and have negative effects on firm profits \citep{draca2011minimum,harasztosi2019pays,drucker2019pays}. The model incorporates both sets of predictions. The same responses in applications that dampen the employment effects generate spillovers to firms that pay higher wages through changes in their sub-markets' tightness, and to high-skill workers through technological restrictions embedded in the production function.\footnote{The model fails to accommodate other relevant effects of the minimum wage documented in the empirical literature, namely the passthrough of minimum wages to output prices \citep{macurdy2015effective,allegretto2016local,harasztosi2019pays,leung2021minimum,ashenfelter2021wages,renkin2020pass} and their effects on worker- and firm-level productivity \citep{riley2017raising,mayneris2018improving,coviello2020minimum,ruffini,emanuel2022firm,ku2020does}. Appendix \ref{results} argues that these pieces are unlikely to play a central role in the optimal policy analysis.}

\section{Optimal Policy Analysis}
\label{sec3}

This section uses the model of the labor market to characterize optimal redistribution for a social planner with three policy instruments: labor income taxes, a corporate tax rate, and a minimum wage.

\subsection{Social planner's problem}

The notion of \textit{optimal policy} refers to policy parameters that maximize a social welfare function. Following related literature (\citealp{kroft2020optimal}; \citealp{lavecchia2020minimum}), the social planner is assumed to be utilitarian and maximize the sum of expected utilities. I assume the social planner does not observe $c$ and $\psi$ and, therefore, constrains the policy choice to second-best incentive-compatible policy schemes. 

The social welfare function is given by
\begin{eqnarray}
SW(\overline{w},T,t) &=& \left(L_I^l+L_I^h + K_I\right)\cdot G(y_0) + \alpha_l\cdot\int_0^{U^l-y_0}G(U^l-c)dF_l(c)\nonumber\\
&&+\alpha_h\cdot\int_0^{U^h-y_0}G(U^h-c)dF_h(c)+ K\cdot\int_{\psi^*}^{\overline{\psi}}G\left((1-t)\cdot \Pi(\psi,t) - \xi\right)dO(\psi), \label{SW_tax}
\end{eqnarray}
where $(\overline{w},T,t)$ are the policy parameters -- the minimum wage, the (possibly non-linear) income tax schedule, and the flat corporate tax rate -- and $G$ is an increasing and concave function that accounts for preferences for redistribution. $G$ induces curvature to the individual money-metric utilities, thus allowing social gains from redistributing from high- to low-utility individuals. The incentive compatibility constraints are included in the limits of integration since the planner internalizes that the policy parameters affect the participation decisions through $U^l$, $U^h$, and $\psi^*$. The first term of equation \eqref{SW_tax} accounts for the utility of inactive workers and inactive capitalists who get income equal to $y_0 = -T(0)$. The second and third terms account for the expected utility of low- and high-skill workers that enter the labor market, also referred to as active workers. Finally, the last term accounts for the utility of active capitalists.\footnote{The average expected utility of active workers of skill $s$ is $\int_0^{U^s-y_0}G(U^s-c)d\tilde{F}_s(c)$, where $\tilde{F}_s(c) = F_s(c)/F_s(U^s - y_0)$. Then, total expected utility is given by $ L_A^s\cdot \int_0^{U^s-y_0}G(U^s-c)d\tilde{F}_s(c)$, which yields the expressions above noting that $L_A^s = \alpha_s\cdot F(U^s-y_0)$. The average utility of capitalists is $\int_{\psi^*}^{\overline{\psi}}G((1-t)\cdot\Pi(\psi,t)-\xi)d\tilde{O}(\psi)$, with $\tilde{O}(\psi) = O(\psi)/(1-O(\psi^*))$. Their total utility is therefore $K_A\cdot\int_{\psi^*}^{\overline{\psi}}G((1-t)\cdot\Pi(\psi,t)-\xi)d\tilde{O}(\psi)$, which yields the expression above noting that $K_A = K\cdot (1-O(\psi^*))$.}

Assuming no exogenous spending requirement, the planner's budget constraint is given by
\begin{eqnarray}
\left(L_I^l+L_I^h + K_I + \rho^l\cdot L_A^l + \rho^h\cdot L_A^h\right)\cdot y_0 &\leq & \int \left(E_m^l\cdot T(w_m^l)+E_m^h\cdot T(w_m^h)\right) dm\nonumber\\
&&+ t\cdot K\cdot\int_{\psi^*}^{\overline{\psi}}\Pi(\psi,t)dO(\psi), \label{BC}
\end{eqnarray}
where $E_m^s = p_m^s\cdot L_m^s$ is the mass of employed workers of skill $s$ in sub-market $m$ and $\rho^s$ is the skill-specific unemployment rate given by $\left(L_A^s - \int E_m^sdm\right)/L_A^s$. The budget constraint establishes that the transfer paid to individuals with no market income must be funded by the tax collection on employed workers and active capitalists. 

\paragraph{Understanding $G$} To better understand the role of $G$, define the average social marginal welfare weights (SMWWs) of inactive workers, active workers of skill type $s$, and active capitalists of type $\psi$ as
\begin{eqnarray}
g_0 = \frac{G'(y_0)}{\gamma},\qquad g_1^s = \frac{\alpha_s\cdot \int_0^{U^s-y_0}G'(U^s-c)dF_s(c)}{\gamma\cdot L_A^s},\qquad g_{\psi} = \frac{G'((1-t)\cdot\Pi(\psi,t)-\xi)}{\gamma}, \label{smwws}
\end{eqnarray}
where $\gamma>0$ is the budget constraint multiplier. Average SMWWs represent the social value of the marginal utility of consumption normalized by the social cost of raising funds, thus measuring the social value of redistributing one dollar uniformly across a group of individuals. When the SMWWs are above one, the planner benefits from redistribution since the gains outweight the distortions induced by the increase in revenues. A given value of $g_X$ indicates that the government is indifferent between $g_X$ more dollars of public funds and 1 dollar of additional consumption of individuals of group $X$ \citep{saez2001using}.  

The utilitarian assumption used in equation \eqref{SW_tax} implies that the SMWWs are endogenous to final allocations (and, therefore, to the policy parameters) since social welfare only depends on the concave transformation of individual money-metric utilities. Alternative formulations of the problem can generate different microfoundations for the SMWWs, for example, through exogenous Pareto weights or generalized SMWWs \citep{saez2016generalized}. More generally, SMWWs are sufficient statistics for preferences for redistribution since their values inform the willingness to transfer incomes between different groups of individuals. I return to this when discussing the results of the optimal policy analysis.

\paragraph{Rationing assumptions} Since the social planner cares about expected utilities, rationing assumptions conditional on entering the labor market do not affect the welfare analysis: all workers have equal ex-ante expected utilities, so the allocation to jobs and unemployment after policy changes does not condition the planner's problem. By contrast, rationing assumptions are central in optimal policy analyses based on competitive labor markets \citep{lee2012optimal}. Rationing matters if sorting to firms conditional on participation depends on $c$, and would play a role if adding additional layers of worker-level heterogeneity imply that some groups are more likely to work at low-wage firms or to be unemployed. This would affect the analysis since the presence of winners and losers within skill-group may distort the assessment of the distributional effects of minimum wage increases \citep{hurst}. I return to this question in Section \ref{sec5} when testing for heterogeneities in the empirical estimation of the worker-level sufficient statistics.

\subsection{Case with no taxes} 

I now proceed to analyze the redistributive properties of the minimum wage using the framework described above. I start abstracting from the tax system to isolate the effects on the relative tradeoff between low-skill workers, high-skill workers, and capitalists. Taxes and transfers are introduced in the next subsection.\\

\noindent\textsc{Proposition I}: \textit{In the absence of taxes, increasing the minimum wage is welfare improving if}
\begin{eqnarray}
\frac{dU^l}{d\overline{w}}\cdot L_A^l \cdot g_1^l +\frac{dU^h}{d\overline{w}}\cdot L_A^h \cdot g_1^h+ K\cdot\int_{\psi^*}^{\overline{\psi}}g_{\psi} \frac{d\Pi(\psi)}{d\overline{w}}dO(\psi) &>& 0.  \label{prop1_main}
\end{eqnarray}
\noindent\textit{Proof: See Appendix \ref{proofs}.}\\

Proposition I shows that a small increase of the minimum wage can affect the welfare of active low-skill workers (first term), active high-skill workers (second term), and active capitalists (third term). Depending on the change in utility for the different groups ($dU^s/d\overline{w}$ and $d\Pi(\psi)/d\overline{w}$), the social value of those changes ($g_1^s$ and $g_{\psi}$), and the size of the groups ($L_A^s$ and $K\cdot o(\psi)$), increasing the minimum wage may be desirable or not for the social planner.\footnote{While changes in $U^s$ and $\psi^*$ also affect extensive margin decisions, those margins do not induce first-order welfare effects because marginal workers and capitalists are initially indifferent between states.} This implies that the minimum wage can affect both the relative welfare within labor income earners and between labor and capital income earners.

\paragraph{Welfare weights} To understand why the analysis emphasizes the distributional effects of the minimum wage, consider a situation where $g_1^l = g_1^h = g_{\psi} = 1$, for all $\psi$. Then, the planner's problem is reduced to assessing changes in total output. The analysis changes when SMWWs are unrestricted. Total output could decrease after minimum wage increases, but if the gains for winners are more socially valuable than the losses for losers, then increasing the minimum wage can be welfare-improving. For example, if the social planner does not care about the utility of capitalists and high-skill workers, there could be scope to increase the minimum wage if the utility of low-skill workers increases after the policy change. The utilitarian assumption implies that SMWWs are endogenous to final allocations, so they are inversely proportional to after-tax incomes. The steepness of the relationship depends on the concavity of $G$. 

\paragraph{Sufficient statistics} Given values for the SMWWs, if the sizes of the groups are observed, taking equation \eqref{prop1_main} to the data requires values for $dU^s/d\overline{w}$ and $d\Pi(\psi)/d\overline{w}$. Reduced-form estimates of these elasticities facilitate the quantitative assessment of Proposition I without needing to impose structural restrictions on the primitives of the model of the labor market. That is, empirical counterparts of $dU^s/d\overline{w}$ and $d\Pi(\psi)/d\overline{w}$ work as sufficient statistics \citep{chetty2009sufficient,kleven} for assessing the welfare implications of minimum wage changes.

Profits are, in principle observable, so it is feasible to have reduced-form estimates of $d\Pi(\psi)/d\overline{w}$. Regarding $U^s$, recall that, in the absence of taxes, $U^s = p_m\cdot w_m$. Multiplying both sides by the sub-market mass of applicants, $L_m^s$, and integrating over $m$, yields
\begin{eqnarray}
U^s = \frac{\int E_m^sw_m^sdm}{L_A^s} = (1 - \rho^s)\cdot\mathbb{E}_m[w_m^s] + \rho^s\cdot 0 , \label{suffstat}
\end{eqnarray}
where $\rho^s$ is the skill-specific unemployment rate and $\mathbb{E}_m[w_m^s] = \int \nu_m^s w_m^sdm$, with $\nu_m^s = E_m^s/\int E_m^s dm$, is the average wage of employed workers. This implies that $U^s$ is equal to the average wage of active workers including the unemployed. In the case with taxes, $U^s$ is equal to the average pre-tax wage of active workers including the unemployed net of the their average tax liabilities.\footnote{Recall that, in the case with taxes, $U^s = p_m^s\cdot y_m^s + (1-p_m^s)\cdot y_0$. Multiplying both sides by the sub-market mass of applicants, $L_m^s$, and integrating over $m$, gives
\begin{eqnarray}
U^s = \frac{\int E_m^s(w_m^s-T(w_m^s) - y_0)dm}{L_A^s} + y_0 = \frac{\int E_m^sw_m^sdm}{L_A^s} - \frac{\int E_m^s(T(w_m^s) + y_0)dm}{L_A^s} + y_0,
\end{eqnarray}
where $E_m^s = p_m^s\cdot L_m^s$. If the tax schedule is constant, then
\begin{eqnarray}
\frac{dU^s}{d\overline{w}} = \frac{d}{d\overline{w}}\left(\frac{\int E_m^sw_m^sdm}{L_A^s}\right) - \frac{d}{d\overline{w}}\left(\frac{\int E_m^s(T(w_m^s) + y_0)dm}{L_A^s}\right). \label{suffstat_2}
\end{eqnarray}
The first term represents the change in the average pre-tax wage among active workers (see equation \eqref{suffstat}). The second term represents the change in average tax liabilities net of transfers among active workers.} In both cases, $U^s$ can be computed using data on wages, tax liabilities, employment and participation rates. Then, $dU^s/d\overline{w}$ can be estimated to quantitatively assess equation \eqref{prop1_main}. Section \ref{sec5} illustrates this exercise.

Two things are worth discussing about the sufficient statistic for workers, $dU^s/d\overline{w}$. First, $dU^s/d\overline{w}$ captures all general equilibrium effects that affect workers' utility, including effects on wages, employment, and participation. There is an unsettled discussion in the public debate about the appropriate way of weighting these different effects. The proposed framework offers an avenue for aggregating them into a single elasticity.\footnote{While the sign of $dU^s/d\overline{w}$ is in principle ambiguous, it is not determined by the sign of the employment effects. Appendix \ref{results} shows the disemployment effects that can be tolerated for the minimum wage to increase average workers' welfare given positive wage effects. If employment and wage effects are positive, welfare effects on workers are unambiguously positive.} Second, equation \eqref{suffstat} relies on the risk-neutrality assumption made in Section \ref{sec2}. If workers are risk-averse, then Proposition I remains valid but $U^s$ no longer equals the average wage among active workers including the unemployed, so it cannot be estimated without further assumptions. One way to assess the concerns of using the risk-neutral sufficient statistic is to decompose the empirical estimate across the different margins. If changes in employment are negligible relative to changes in wages, then the risk-neutrality assumption should not have first-order effects on the interpretation of the estimated elasticities. I come back to this discussion in Section \ref{sec5}. 

\subsection{Case with taxes}

The case without taxes informs about the direct welfare effects of the minimum wage. However, in the presence of taxes, changes in labor market outcomes and profits affect tax collection and transfer spending. These fiscal externalities matter for assessing whether increasing the minimum wage is desirable. 

\paragraph{Fixed taxes} I first consider a case where the social planner takes the tax system as given and chooses $\overline{w}$ to maximize equation \eqref{SW_tax}. This extension characterizes the mechanical interactions between the minimum wage and the tax system. When unmodeled constraints restrict the scope for simultaneous tax reforms, this case may be the policy-relevant scenario for assessing the desirability of minimum wage reforms.\\

\noindent\textsc{Proposition II}: \textit{If taxes are fixed, increasing the minimum wage is welfare improving if}
\begin{eqnarray}
\frac{dU^l}{d\overline{w}}\cdot L_A^l\cdot g_1^l+\frac{dU^h}{d\overline{w}}\cdot L_A^h\cdot g_1^h+ K\cdot (1-t)\cdot \int_{\psi^*}^{\overline{\psi}}g_{\psi}\frac{d\Pi(\psi,t)}{d\overline{w}}dO(\psi)&&\nonumber \\
+\int\left(\frac{dE_m^l}{d\overline{w}}\left(T(w_m^l) + y_0\right) + E_m^lT'(w_m^l)\frac{dw_m^l}{d\overline{w}}\right)dm&&\nonumber \\ 
+ \int\left(\frac{dE_m^h}{d\overline{w}}\left(T(w_m^h) + y_0\right) + E_m^hT'(w_m^h)\frac{dw_m^h}{d\overline{w}}\right)dm&&\nonumber\\
 + t\cdot K\cdot \int_{\psi^*}^{\overline{\psi}}\frac{d\Pi(\psi,t)}{d\overline{w}}dO(\psi) - \frac{dK_I}{d\overline{w}}\cdot\left(t\cdot \Pi(\psi^*,t) + y_0\right) &>& 0.\label{prop2}
\end{eqnarray}
\noindent\textit{Proof: See Appendix \ref{proofs}.}\\

The first line of Proposition II reproduces the welfare tradeoff described in Proposition I. The second to fourth lines summarize the fiscal externalities on both sides of the market. These fiscal externalities matter for the analysis since they either relax or restrict the planner's budget constraint, consequently relaxing or restricting the redistribution already done by the existing tax system.

The second line describes the fiscal externalities on low-skill labor markets. The first term shows that, if low-skill employment increases, there is an increase in tax collection (or expenditure if there are transfers to workers), $T(w_m^l)$, and a decrease in transfers paid to unemployed individuals, $y_0$. The opposite happens when employment decreases. The second term shows that if the wages of employed workers change, income tax collection changes according to the shape of the income tax schedule, $T'(w_m^l)$. The third line represents the same effects but for high-skill labor markets. 

The fourth line describes the fiscal externalities on the capitalists' side. The first term shows that changes in profits affect the corporate tax revenue. If profits decrease, the social planner collects less revenue. The second term shows that firms that exit the market generate a negative fiscal externality since they switch from paying taxes to receiving a transfer. Both effects are increasing in the corporate tax rate: the larger $t$, the larger the revenue loss produced by smaller profits and extensive margin responses.

Firm-level fiscal externalities seem particularly relevant in the current state of international tax competition \citep{devereux2008countries,devereux2021taxing}. Under international capital mobility, it may be difficult to enforce large corporate tax rates because capital can fly to low-tax countries. If corporate taxes are low, then the rationale for using the minimum wage becomes stronger. One concern with this argument is that the same reasons that limit corporate tax rates could apply to the minimum wage: international capital could also react to minimum wage changes. In Section \ref{sec5} I document that the profit effects are concentrated in labor-intensive industries whose capital is presumably less mobile relative to other industries. By contrast, the effects of corporate tax changes on pre-tax profits seem to be concentrated in capital-intensive industries \citep{pat}. This suggests that the economic reasons that push corporate tax rates down are not extendable to minimum wages. I formalize this intuition in Section \ref{sec4}.

\paragraph{Optimal taxes} The previous analysis illustrates the mechanical interaction between the minimum wage and the tax schedule but does not answer if both policies are desirable at an hypothetical joint optimum. The following proposition explores the desirability of the minimum wage when the social planner jointly optimizes the tax system and the minimum wage. For analytical simplicity, I assume that either $\max_i w_i^l < \min_j w_j^h$, or that the social planner can implement skill-specific income tax schedules.\footnote{This allows me to solve the planner's problem doing pointwise maximization. These assumptions increase the attractiveness of the tax system, making more restrictive the case for a binding minimum wage.}\\

\noindent\textsc{Proposition III}: \textit{If taxes are optimal, increasing the minimum wage is welfare improving if}
\begin{eqnarray}
\frac{\partial U^l}{\partial\overline{w}}\cdot L_A^l\cdot g_1^l+\frac{\partial U^h}{\partial\overline{w}}\cdot L_A^h\cdot g_1^h+ K\cdot (1-t)\cdot \int_{\psi^*}^{\overline{\psi}}g_{\psi}\frac{\partial \Pi(\psi,t)}{\partial\overline{w}}dO(\psi) &&\nonumber \\
+\int\left(\frac{\partial E_m^l}{\partial\overline{w}}\left(T(w_m^l) + y_0\right) + E_m^l\frac{\partial w_m^l}{\partial\overline{w}}\right)dm &&\nonumber \\ 
+ \int\left(\frac{\partial E_m^h}{\partial\overline{w}}\left(T(w_m^h) + y_0\right) + E_m^h\frac{\partial w_m^h}{\partial \overline{w}}\right)dm &&\nonumber\\
 + t\cdot K\cdot \int_{\psi^*}^{\overline{\psi}}\frac{\partial \Pi(\psi,t)}{\partial\overline{w}}dO(\psi) - \frac{\partial K_I}{\partial\overline{w}}\cdot\left(t\cdot \Pi(\psi^*,t) + y_0\right) &>0&.\label{prop3}
\end{eqnarray}
\noindent\textit{Furthermore, at the joint optimum: (i) the SMMWs of inactive individuals, active low-skill workers, and active high-skill workers average to 1, and (ii) the average SMMW among active capitalists is below 1.}\\
\noindent\textit{Proof: See Appendix \ref{proofs}.}\\

At a high-level, Proposition III reproduces the same intuition as Proposition II: the desirability of the minimum wage depends on both the effects on the relative welfare of active low-skill workers, active high-skill workers, and active capitalists, and on the fiscal externalities generated on labor markets and profits. However, when taxes are optimized together with the minimum wage, how the minimum wage affects welfare and generates fiscal effects changes. This is reflected in two important differences between equations \eqref{prop2} and \eqref{prop3} that illustrate the forces that play a role in the joint optimum. 

First, all relevant elasticities are \textit{micro} rather than \textit{macro} elasticities \citep{landais2018macroeconomica,landais2018macroeconomicb,kroft2020optimal,lavecchia2020minimum}, which I denote by partial derivatives. Macro elasticities (Propositions I and II) internalize all general equilibrium effects of the minimum wage, while micro elasticities (Proposition III) mute some of these effects because, at the joint optimum, the minimum wage moves in tandem with taxes. Recall that $U^s = p_m^s\cdot y_m^s+(1-p_m^s)\cdot y_0 \equiv p_m^s\cdot \Delta y_m^s + y_0$, with $\Delta y_m^s = y_m^s - y_0$. When taxes are fixed, both $\Delta y_m^s$ and $p_m^s$ can react to minimum wage changes. However, at the joint optimum, an increase in the minimum wage is accompanied by a change in tax-based subsidies to low-skill workers, possibly leading consumption fixed. Then, the minimum wage directly affects workers' welfare mainly through potential changes in the employment probabilities driven by changes in vacancy posting. This logic also applies to the effects on employment and profits.\footnote{The direct welfare effects on workers are proportional to the (presumably negative) employment effects. If $U^s = p_m^s\cdot \Delta y_m^s + y_0$, multiplying by $L_m^s$ and integrating over $m$ yields $(U^s - y_0)\cdot L_A^s = \int E_m^s \cdot \Delta y_m^s dm$. Then, if $\Delta y_m^s$ is fixed,
\begin{eqnarray}
\frac{\partial U^s}{\partial \overline{w}}\cdot\left(L_A^s + (U^s - y_0)\cdot f^s(U^s - y_0)\right) &=& \int \frac{\partial E_m^s}{\partial\overline{w}}\Delta y_m^s dm.
\end{eqnarray}} 

The fiscal externalities are also affected by optimal taxes. Changes in the minimum wage paired with reductions in low-wage subsidies affect within-firm redistribution. This effect is captured by the term $E_m^s\cdot(\partial w_m^s/\partial \overline{w})$ which, for the minimum wage sub-market, is equal to low-skill employment given that $\partial w_m^l/\partial \overline{w}=1$. Intuitively, there are fiscal gains from minimum wage increases because they switch the burden of redistribution from the government to firms and, therefore, relax the social planner's budget constraint by transferring profits to the social planner. To develop intuition, consider a marginal increase in the subsidy to low-skill workers. This reform increases labor supply, so firms optimally react by lowering pre-tax wages \citep{rothstein2010eitc}. The minimum wage mutes this behavioral response, making the transfer to low-skill workers less costly. This reform cannot be exactly mimicked by the corporate tax since it distorts pre-tax profits. Possibly, both instruments are used to redistribute profits to workers.\footnote{Firm-level heterogeneity, revenue distortions, and entry distortions impede $t$ to fully redistribute from capitalists to workers. That is why the average SMWW of active capitalists is less than 1 at the joint optimum.} 

The desirability of the minimum wage at the joint optimum depends on how these two forces balance; the assessment of equation \eqref{prop3} is ultimately a quantitative question. Distortions in vacancies are likely to be negligible when the minimum wage is just above the market level. Consequently, the fiscal benefit likely dominates the employment costs when being close to the market level. As the minimum wage departs from the market level, the employment costs increase and become more likely to outweigh the fiscal benefits, hinting at the existence of an interior solution for the optimal minimum wage policy. The next section considers a restricted version of the model to provide concrete analytical conditions to justify binding minimum wages under optimal taxes.\footnote{The proposition also states that, in the joint optimum, the SMMWs of inactive individuals, active low-skill workers, and active high-skill workers average to 1, which is a standard result of optimal tax analyses with quasi-linear utility functions.} 

\paragraph{Caveats} I briefly discuss two elements that may affect the optimal policy analysis whose formal treatment is beyond the scope of this paper. 

First, the theoretical attractiveness of the income tax system relies on its flexibility. In the real world, income tax schedules are not fully non-linear, are not perfectly enforced, and are costly to administrate because, for example, tax evasion, tax avoidance, and imperfect benefit take up.\footnote{See, for example, \cite{andreoni1998tax}, \cite{slemrod2002tax}, \cite{kleven2011unwilling}, \cite{currie2006take}, \cite{kopczuk2007electronic}, \cite{chetty2013using}, \cite{bhargava2015psychological}, \cite{guyton2017reminders}, \cite{goldin2018tax}, \cite{cranor2019does}, \cite{finkelstein2019take}, \cite{guyton2021tax}, and \cite{linos2020can}.} These frictions generate additional efficiency costs to the tax system.\footnote{Abstracting from tax evasion also rules out additional complementarities between the minimum wage and the tax system. For example, if workers under report their incomes, then the minimum wage can increase tax collection by setting a floor on reported labor income \citep{biro2021minimum,javi}.} Minimum wages can also be difficult to enforce \citep{stansbury2021us,clemens2022understanding}, so a more general analysis should consider the relative enforcement costs of the two instruments. On the other hand, tax and transfer systems can tag on additional variables such as family size. That type of flexibility is unlikely to apply to the minimum wage policy \citep{stigler1946economics}. This benefit from using the tax system is not present in the proposed analysis.

Second, the minimum wage affects the distribution before taxes and transfers while taxes and transfers alter pre-tax values to generate the after-tax distribution.\footnote{This claim is true only to a first-approximation since changes in taxes can also affect the pre-tax income distribution \citep{roine2009long,alvaredo2013top,piketty2014optimal,vergara2022policies}.} The optimal policy analysis assumes that the social value of after-tax allocations does not depend on the composition between pre-tax incomes and taxes and transfers. However, recent evidence suggests that affecting the pre- and the post-tax and transfer distribution has different implications for long-run trends in inequality \citep{bozio2020predistribution,blanchetlancel}. Also, social preferences may put different weights on the two types of interventions. For example, \cite{mccall2013undeserving} provides survey evidence that suggests that the US public cares about inequality and redistribution, but prefers policies that address inequality within the firm rather than with taxes and transfers. This is consistent with the results of state-level ballot initiatives that have favored minimum wage changes relative to reforms to top marginal income tax rates \citep{saez2021public}. Such social preferences could be incorporated by generalizing the SMWWs \citep{saez2016generalized}.

\section{Policy Applications}
\label{sec4}

The policy analysis developed in the previous section uses an equilibrium framework to study, at a high-level, the desirability of the minimum wage. When taxes are fixed, the desirability of the minimum wage depends on the relative weight of the direct effects on workers and capitalists -- net of fiscal externalities -- which can be empirically measured by sufficient statistics. However, when the social planner optimizes both the minimum wage and the tax system, the generality of the model puts limits to the analytical insights that can be obtained in terms of concrete policy recommendations. 

In this section, I consider a restricted version of the model to get additional analytical results on the minimum wage desirability under optimal taxes. I consider three policy applications. First, I explore conditions under which it is optimal to have a binding minimum wage that complements a tax-based subsidy to low-skill workers. Second, I focus on the interaction between the minimum wage and the corporate tax rate when the corporate tax distortions differ between firms. Finally, I develop a suggestive numerical exercise to further explore the interactions between the policy instruments at the joint optimum.

\subsection{When is it optimal to have a binding minimum wage complemented by a tax-based subsidy to low-skill workers?}

Proposition III above gives conditions under which increasing the minimum wage when the tax system is optimal is welfare improving. Equation \eqref{prop3}, however, is difficult to assess both in empirical and analytical terms.\footnote{To identify micro elasticities, it is needed variation in minimum wages while holding after-tax allocations fixed.} In what follows, I restrict the model to gain analytical tractability and derive additional results that inform the optimal policy at the joint optimum. As noted below, some of these restrictions are supported by the empirical evidence presented in the next section.

The findings presented in the next section suggest that the profit incidence of the minimum wage is concentrated in low-skill labor-intensive services industries, while no effect on profits is found in other industries. This finding is consistent with the idea that minimum wage workers are concentrated in industries such as food and accommodation, low-skill health services, and retail. The empirical results also show no effect of minimum wage reforms on the number of establishments -- even in affected industries -- and on high-skill workers' outcomes. Given these facts, the first assumption restricts firm-level heterogeneity to represent a two-industry economy with inframarginal firms and no within-industry heterogeneity, where firms in one sector only hire low-skill workers and firms in the other sector only hire high-skill workers. To fix ideas, one sector will represent ``services'' and the other ``manufacturing''. This simplification puts restrictions on the wage distribution but accommodates the fact that the firms and workers that are affected by the minimum wage may be different from the non-affected ones.\footnote{Under this assumption, the wage distribution consists on two mass points, a low-skill wage (the minimum wage) and a high-skill wage. The results presented in the next section suggest no heterogeneous effects of minimum wage reforms on low-skill workers, suggesting that wage dispersion within skill types is unlikely to play an important role in these applications.}\\

\noindent \textsc{Assumption 1 (A1)}: \textit{There are two fixed populations of inframarginal capitalists indexed by $I = \{S,M\}$ with sizes $K_I$, where capitalists of type $I=S$ -- ``services'' -- only employ low-skill workers, and capitalists of type $I=M$ -- ``manufacturing'' -- only employ high-skill workers. Their respective production functions are given by $\phi^S(n^l,t)$ and $\phi^M(n^h,t)$, and their respective SMWWs are denoted by $g_K^S$ and $g_K^M$.}\\

I also assume that the technological second-order effects captured by $\phi_{nn}^I$ are negligible. This is done for analytical tractability. Abstracting from these second-order effects works against the minimum wage desirability. Intuitively, if a minimum wage shock pushes firm size downwards, there is an unintended benefit to firms because the marginal product of labor increases if technology features decreasing returns to scale. This effect attenuates the potential employment effects of the minimum wage.\\

\noindent \textsc{Assumption 2 (A2)}: \textit{$\phi_{nn}^I = 0$, for $I\in\{M,S\}$. }\\

Given these assumptions, the following proposition shows that when the social planner jointly optimizes the minimum wage and the income tax system, the minimum wage must bind if both the vacancy distortions and the social value of profits are small. The proposition also provides sufficient conditions on the SMWW of active low-skill workers under which the optimal binding minimum wage is complemented by a negative marginal tax rate on employed low-skill workers. The proposition extends existing results on the complementarity of the minimum wage and policies such as the EITC to a more general labor market framework with search and matching frictions and firm profits \citep{lee2012optimal}.\\

\noindent\textsc{Proposition IV}: \textit{Assume A1 and A2 hold. Consider the allocation induced by the optimal tax system with no minimum wage. Let $\varepsilon_{\theta,\overline{w}}^l$ denote the elasticity of low-skill labor market tightness with respect to changes in the minimum wage when after-tax allocations are fixed.}

\textit{(i) If $\varepsilon_{\theta,\overline{w}}^l\to 0$ when $\overline{w}$ is set at the market-level, having a binding minimum wage is optimal if $g_K^S<1$.}

\textit{(ii) Under the optimal binding minimum wage, the optimal marginal tax rate on employed low-skill workers is negative if}
\begin{eqnarray}
g_1^l &>& \frac{1 - C\cdot \varepsilon_{\theta,\Delta}^l\cdot\left[(1-t)\cdot g_K^S + t\right]}{1 - B\cdot\varepsilon_{\theta,\Delta}^l},\label{prop5}
\end{eqnarray}
\textit{where $B\in(0,1)$, $C\in(0,1)$, and $\varepsilon_{\theta,\Delta}^l$ is (the absolute value of) the elasticity of low-skill labor market tightness with respect to changes in low-skill net-of-tax wage when the minimum wage is fixed.}

\noindent\textit{Proof: See Appendix \ref{proofs}.}\\

The first part of the proposition provides conditions under which having a binding minimum wage on top of the optimal tax system is desirable. A binding minimum wage under optimal taxes generates three effects that are illustrated by the following hypothetical reform. Suppose the planner increases the minimum wage and simultaneously increases the net-tax on employed low-skill workers (possibly, by reducing a subsidy), to hold after-tax incomes fixed. This generates a positive fiscal gain for the planner proportional to low-skill employment. This fiscal externality is paid by employers through higher wages, so there is a mechanical decrease in profits also proportional to low-skill employment. Finally, while wages are fixed and labor supply is invariant to the change in the wage given the decrease in the subsidy, firms may have incentives to decrease vacancies. The distortion in posted vacancies generates a congestion externality that, most likely, has a negative effect on social welfare.\footnote{On one hand, the smaller employment probabilities affect the expected utility of inframarginal low-skill workers. On the other hand, the decrease in posted vacancies generates a marginal increase in profits and a potential positive fiscal externality, since the planner has to pay the subsidy to employed low-skill workers to a small mass of individuals. If the former effect dominates, then the congestion externality decreases social welfare.}

Whether the social planner wants to have a binding minimum wage depends on how these three forces balance. When the minimum wage is set at the market wage and, therefore, the policy respects the first order conditions of the firm, a marginal increase in the minimum wage is likely to have negligible effects on vacancy posting. In those cases, the marginal increase in the minimum wage mimics a one-to-one transfer from profits to the social planner. This transfer increases social welfare only if the SMWW on affected profits, $g_K^S$, is smaller than 1. When vacancy distortions are not negligible, the fiscal externality has to compensate for both the decrease in profits and the overall effects of the congestion externality, implicitly requiring an even smaller SMWW on affected capitalists.\footnote{\cite{lee2012optimal} find that, under efficient rationing and optimal taxes, a binding minimum wage is always desirable if the SMWW on active low-skill workers is greater than 1. It is tempting to think that firms setting wages work as an analog of efficient rationing. However, firms can also adjust vacancy posting which generates the congestion externality. Formally, my model cannot be written as a particular case of \cite{diamond1971optimal}, so I cannot apply the results on quantity controls in second-best economies proposed by \cite{guesnerie1981could} and \cite{guesnerie1984effective}.}

The condition above requires that the SMWW on affected capitalists is smaller than 1 \textit{under the optimal tax allocation}. That is, $g_K^S$ incorporates the effect of the optimal corporate tax rate. If the corporate tax rate is non-distortionary, its optimal value is possibly large, making $g_K^S$ approach 1. On the contrary, if the corporate tax rate is very distortionary, its optimal value is possibly small, making $g_K^S$ approach 0 when pre-tax profits are large. Then, the corporate tax rate matters for the desirability of the minimum wage. If the distortion of the corporate tax increases with the square of the tax rate, the optimum possibly involves both a corporate tax rate and a binding minimum wage.

The second part of the proposition provides conditions under which the optimal minimum wage is complemented by tax-based transfers to employed low-skill workers such as the EITC. The condition specifies a threshold on the SMWW of active low-skill workers that depends on the effects of the EITC on labor market tightness.\footnote{With no matching frictions, $\varepsilon_{\theta,\Delta}^l = 0$, so equation \eqref{prop5} is reduced to $g_1^l > 1$, which is the standard result on the EITC desirability in frictionless labor markets with extensive margin responses \citep{lee2012optimal,piketty2013optimal}.} Since the EITC generates an increase in labor supply and wages are fixed, firms react by decreasing vacancies, thus generating a congestion externality. This effect is captured by $\varepsilon_{\theta,\Delta}^l$ which, as captured in the denominator, generates a market-level inefficiency that makes the critical SMWW higher: the transfer to active low-skill workers needs to be socially valuable beyond the generated distortion. Also, the congestion effect generates an increase in profits that may slack the critical SMWW because of two reasons. First, if the planner values redistribution toward firms, the increase in after-tax profits is socially valuable. Second, even if $g_K^S\to 0$, the transfer to firms allows the social planner to enforce larger corporate tax rates by reducing its distortions on pre-tax profits.

\subsection{Minimum wages and corporate tax rates under international capital mobility}

The second policy application focus on the interaction between the minimum wage and the corporate tax rate. The two-sector model specified in A1 allows to tractably incorporate additional differences between affected and non-affected firms. In particular, services and manufacturing not only differ in their exposure to minimum wage workers but also their capital intensity and capital mobility. Manufacturing is more capital intensive, and its capital is presumably more internationally mobile than the one employed in services industries. This observation suggests that the behavioral response of profits to changes in corporate taxes is also likely to differ between industries. 

I allow for this possibility by incorporating a capital-allocation microfoundation of the dependence of $\phi$ on $t$ discussed in Appendix \ref{results} (see discussion in Section \ref{sec2}). Capitalists are endowed with a fixed stock of capital that has to be allocated between domestic and foreign investment. The domestic production function is given by $\widetilde{\phi}^I(n,k)$, where $k$ is capital and $\widetilde{\phi}^I(n,k(t)) = \phi^I(n,t)$. The domestic corporate tax rate distorts the amount of capital invested in the domestic firm. Under this structure, the response of pre-tax profits to changes in the corporate tax rate is proportional to the degree of (sector-specific) capital mobility, which I denote by $\varepsilon_{k,t}^I$, for $I\in\{S,M\}$ (see Appendix \ref{results}). 

The following proposition shows that, if capital and labor are complements ($\widetilde{\phi}_{kn}>0$), the desirability of the minimum wage is increasing in the capital mobility of the non-affected sector -- $M$ -- while the capital mobility of the affected sector -- $S$ -- has an ambiguous effect on the optimal minimum wage. For analytical simplicity, and without loss of generality, the analysis abstracts from the income tax system.\\

\noindent\textsc{Proposition V}: \textit{Assume that A1 and A2 holds, that there is no income tax system, and that capital and labor are complements.}

\textit{(i) The marginal social welfare of increasing $\overline{w}$ when $t$ is optimal is increasing in $\varepsilon_{k,t}^M$.} 

\textit{(ii) The effect of $\varepsilon_{k,t}^S$ on the optimal $\overline{w}$ is ambiguous.}

\noindent\textit{Proof: See Appendix \ref{proofs}.}\\

This proposition provides additional insights into the optimal combination of a minimum wage and a corporate tax rate when the planner wants to tax profits. Capital mobility has a negative impact on the optimal corporate tax rate. The higher the mobility, the more distortionary is $t$, and therefore the lower its optimal level. This increases the optimal minimum wage, $\overline{w}$, because both policies redistribute profits: when $t$ is smaller, then the net benefits from increasing $\overline{w}$ are larger. However, capital mobility in the affected sector decreases the optimal minimum wage because of similar distortions on domestic capital. This implies that increasing $\varepsilon_{k,t}^M$ unambiguously increases the optimal minimum wage, because it increases the distortion of $t$ without affecting the distortion of $\overline{w}$, but increasing $\varepsilon_{k,t}^S$ has an ambiguous effect on its level given the two forces that work in opposite directions.

This result gives more policy-relevance to the empirical results documented in the next section. Governments have trouble enforcing large corporate tax rates because of international capital mobility. This is especially driven by capital-intensive industries, such as manufacturing, where corporate tax rates are more likely to generate real productive distortions \citep{pat}. However, minimum wages do not affect profits in capital-intensive industries. By contrast, low-skill labor-intensive industries are unlikely to be affected by capital mobility distortions but are affected by the minimum wage through the direct effects on wages and profits. That is, in practice, $\varepsilon_{k,t}^M$ is possibly large and $\varepsilon_{k,t}^S$ is possibly low. 

This result suggests that the minimum wage can be interpreted as an industry-specific corporate tax rate that minimizes distortions related to capital mobility. Said differently, the minimum wage can effectively tax profits in affected sectors without distorting capital allocation in non-affected sectors. This insight arises as particularly policy-relevant given the documented decline in effective capital taxation in developed countries due to globalization \citep{bachas2022globalization}.

\subsection{Numerical exercise}

Finally, I perform a suggestive numerical exercise that illustrates the intuitions developed in the optimal policy analysis. I calibrate a version of the simplified model to match empirical moments of the US labor market and compute the optimal minimum wage under different tax systems.\footnote{I consider the two-sector model summarized in A1 with the capital allocation problem to model the distortions of the corporate tax rate on pre-tax profits. I do not impose A2, allowing for decreasing returns to scale.} This exercise further informs about the interactions between the policy instruments and allows a better characterization of the joint optimum away from the local analysis. This subsection discusses the main conclusions of the analysis. All the details about the simulations, including calibration and results, are presented in Appendix \ref{simulation}. 

There are three conclusions from the numerical analysis. First, given a tax system, social welfare is generally a globally concave function of the minimum wage, so the model generates an interior solution for its optimal value. This result is explained by the fact that wage effects tend to dominate employment effects at low levels because vacancy posting distortions are small, but employment effects become larger and eventually dominate wage effects as the minimum wage departs from the market level. This result, while expected and present in many labor market models, works as a sanity check for the proposed framework. Second, the optimal minimum wage under fixed taxes varies with the tax parameters. The optimal minimum wage is larger when the EITC is larger -- that is, the minimum wage seems to complement the EITC -- and is also larger when the corporate tax rate is smaller, suggesting that the minimum wage serves as a substitute for corporate taxation. Third, the joint optimum seems to use all policies in tandem. That is, optimal redistribution consists of a large EITC, a binding minimum wage far above the market wage, and a non-trivial corporate tax rate. Among the cases considered, the optimal policy consists of an EITC of 100\%, an hourly minimum wage of \$12, and a corporate tax rate of 35\%. The optimal minimum wage is substantially larger than the market wage -- which is simulated to be below \$7. Efficiency considerations not included in the model could lead to higher optimal minimum wages (e.g., \citealp{burdett1998wage}; \citealp{acemoglu2001good}; \citealp{bergermw}; see discussion in Section \ref{sec2}), while risk aversion due to potential long term unemployment could lead to lower optimal minimum wages (e.g., \citealp{sorkin2015there,hurst}). While the minimum wage and the corporate tax rate partially work as substitutes, the planner prefers to use them both because each policy's distortion is increasing in its level. These results reinforce the idea that the minimum wage can increase the efficiency of tax-based redistribution: optimally combining all instruments can lead to larger social welfare.

\section{Sufficient Statistics Estimation}
\label{sec5}

Section \ref{sec3} develops a general policy analysis and Section \ref{sec4} delves into the minimum wage desirability when taxes are optimal. One lesson from Section \ref{sec3} is that the desirability of the minimum wage can be expressed as a function of sufficient statistics, a feature that is especially useful in cases with fixed taxes.\footnote{Estimating micro elasticities -- optimal taxes -- requires a stricter empirical design that exploits variation in minimum wages while keeping after-tax allocations fixed.} This section estimates the sufficient statistics that inform about the welfare effects of minimum wage reforms when taxes are fixed using publicly available US data. The estimated causal effects can be of interest by themselves, and also allow me to explore whether the current minimum wage is too high or too low under the current tax system.

\subsection{Empirical strategy}

The empirical strategy exploits state-level variation in minimum wages to estimate stacked event studies. 

\paragraph{Events} I follow \cite{cengiz2019effect,cengiz2021ml}'s strategy to define state-level events. A state-by-year minimum wage is defined as the maximum between the statutory values of the federal and state minimum wages throughout the calendar year. I use data from \cite{vaghul2016historical} for the 1997-2019 period for which I can observe all the outcomes of interest within eight-year balanced windows. Nominal values are transformed to 2016 dollars using the R-CPI-U-RS index including all items. An event is defined as a state-level hourly minimum wage increase above the federal minimum wage of at least \$0.25 (in 2016 dollars) in a state with at least 2\% of the employed population affected, where the affected population is computed using the NBER Merged Outgoing Rotation Group of the CPS (henceforth, CPS-MORG).\footnote{This is done by computing employment counts by wage bins and checking whether, on average, the previous year share of workers with wages below the new minimum wage is above 2\% \citep{cengiz2019effect}.} These restrictions are imposed to focus on minimum wage increases that are likely to have effects on the labor market. Small state-level or binding federal minimum wage increases are not recorded as events, however, regressions control for small state-level and federal minimum wage increases. I also restrict the attention to events where treated states do not experience other events in the three years previous to the event and whose timing allows me to observe the outcomes from three years before to four years after. This results in 50 valid state-level events, whose time distribution is plotted in Figure \ref{events} of Appendix \ref{robust}. Table \ref{list_events} of Appendix \ref{robust} display the list of the considered events with their corresponding treated states. 

\paragraph{Estimating equation} Estimating event studies in this setting is challenging for two reasons. First, states may increase their minimum wages several times over the period considered. Second, treatment effect heterogeneity may induce bias when treatment adoption is staggered \citep{de2022two,roth2022s}. 

To deal with these issues, I implement stacked event studies \citep{cengiz2019effect,cengiz2021ml,gardnertwo,baker2021much} as follows. For each event, I define a time window that goes from 3 years before the event to 4 years after. All states that do not experience events in the event-specific time window define an event-specific control group. This, in turn, defines an event-specific dataset. Finally, all event-specific datasets are appended and used to estimate a standard event study with event-specific fixed effects. This leads to the following estimating equation:
\begin{eqnarray}
\log Y_{ite} = \sum_{\tau=-3}^4 \beta_{\tau} D_{ite}^{\tau} + \alpha_{ie} + \gamma_{te} + \rho_{ite} + \epsilon_{ite}, \label{reg}
\end{eqnarray}
where $i$, $t$, and $e$ index state, year, and event, respectively,  $Y_{ite}$ is an outcome of interest (see next subsection), $D_{ite}^{\tau}$ are event indicators with $\tau$ the distance from the event (in years), $\alpha_{ie}$ are state-by-event fixed effects, $\gamma_{te}$ are year-by-event fixed effects, and $\rho_{iqe}$ are state-by-year-by-event varying controls that include small state-level minimum wage increases and binding federal minimum wage increases.\footnote{Following \cite{cengiz2019effect,cengiz2021ml}, controls for small state-level and binding federal minimum wage increases are included as follows. Let $\hat{t}$ be the year in which the small state-level or binding federal minimum wage increase takes place. Then, define $Early_t = 1\{t\in\{\hat{t}-3,\hat{t}-2\}\}$, $Pre_t = 1\{t = \hat{t}-1\}$ and $Post_t = 1\{t\in\{\hat{t},\hat{t}+1,\hat{t}+2, \hat{t}+3,\hat{t}+4\}\}$, and let $Small_i$ and $Fed_i$ be indicators of states that face small state-level and binding federal minimum wage increases, respectively. Then $\rho_{ite}$ includes all the interactions between $\{Early_t,Pre_t,Post_t\}\times \{Small_i,Fed_i\}$ for each event separately.} I also consider specifications where the year-by-event fixed effects are allowed to vary across census regions and census divisions. $\beta_{-1}$ is normalized to 0. To allow for correlation within states across events, standard errors are clustered at the state level. Regressions are weighted by state-by-year average total population. Regressions for capitalists' outcomes that vary at the industry-level allow for state-by-industry-by-event fixed-effects, cluster standard errors at the state-by-industry level, and weight observations using the average state-by-industry employment in the pre-period reported in the QCEW files.

I also consider standard differences-in-differences regressions:
\begin{eqnarray}
\log Y_{ite} = \beta T_{ie}\text{Post}_{te} + \alpha_{ie} + \gamma_{te} + \rho_{ite} + \epsilon_{ite}, \label{reg_did}
\end{eqnarray}
where $T_{ie}$ is an indicator variable that takes value 1 if state $i$ is treated in event $e$, $\text{Post}_{te}$ is an indicator variable that takes value 1 if year $t$ is larger or equal than the treatment year in event $e$, and all other variables are defined as in equation \eqref{reg}. The coefficient of interest is $\beta$, which captures the average treatment effect in the post-event years (from $\tau= 0$ to $\tau = 4$). 

\subsection{Data} 

Outcomes consist of state-level aggregates for 1997-2019 computed using publicly available data. 

\paragraph{Workers} The sufficient statistic for changes in active workers' welfare is $dU^s/d\overline{w}$, for $s\in\{l,h\}$. Equations \eqref{suffstat} and \eqref{suffstat_2} show that $dU^s/d\overline{w}$ equals the change in the average post-tax wage of active workers including the unemployed, which can be decomposed into changes in their average pre-tax wage (including the unemployed) and changes in their average net tax liabilities.

I use the CPS-MORG data to compute average pre-tax hourly wages and the Basic CPS monthly files to compute employment and participation rates at the state-by-year-by-skill level. The pre-tax component of the sufficient statistic, then, can be computed as the average wage times the employment rate. Low-skill (high-skill) workers are defined as not having (having) a college degree. Hourly wages are either directly reported or indirectly computed by dividing reported weekly earnings by weekly hours worked. I drop individuals aged 15 or less, self-employed individuals, and veterans. Nominal wages are transformed to 2016 dollars using the R-CPI-U-RS index including all items. Observations whose hourly wage is computed using imputed data (on wages, earnings, and/or hours) are excluded to minimize the scope for measurement error. To avoid distorting low-skill workers' statistics with non-affected individuals at the top of the wage distribution, I restrict the low-skill workers' sample to workers that are either out of the labor force, unemployed, or in the bottom half of the wage distribution when employed. I test how results change when considering different wage percentile thresholds. To compute changes in net tax liabilities at the state-by-year level, I use data from the BEA regional accounts. I consider income maintenance benefits, medical benefits, and gross federal income tax liabilities.\footnote{The BEA definition of income maintenance benefits is as follows: ``Income maintenance benefits consists largely of Supplemental Security Income (SSI) benefits, Earned Income Tax Credit (EITC), Additional Child Tax Credit, Supplemental Nutrition Assistance Program (SNAP) benefits, family assistance, and other income maintenance benefits, including general assistance.'' Medical benefits consider both Medicaid and Medicare programs.} 

\paragraph{Capitalists} The sufficient statistic that summarizes changes in active capitalists' welfare driven by minimum wage changes is the change in firm profits, $d\Pi(\psi,t)/d\overline{w}$, for $\psi\in[\psi^*,\overline{\psi}]$. 

Absent firm-level microdata, I compute a measure of average profits per firm at the industry-by-state-by-year level. I use the Gross Operating Surplus (GOS) estimates from the BEA regional accounts as a proxy of state-level aggregate profits and divide them by the average number of private establishments reported in the QCEW data files.\footnote{The BEA definition of gross operating surplus is as follows: ``Value derived as a residual for most industries after subtracting total intermediate inputs, compensation of employees, and taxes on production and imports less subsidies from total industry output. Gross operating surplus includes consumption of fixed capital (CFC), proprietors' income, corporate profits, and business current transfer payments (net).''} Nominal profits are transformed to 2016 dollars using the R-CPI-U-RS index including all items. I consider 25 industries that have a relatively large coverage across states and years. Noting that minimum wage workers are not evenly distributed across industries (e.g., \citealp{bls}), I group industries into three large groups: manufacturing, exposed services, and non-exposed services.\footnote{I exclude agriculture and mining. I also exclude construction and finance since they experience particularly abnormal profit dynamics around the 2009 financial crisis. Manufacturing industries include SIC codes 41, 43, 44, 46, 50, 54, 56, and 57, that is, nonmetallic mineral products, fabricated metal products, machinery, electrical equipment, food and beverages and tobacco, printing and related support activities, chemical manufacturing, and plastics and rubber products. Exposed services include SIC codes 9, 19, 21, 27, 28, and 34, that is, retail trade, ambulatory health services, nursing and residential care facilities, food, accommodation, and social services and other services. Non-exposed services include SIC codes 8, 10, 11, 13, 14, 15, 16, 17, 20, 24, and 25, that is, wholesale trade, transport, information, real estate, professional services, management of businesses, administrative support, educational services, hospitals, arts, and recreation industries.} Fiscal effects are proportional to the effect on profits. I also use data on taxes on production and imports net of subsidies reported on the BEA regional accounts at the industry-level, and data on business and dividend income reported in the state-level SOI tables to test for additional fiscal externalities.

\paragraph{Descriptive statistics} Table \ref{ds} shows descriptive statistics for the non-stacked panel. The total number of observations is 1,173 (51 states times 23 years). All monetary values are annual and in 2016 dollars.\footnote{While the theoretical and empirical analysis on workers' outcomes is based on average hourly wages, I annualize these values by multiplying them by 52 weeks and the average number of hours worked by skill group. I below show that weekly hours worked conditional on employment are not affected by minimum wage changes.} Average pre-tax incomes (including the unemployed) are more than 3 times larger for high-skill workers relative to low-skill workers. This is explained by higher hourly wages and weekly hours conditional on employment, and also by higher employment rates. Average income maintenance benefits per working-age individual are 1,051 dollars, which represents around 5\% of low-skill workers' pre-tax income. Average medical benefits and gross federal income taxes per working-age individual are 4,541 and 7,179 dollars, respectively. Average pre-tax profits per establishment are substantially larger than disposable incomes for workers. In exposed services, the average pre-tax profit per establishment is almost 9 times the average pre-tax income of low-skill workers including the unemployed. The ratio increases to almost 50 times and to 100 times when looking at non-exposed services and manufacturing.

\subsection{Results}

\begin{figure}[t!]
\centering
\caption{Changes in workers' welfare after minimum wage increases }
\label{es_w}
\subfigure[Low-skill workers]{\includegraphics[width=0.49\textwidth]{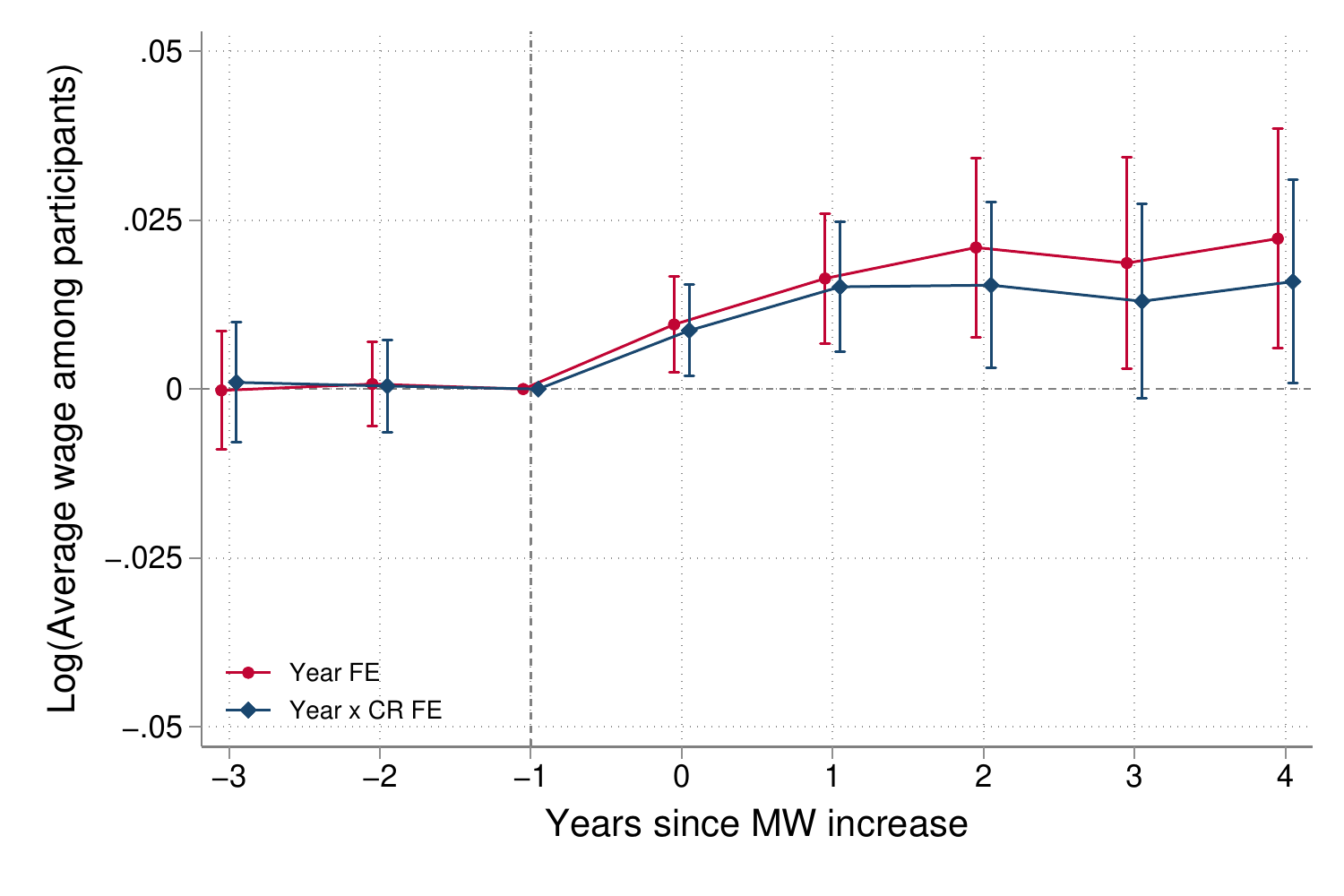}}
\subfigure[High-skill workers]{\includegraphics[width=0.49\textwidth]{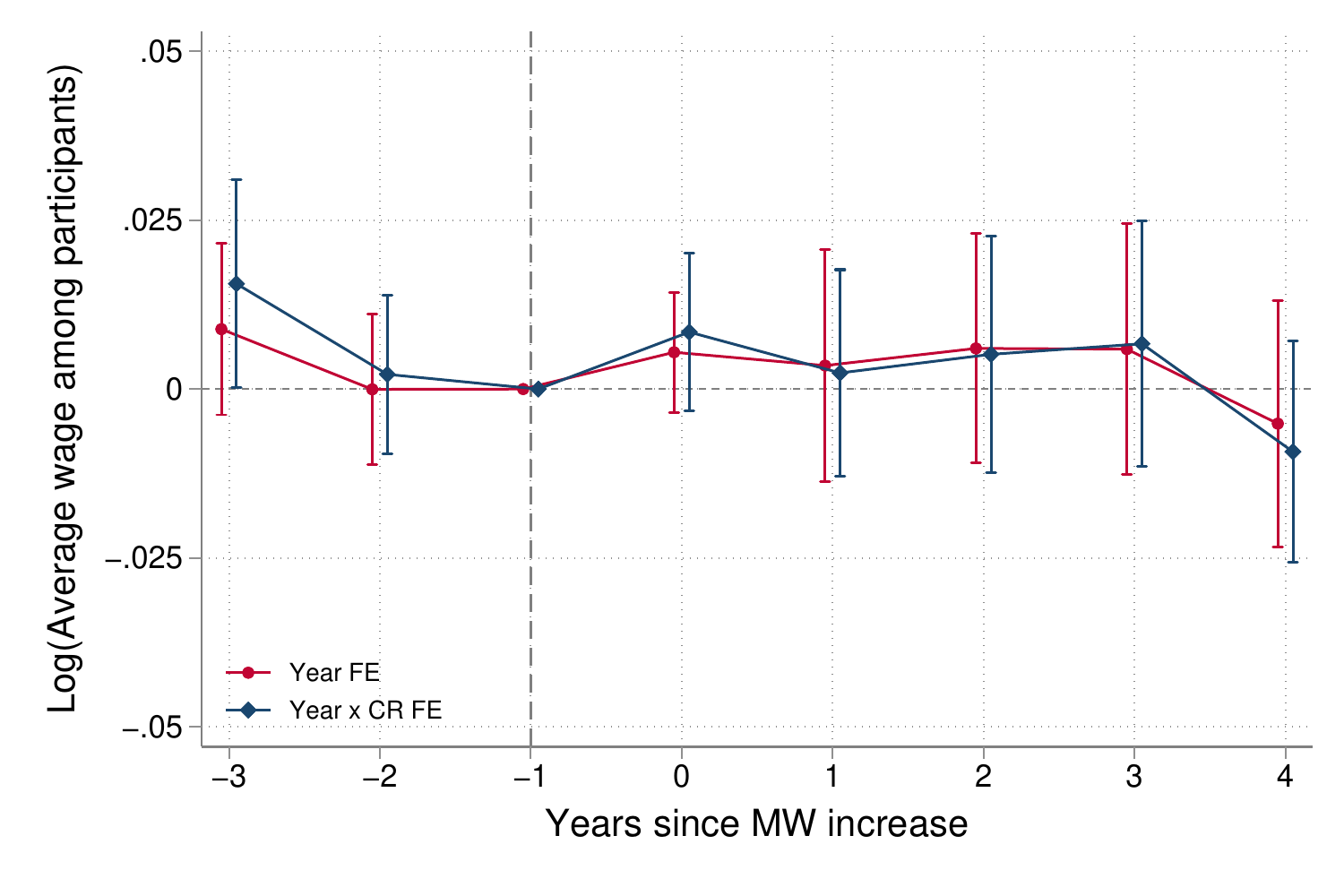}}

\begin{minipage}{\textwidth} 
{\scriptsize \vspace{0.5cm} Notes: These figures plot the estimated $\beta_{\tau}$ coefficients with their corresponding 95\% confidence intervals from equation \eqref{reg}. Panel (a) uses the average pre-tax wage of active low-skill workers including the unemployed as dependent variable. Panel (b) uses the average pre-tax wage of active high-skill workers including the unemployed as dependent variable. Low- and high- skill workers are defined as not having (having) a college degree. Red lines represent specifications that control by year-by-event fixed effects. Blue lines represent specifications that control by census-region-by-year-by-event fixed effects. Standard errors are clustered at the state level, and regressions are weighted by state-by-year average population.\par}
\end{minipage}
\end{figure}

\paragraph{Worker-level pre-tax outcomes} Figure \ref{es_w} plots the estimated coefficients $\{\beta_{\tau}\}_{\tau=-3}^{4}$ of equation \eqref{reg} with their corresponding 95\% confidence intervals using the average pre-tax hourly wage of active low- and high-skill workers including the unemployed as dependent variables to proxy for $U^l$ and $U^h$. Each figure plots regressions with two different types of time fixed-effects: year-by-event fixed effects and census-region-by-year-by-event fixed effects. Table \ref{T_w} presents the estimated coefficients $\beta$ of equation \eqref{reg_did} that summarize the average treatment effect in the post-event period, and also includes specifications that control for census-division-by-year-by-event fixed effects. Panel (a) of Figure \ref{es_w} shows that state-level minimum wage increases have increased active low-skill workers' welfare. Table \ref{T_w} shows that the implied elasticity, $d\log U^l/d\log \overline{w}$, ranges between 0.10 and 0.13. Panel (b) of Figure \ref{es_w} shows that these minimum wage increases have had null effects on high-skill workers' welfare. Table \ref{T_w} shows that the estimate of $d\log U^h/d\log \overline{w}$ is a precise zero, regardless of the fixed-effects considered. These results suggest that state-level minimum wages have reduced welfare gaps between low- and high-skill active workers.\footnote{Figure \ref{wage_perc} of Appendix \ref{robust} tests the sensitivity of the result on low-skill workers to the choice of the wage percentile used to trim the sample of employed low-skill workers. Results are robust to using more restrictive samples and to the incorporation of low-skill workers until percentile 80. Including low-skill workers belonging to the top 20\% attenuates the estimated result, which is expected given the unlikely response of top wages to changes in the minimum wage.}

\begin{figure}[t!]
\centering
\caption{Decomposing the effect of minimum wages on low-skill workers}
\label{margins_fig}
\subfigure[Wage (conditional on employment)]{\includegraphics[width=0.49\textwidth]{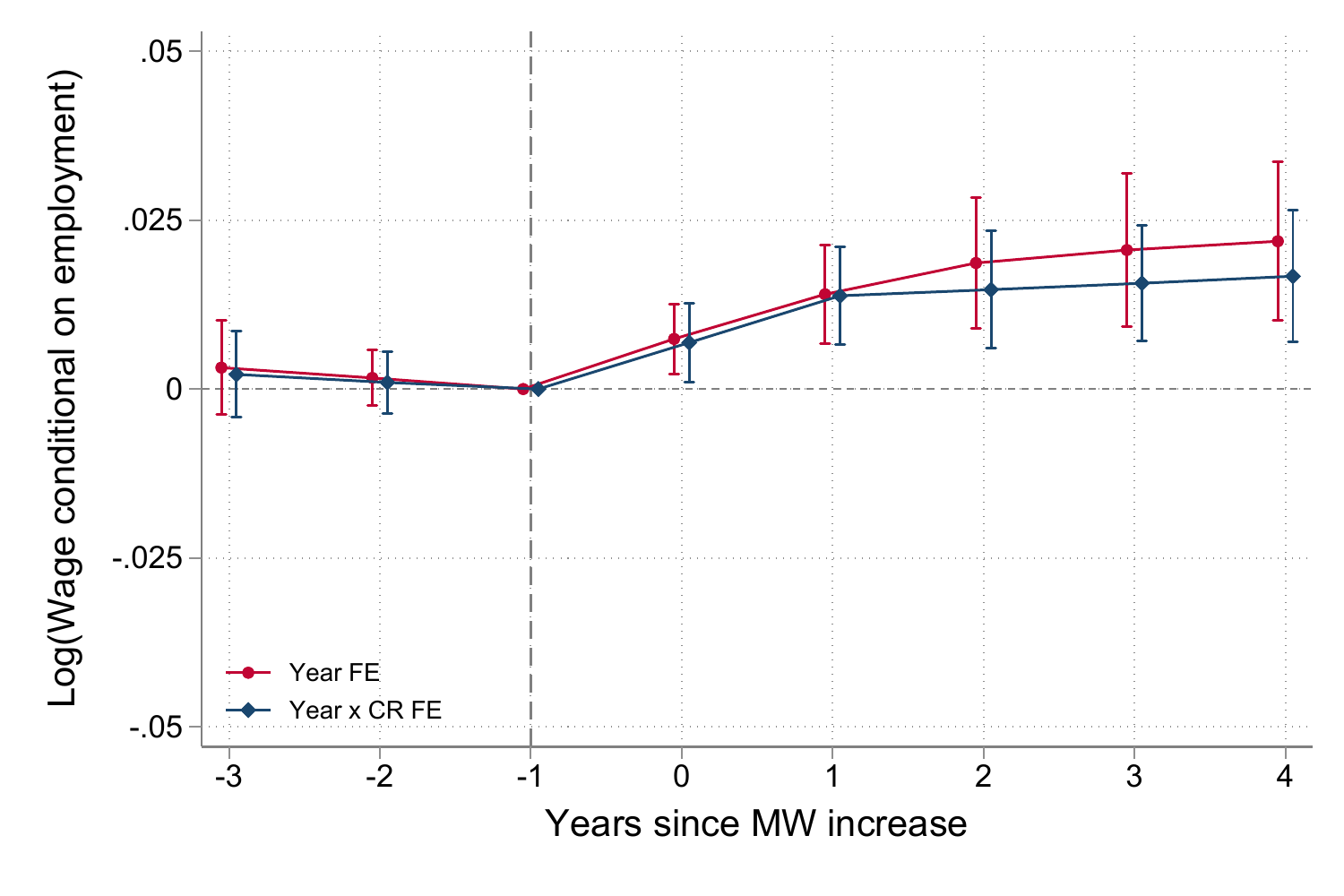}}
\subfigure[Employment rate]{\includegraphics[width=0.49\textwidth]{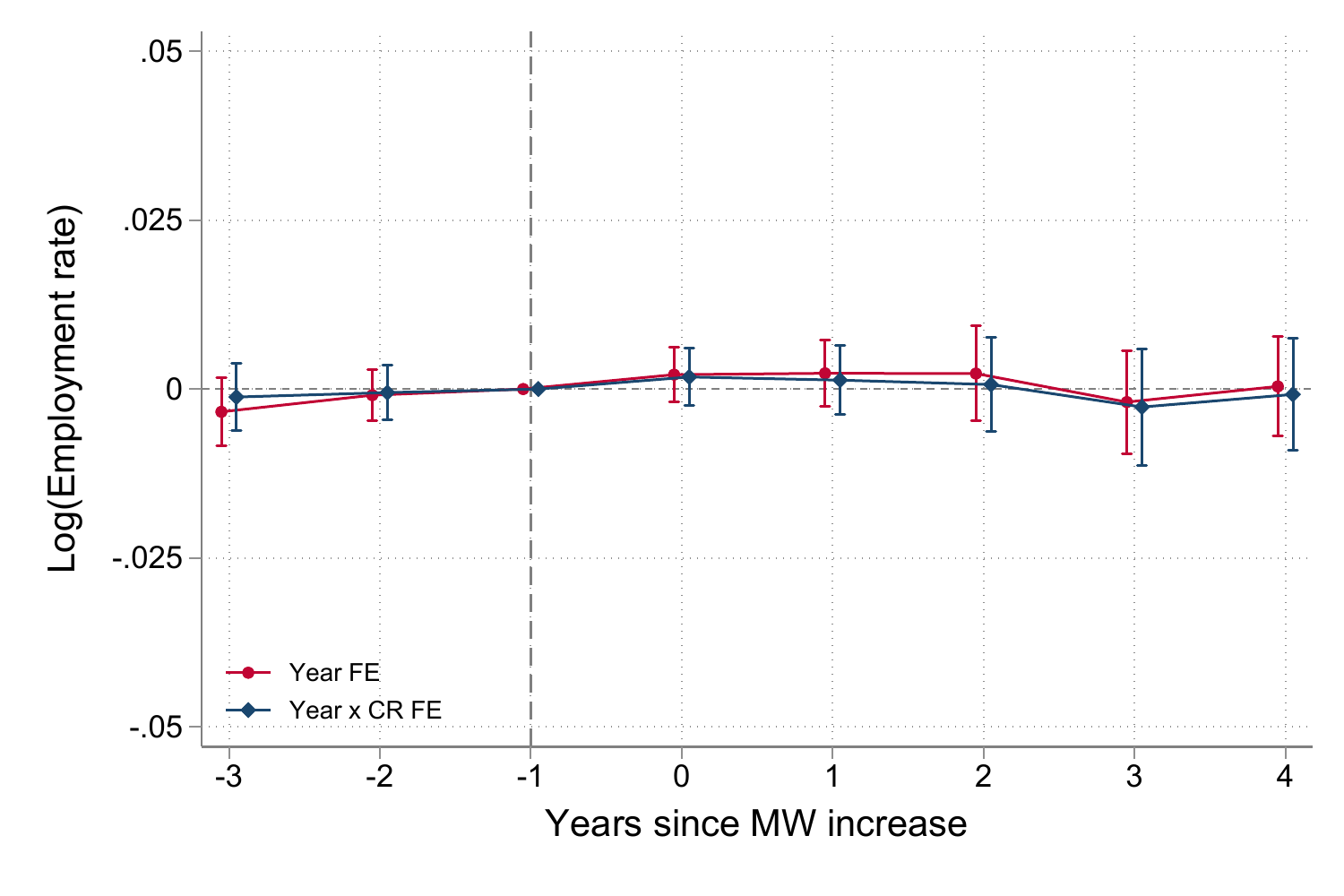}}
\subfigure[Weekly hours (conditional on employment)]{\includegraphics[width=0.49\textwidth]{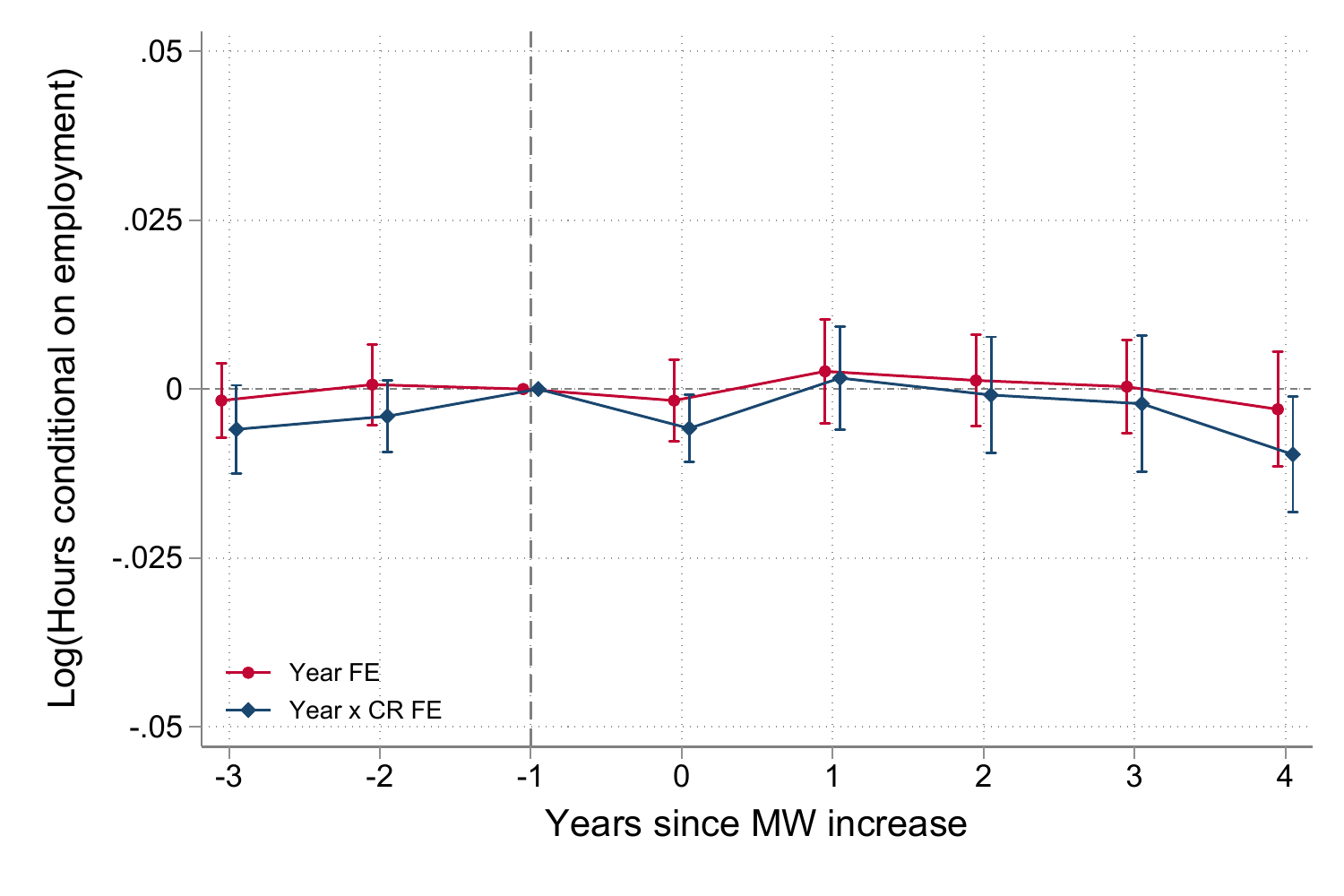}}
\subfigure[Participation rate]{\includegraphics[width=0.49\textwidth]{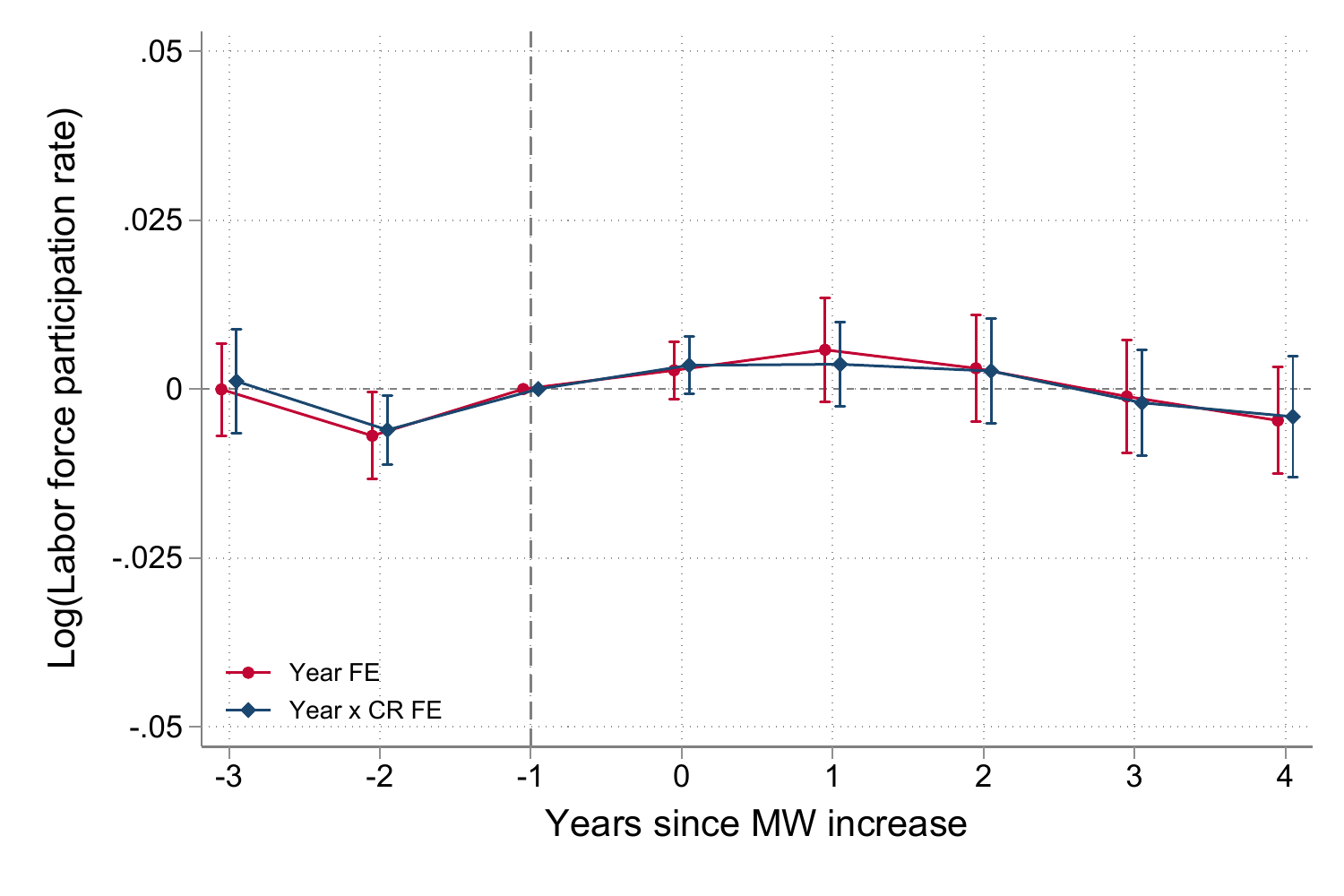}}
\begin{minipage}{\textwidth} 
{\scriptsize \vspace{0.5cm} Notes: These figures plot the estimated $\beta_{\tau}$ coefficients with their corresponding 95\% confidence intervals from equation \eqref{reg}. Panel (a) uses the average pre-tax hourly wage of low-skill workers conditional on employment as dependent variable. Panel (b) uses the average employment rate of low-skill workers as dependent variable. Panel (c) uses the average weekly hours worked of low-skill workers conditional on employment as dependent variable. Panel (d) uses the average participation rate of low-skill workers as dependent variable. Low-skill workers are defined as not having a college degree. Red lines represent specifications that control by year-by-event fixed effects. Blue lines represent specifications that control by census-region-by-year-by-event fixed effects. Standard errors are clustered at the state level, and regressions are weighted by state-by-year average population.\par}
\end{minipage}
\end{figure}

To better understand how minimum wages have affected low-skill workers, Figure \ref{margins_fig} presents separate results for each of the margins that can play a role in the evolution of the sufficient statistic: hourly wages, weekly hours (both conditional on employment), employment rates, and participation rates. Results indicate that all the effect of minimum wage increases on $U^l$ is driven by an increase in the wage conditional on employment, with no effect on hours, employment, or participation.\footnote{The lack of employment responses suggests that the result is robust to including curvature in the flow utility of workers.}\footnote{My results differ from \cite{gandhi2022minimum} and \cite{jardim2022minimum} who find effects on hours worked.} 

\begin{figure}[t!]
\centering
\caption{Minimum wage effects on low-skill workers: heterogeneity}
\label{het_fig}
\includegraphics[width=\textwidth]{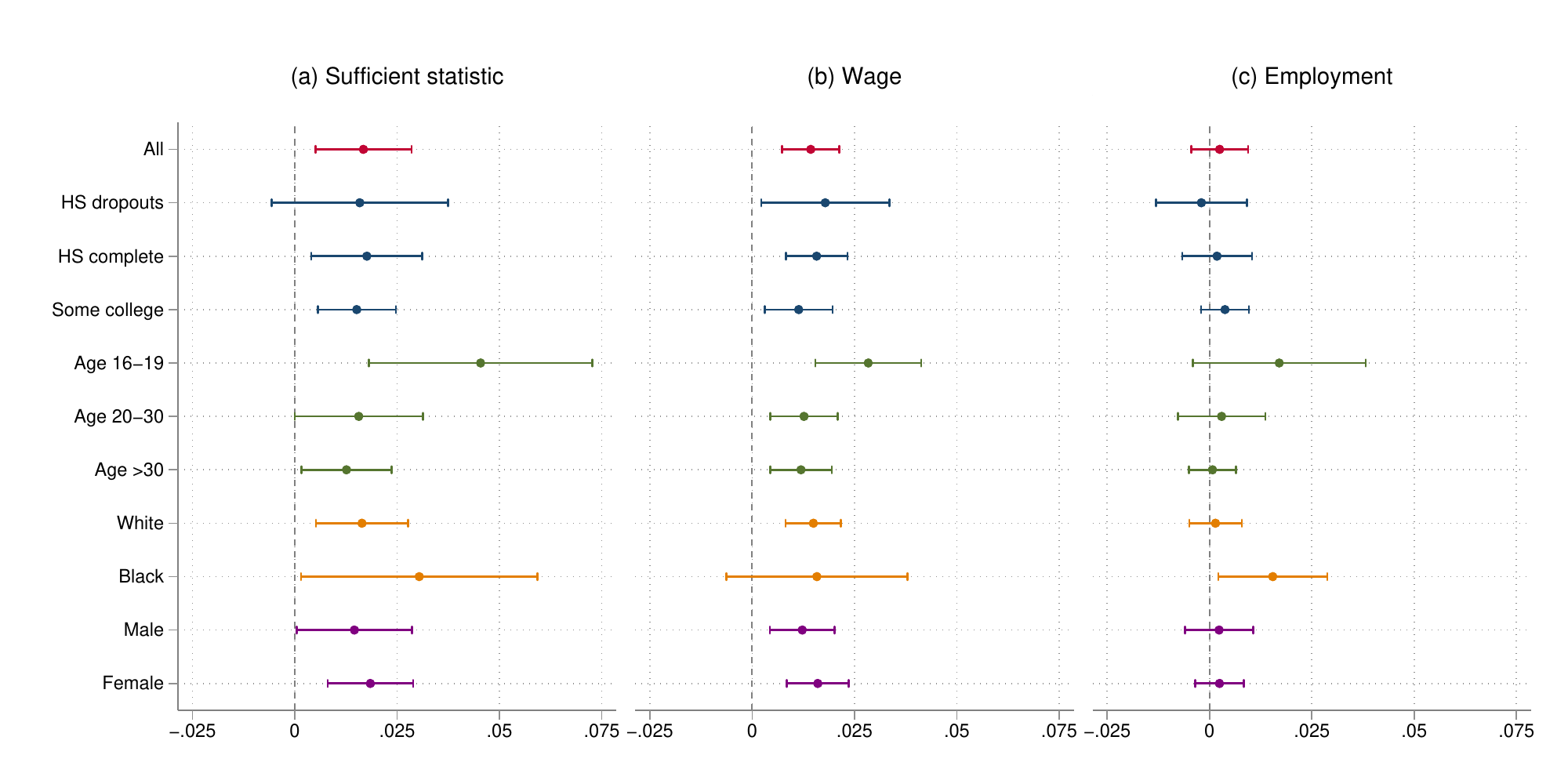}
\begin{minipage}{\textwidth} 
{\scriptsize \vspace{0.5cm} Notes: This figure plots the estimated $\beta$ coefficient with its corresponding 95\% confidence intervals from equation \eqref{reg_did} for different groups of low-skill workers and different dependent variables. Panel (a) uses the average pre-tax wage of active low-skill workers including the unemployed as dependent variable. Panel (b) uses the average pre-tax hourly wage of low-skill workers conditional on employment as dependent variable. Panel (c) uses the average employment rate of low-skill workers as dependent variable. Low-skill workers are defined as not having a college degree. Red coefficients reproduce the analysis with the complete sample. Blue coefficients split low-skill workers by education (high-school dropouts, high-school complete, and college incomplete). Green coefficients split low-skill workers by age (16-19, 20-30, and more than 30). Orange coefficients split low-skill workers by race (white and black). Purple coefficients split low-skill workers by sex (male and female). All regressions include year-by-event fixed effects. Standard errors are clustered at the state level, and regressions are weighted by state-by-year average population.\par}
\end{minipage}
\end{figure}

To test for patterns of heterogeneity, Figure \ref{het_fig} plots the estimated $\beta$ coefficient of equation \eqref{reg_did} with its corresponding 95\% confidence interval using different groups of low-skill workers. Panel (a) uses the average pre-tax wage of active low-skill workers including the unemployed ($U^l$) as dependent variable. Panel (b) uses the average pre-tax hourly wage of low-skill workers conditional on employment as dependent variable. Panel (c) uses the average employment rate of low-skill workers as dependent variable. The effects are very stable across groups: all groups experience an increase in welfare driven by changes in wages conditional on employment with no effects on employment. If anything, teen (aged 16-19) and black low-skill workers seem to experience larger welfare gains. This result suggests there are no clear groups of winners and losers within the broad population of low-skill workers.\footnote{Results are consistent with \cite{cengiz2019effect,cengiz2021ml} who find positive wage effects, limited employment effects, and limited participation effects on low-wage workers, i.e., the part of the distribution close to the minimum wage, using similar data and empirical strategy. My results differ from theirs in two dimensions. First, I focus on broad skill groups that are not exclusively composed of minimum wage workers. Second, the main focus of my analysis is the estimation of the composite sufficient statistic rather than the effect on the different margins.}

\begin{figure}[t!]
\centering
\caption{Worker-level fiscal effects after minimum wage increases}
\label{es_fiscal}
\subfigure[Income maintenance transfers]{\includegraphics[width=0.49\textwidth]{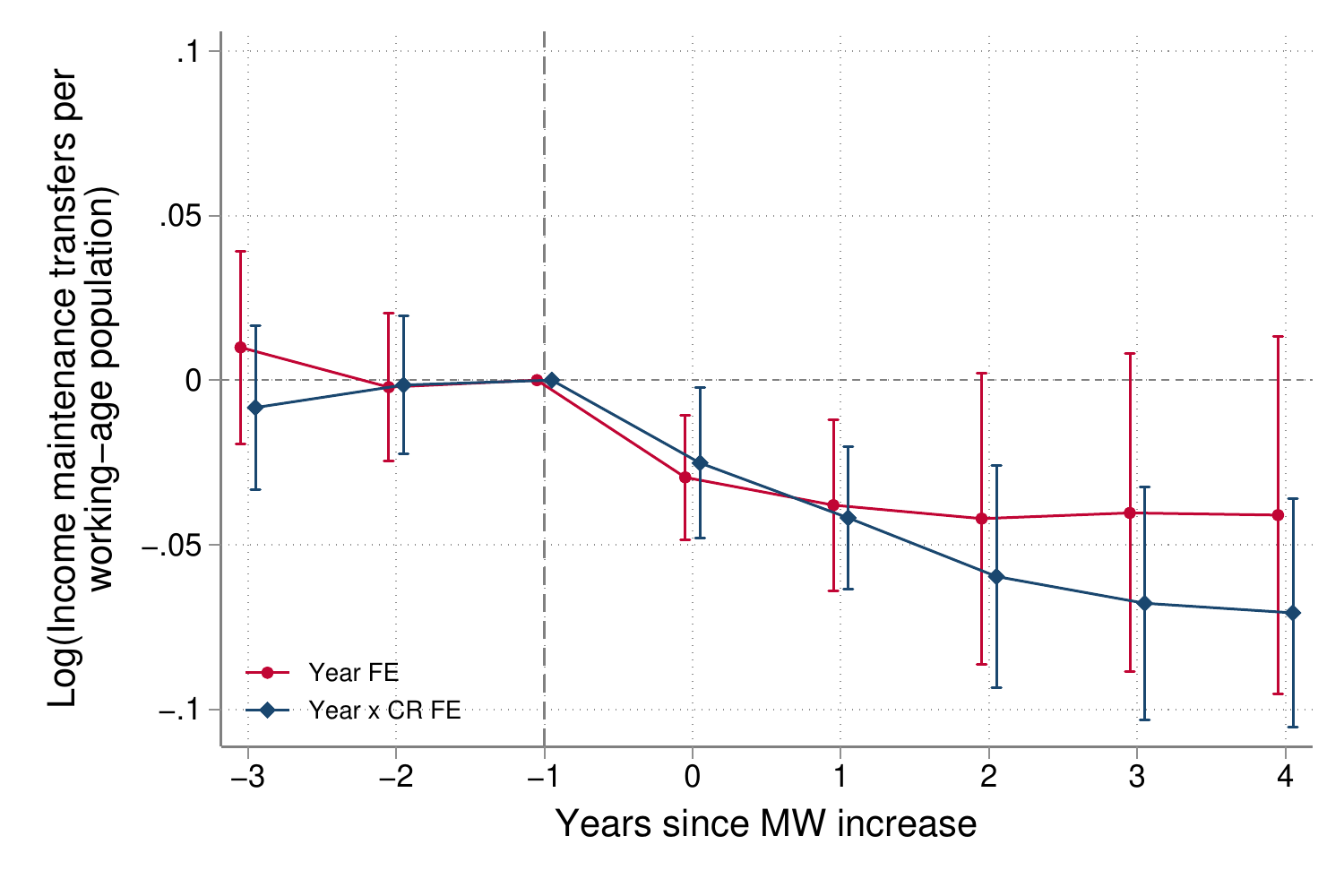}}
\subfigure[Medical benefits]{\includegraphics[width=0.49\textwidth]{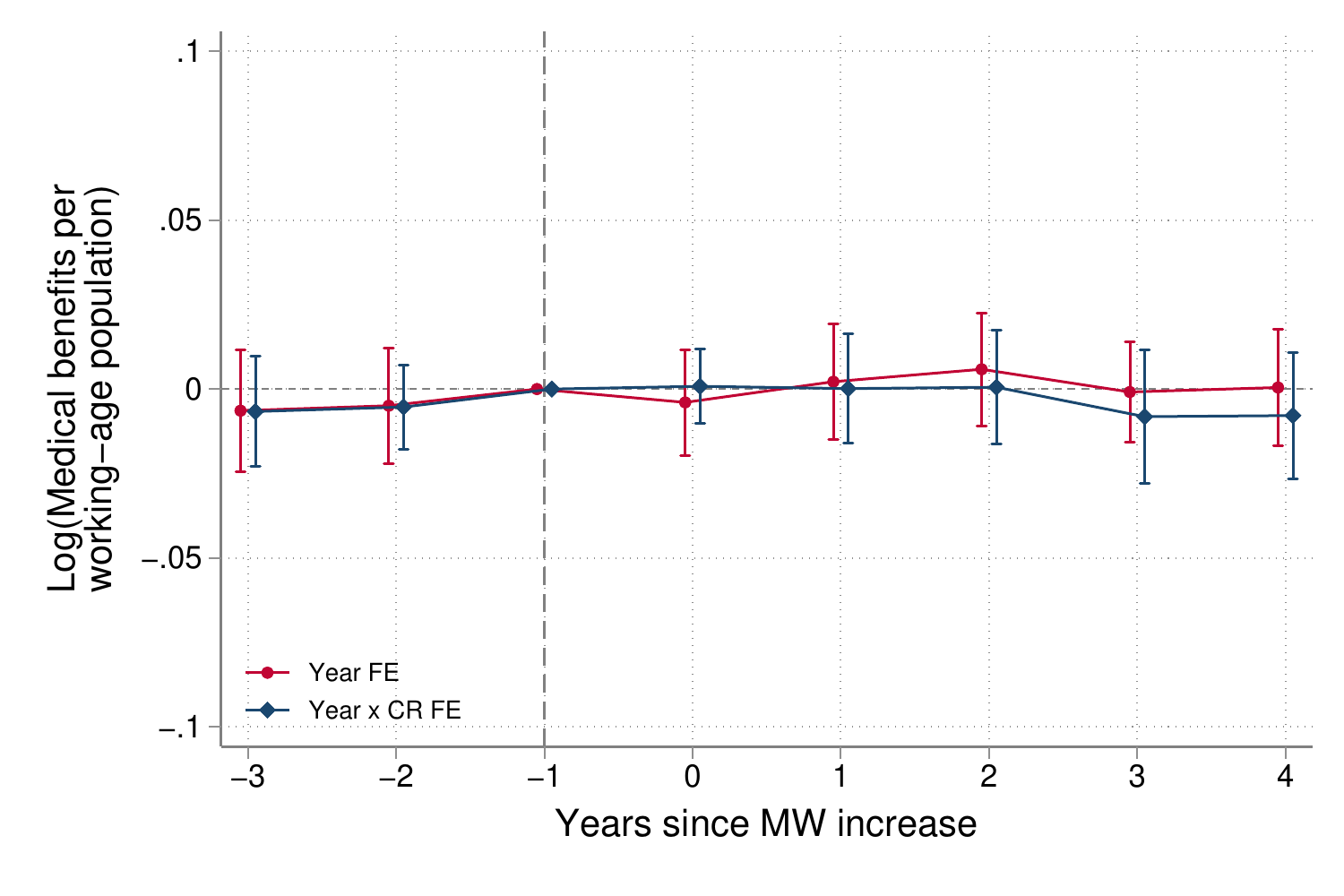}}
\subfigure[Gross federal income taxes]{\includegraphics[width=0.49\textwidth]{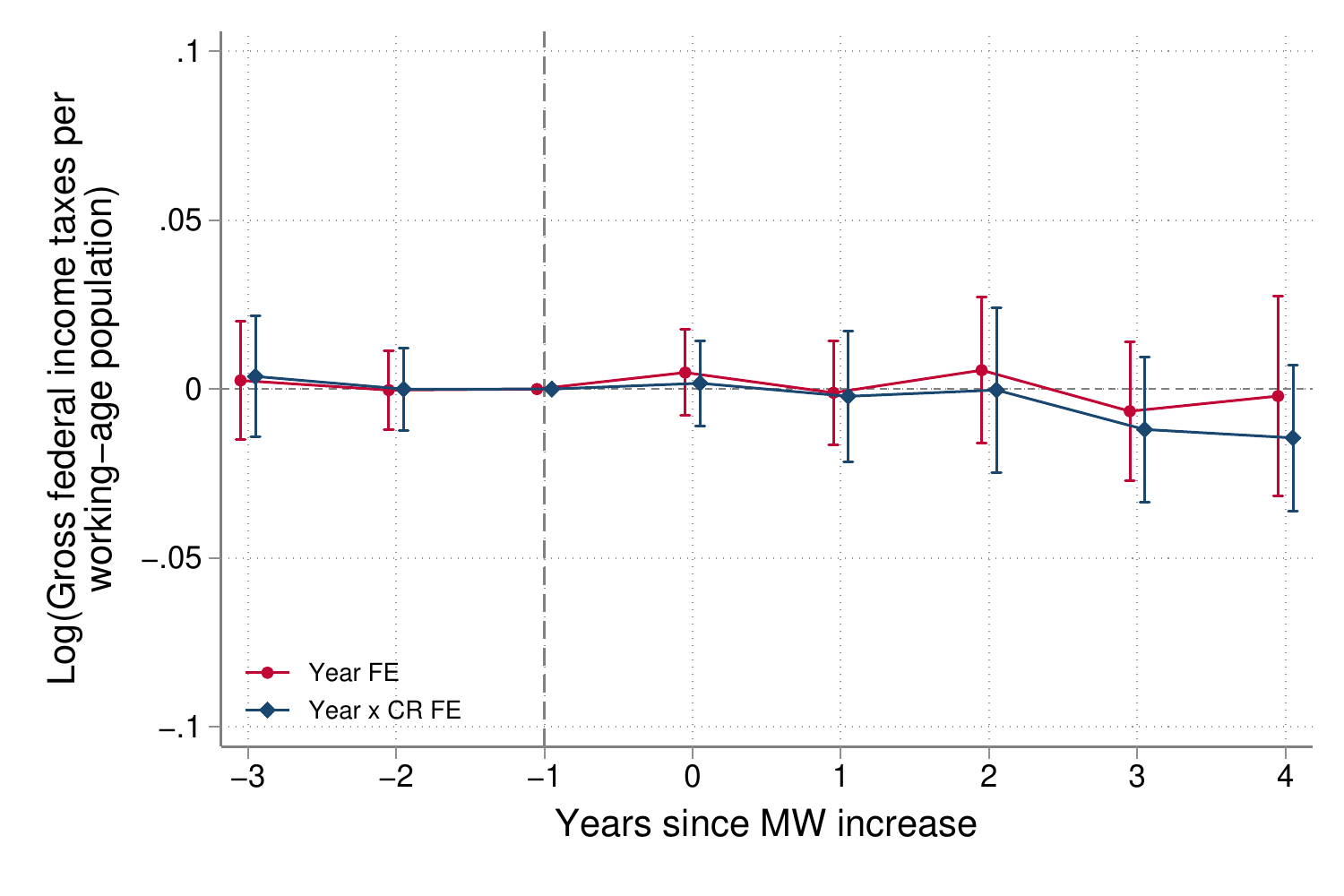}}
\begin{minipage}{\textwidth} 
{\scriptsize \vspace{0.5cm} Notes: These figures plot the estimated $\beta_{\tau}$ coefficients with their corresponding 95\% confidence intervals from equation \eqref{reg}. Panel (a) uses total income maintenance transfers per working-age individual as a dependent variable. Panel (b) uses total medical benefits per working-age individual as a dependent variable. Panel (c) uses total gross federal income taxes per working-age individual as a dependent variable. Red lines represent specifications that control by year-by-event fixed effects. Blue lines represent specifications that control by census-region-by-year-by-event fixed effects. Standard errors are clustered at the state level, and regressions are weighted by state-by-year average population.\par}
\end{minipage}
\end{figure}

\paragraph{Worker-level fiscal effects} The previous analysis focuses on pre-tax workers' outcomes. Figure \ref{es_fiscal} plots the estimated coefficients $\{\beta_{\tau}\}_{\tau=-3}^{4}$ of equation \eqref{reg} with their corresponding 95\% confidence intervals using fiscal variables as dependent variables to estimate worker-level fiscal externalities. Table \ref{T_fiscal} presents the estimated coefficients $\beta$ of equation \eqref{reg_did} that summarize the average treatment effect in the post-event period, and also includes specifications that control for census-division-by-year-by-event fixed effects. Panel (a) of Figure \ref{es_fiscal} uses total income maintenance transfers per working-age individual as a dependent variable. Consistent with \cite{reich2015effects} and \cite{dube2019minimum}, the results suggest that income maintenance benefits have decreased after state-level minimum wage increases, with the implied elasticity ranging between -0.31 and -0.39. Neither medical benefits (Panel (b)) nor gross federal income taxes (Panel (c)) show a response to changes in minimum wages, suggesting that the worker-level fiscal effects are mediated by targeted transfers based on pre-tax income levels.

\begin{figure}[t!]
\centering
\caption{Changes in capitalists' welfare after minimum wage increases}
\label{es_c}
\subfigure[Profit per establishment (all industries)]{\includegraphics[width=0.49\textwidth]{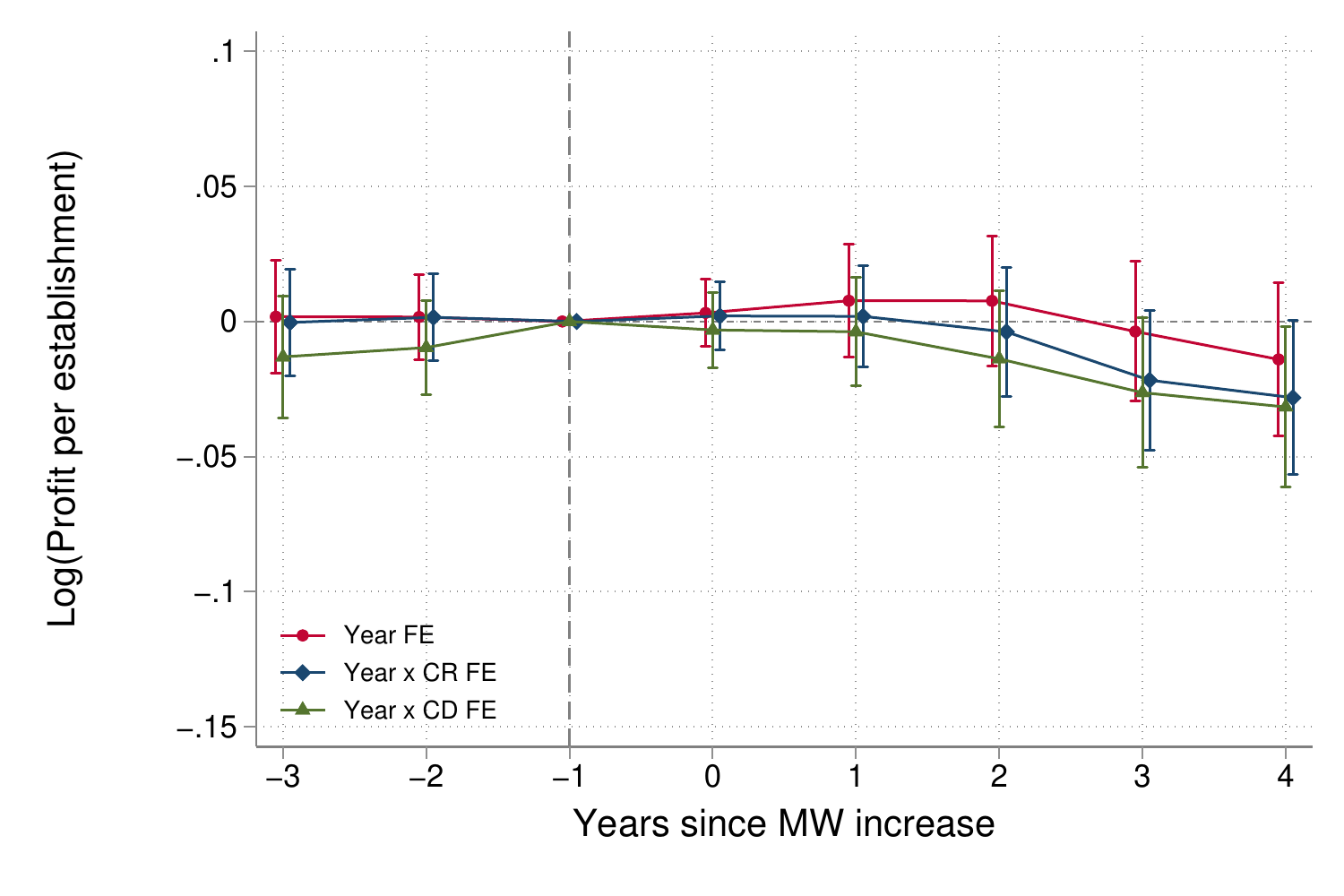}}
\subfigure[Establishments (all industries)]{\includegraphics[width=0.49\textwidth]{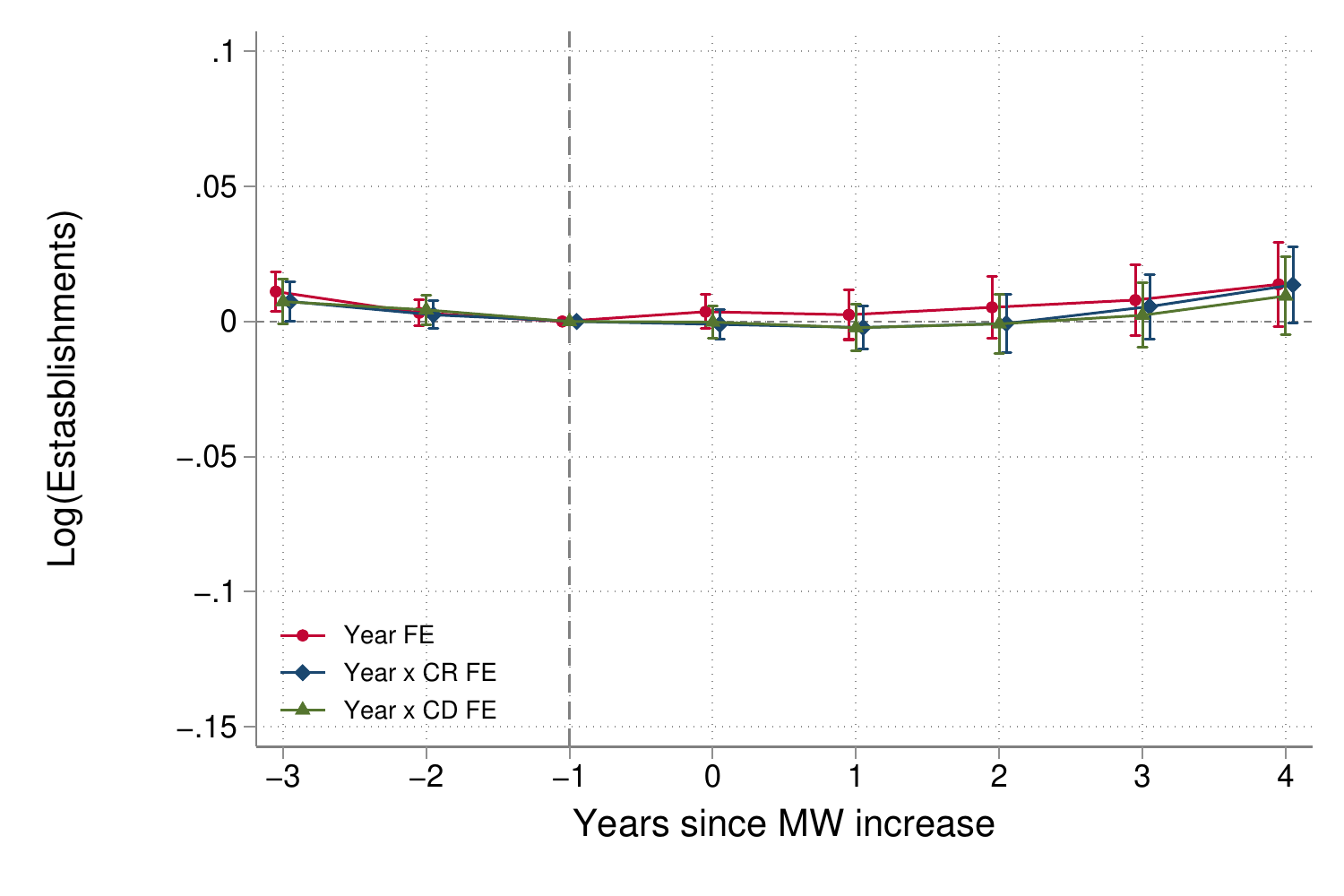}}
\subfigure[Profit per establishment (by  industry)]{\includegraphics[width=0.49\textwidth]{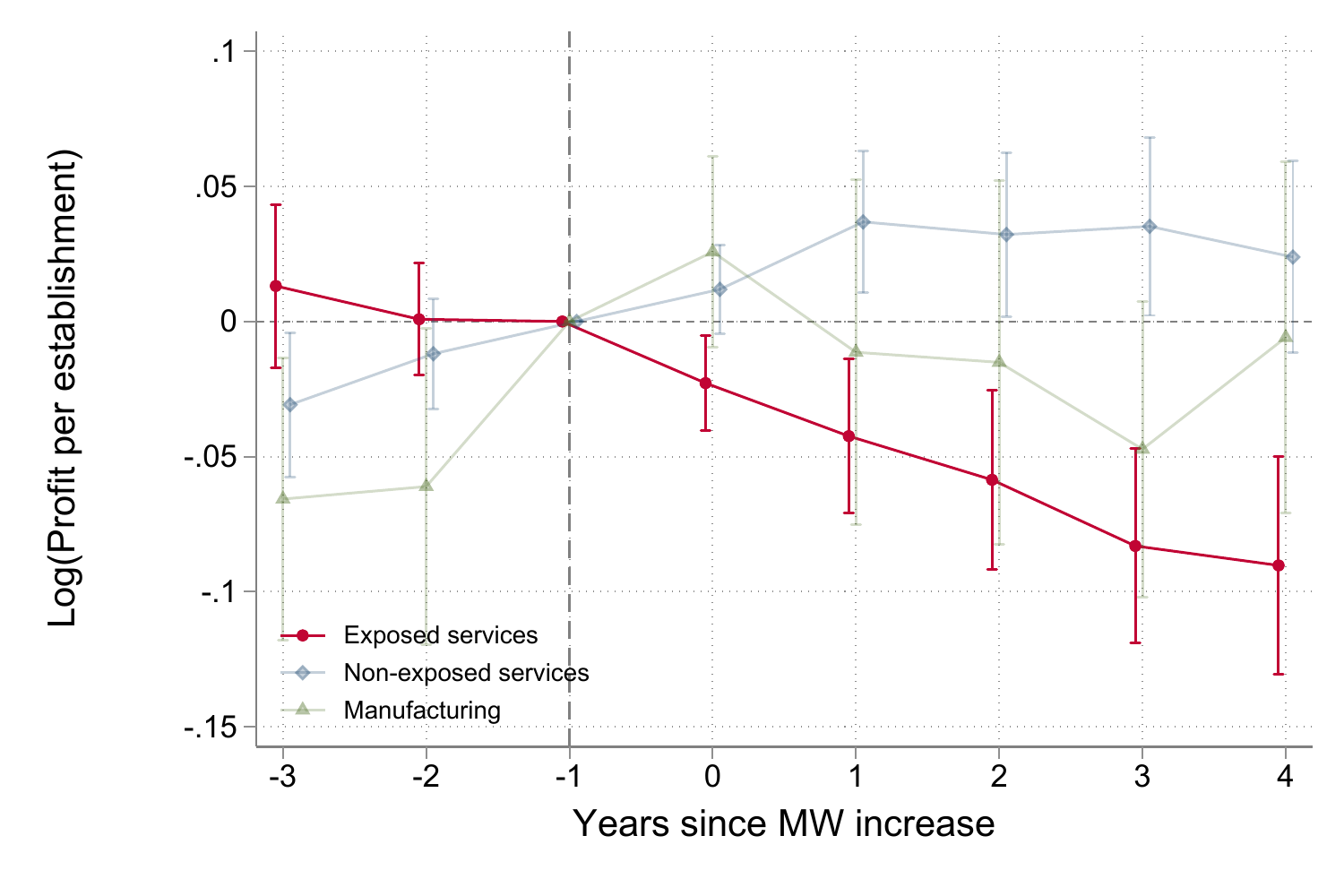}}
\subfigure[Establishments (by industry)]{\includegraphics[width=0.49\textwidth]{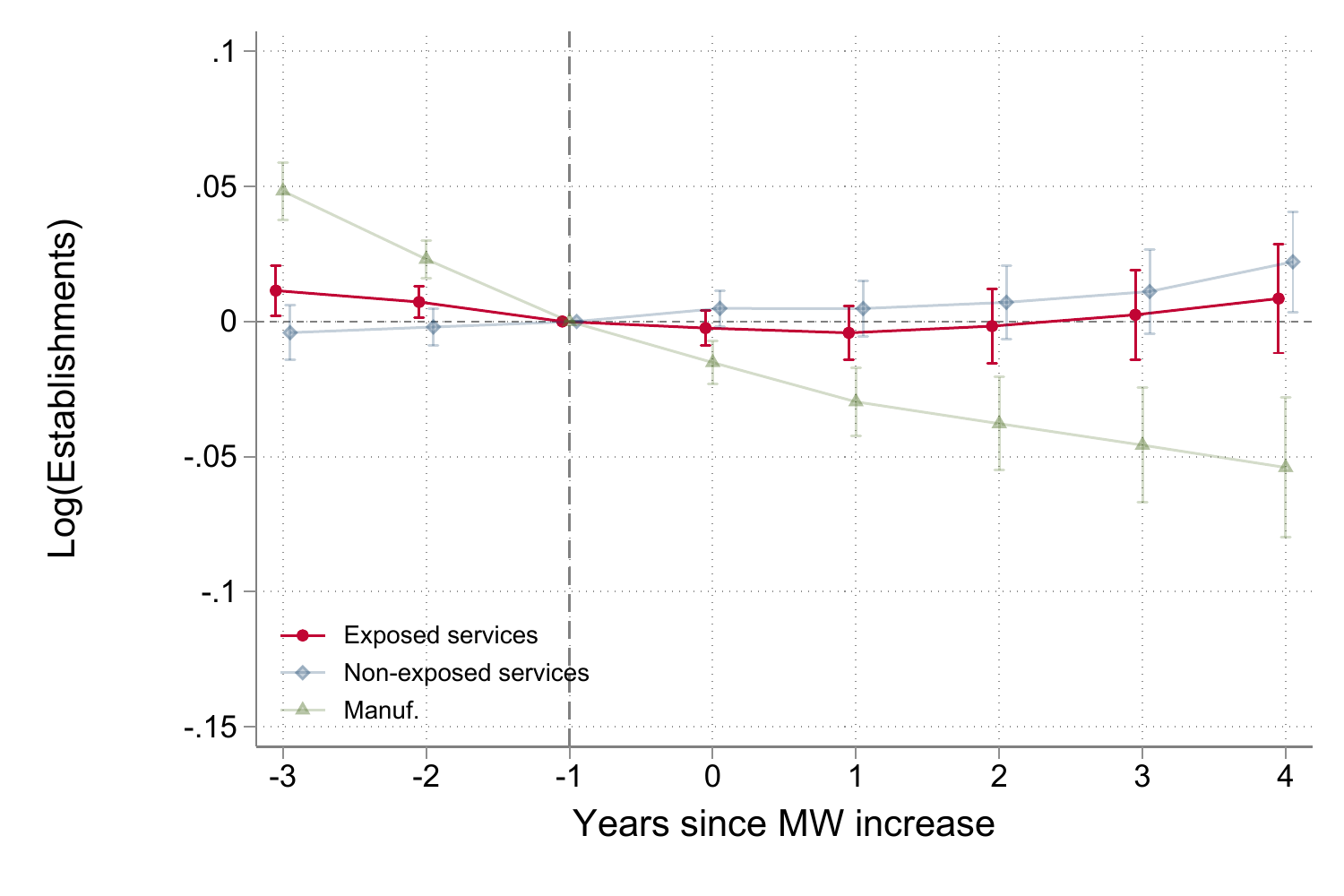}}
\begin{minipage}{\textwidth} 
{\scriptsize \vspace{0.5cm} Notes: These figures plot the estimated $\beta_{\tau}$ coefficients with their corresponding 95\% confidence intervals from equation \eqref{reg}. Panels (a) and (c) use the average profit per establishment as dependent variable. Panels (b) and (d) use the number of establishments as dependent variable. In panels (a) and (b), red lines represent specifications that control by year-by-event fixed effects, blue lines represent specifications that control by census-region-by-year-by-event fixed effects, and green lines represent specifications that control by census-division-by-year-by-event fixed effects. In panels (c) and (d) regressions control by census-division-by-year-by-event fixed effects. Regressions consider a total of 25 industries that are grouped into three categories as follows. Manufacturing industries include SIC codes 41, 43, 44, 46, 50, 54, 56, and 57, that is, nonmetallic mineral products, fabricated metal products, machinery, electrical equipment, food and
beverages and tobacco, printing and related support activities, chemical manufacturing, plastics and rubber products. Exposed services include SIC codes 9, 19, 21, 27, 28, and 34, that is, retail trade, ambulatory health services, nursing and residential care facilities, food, accommodation, and social services and other services. Non-exposed services include SIC codes 8, 10, 11, 13, 14, 15, 16, 17, 20, 24 and 25, that is, wholesale trade, transport, information, real estate, professional services, management of businesses, administrative support, educational services, hospitals, arts, and recreation industries. Standard errors are clustered at the state-by-industry level, and regressions are weighted by the average state-by-industry employment in the pre-period.\par}
\end{minipage}
\end{figure}

\paragraph{Capitalist-level pre-tax outcomes} Figure \ref{es_c} plots the estimated coefficients $\{\beta_{\tau}\}_{\tau=-3}^{4}$ of equation \eqref{reg} with their corresponding 95\% confidence intervals using capitalist-level pre-tax outcomes as dependent variable. Table \ref{T_cap} presents the estimated coefficients $\beta$ of equation \eqref{reg_did} that summarize the average treatment effect in the post-event period. Panels (a) and (b) plot regressions that pool all industries and that control by three different types of time fixed-effects: year-by-event fixed effects, census-region-by-year-by-event fixed effects, and census-division-by-year-by-event fixed effects. Panels (c) and (d) plot regressions splitting by industry group that control by census-division-by-year-by-event fixed effects since they may better capture time-varying shocks at the industry level. 

Panel (a) shows that, when pooling all industries, trends in average profits per establishment seem to be unaffected by minimum wage shocks. Panel (b) shows a similar pattern on the average number of establishments. However, Panel (c) shows a substantial decrease in the average profit per establishment in exposed services, with an implied elasticity of -0.35. Panel (d) shows that this is mainly driven by an intensive margin response since trends in establishments for these industries also seem to be unaffected by minimum wage changes.\footnote{Table \ref{T_cap} suggests a significant elasticity of -0.1 of the number of establishments to changes in minimum wages. However, Panel (d) of Figure \ref{es_c} suggests that the estimated effect is confounded by a differential pre-trend.} While these results should be interpreted with caution since they are based on non-ideal aggregate data of profits and establishments, they suggest that there is substantial profit incidence in industries where this effect is expected. 

\begin{figure}[t!]
\centering
\caption{Capitalist-level fiscal effects after minimum wage increases}
\label{es_fec}
\subfigure[Business income per tax return]{\includegraphics[width=0.49\textwidth]{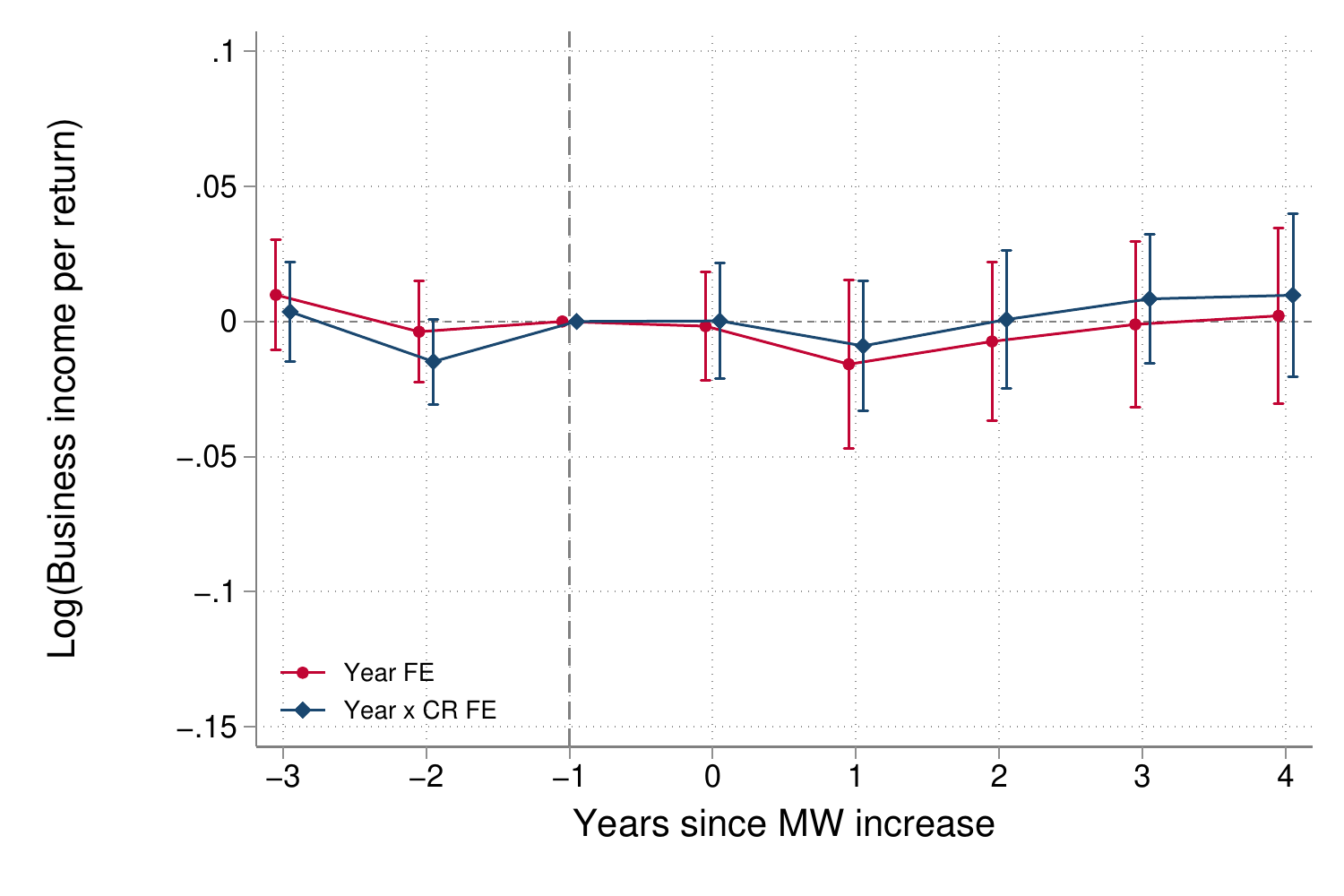}}
\subfigure[Dividend income per tax return]{\includegraphics[width=0.49\textwidth]{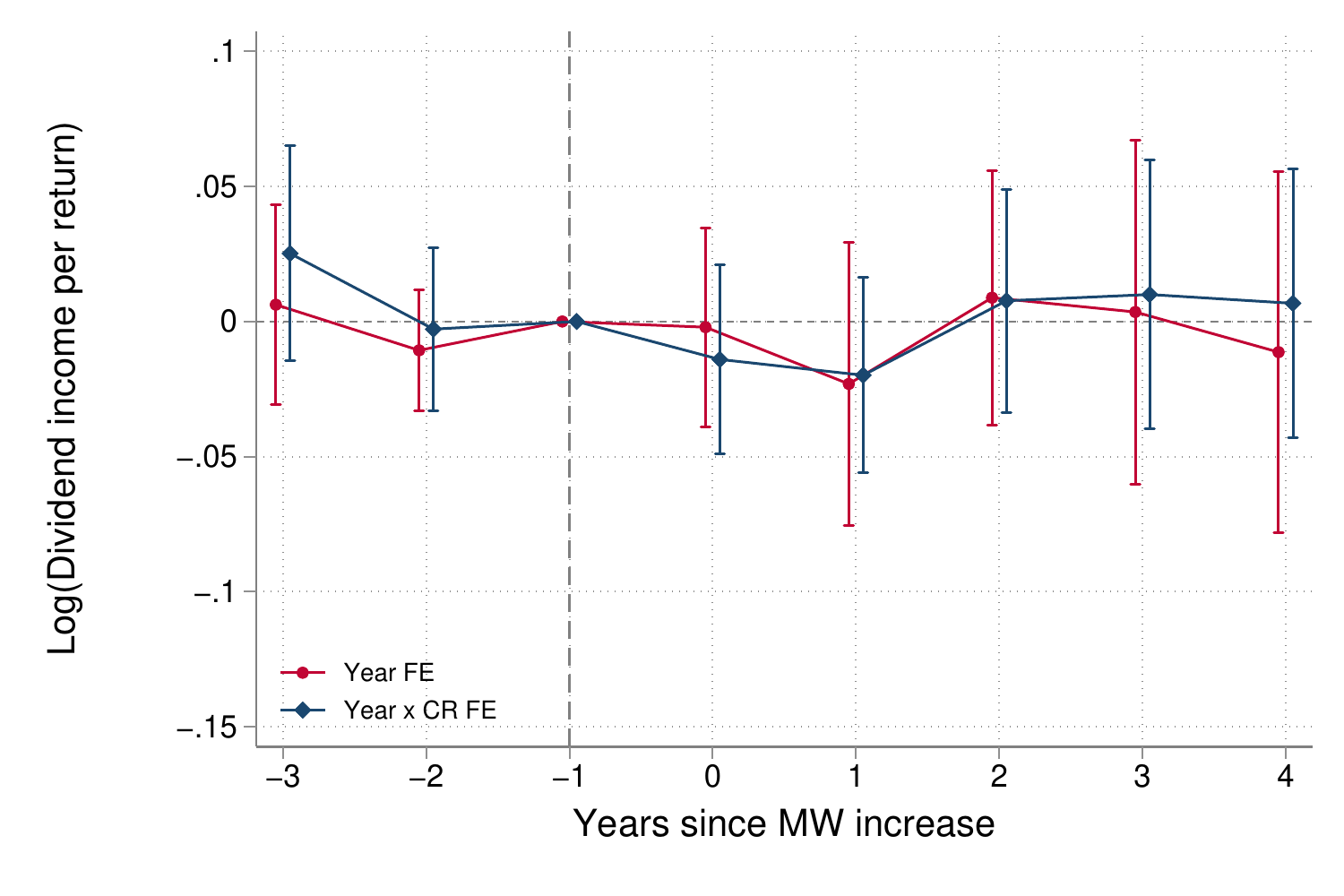}}
\subfigure[Taxes on production and imports net of subsidies]{\includegraphics[width=0.49\textwidth]{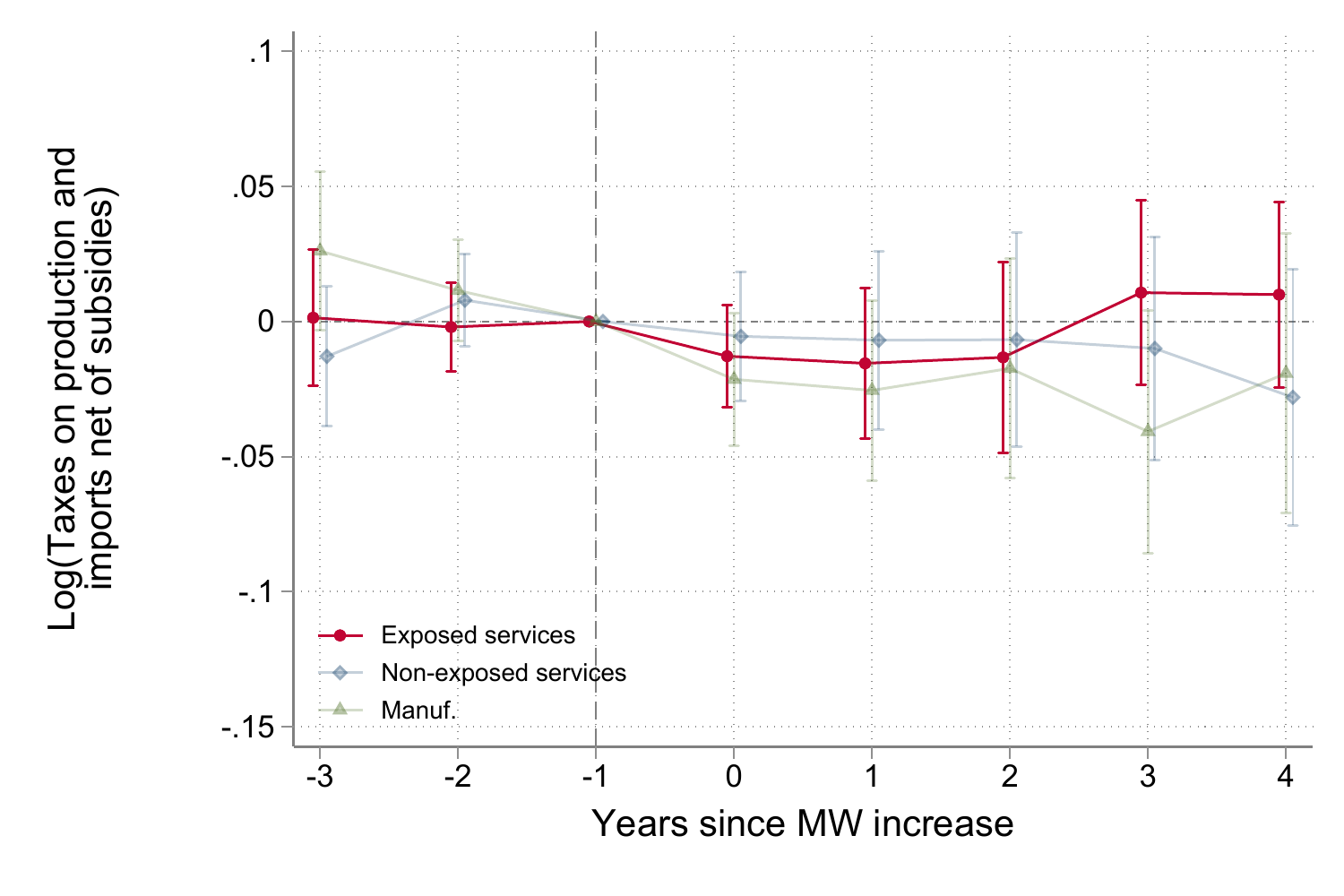}}
\begin{minipage}{\textwidth} 
{\scriptsize \vspace{0.5cm} Notes: These figures plot the estimated $\beta_{\tau}$ coefficients with their corresponding 95\% confidence intervals from equation \eqref{reg}. In Panels (a) and (b), the unit of observation is at the state-by-year level. In Panel (c), the unit of observation is at the state-by-industry-by-year level. Panel (a) uses the average business income per return as dependent variable. Panel (b) uses the average dividend income per return as dependent variable. Panel (c) uses total taxes on production and imports net of subsidies as dependent variable. In panels (a) and (b), red lines represent specifications that control by year-by-event fixed effects, and blue lines represent specifications that control by census-region-by-year-by-event fixed effects. In panels (c), the regression controls by census-division-by-year-by-event fixed effects. Panel (c) considers a total of 25 industries that are grouped into three categories as follows. Manufacturing industries include SIC codes 41, 43, 44, 46, 50, 54, 56, and 57, that is, nonmetallic mineral products, fabricated metal products, machinery, electrical equipment, food and
beverages and tobacco, printing and related support activities, chemical manufacturing, plastics and rubber products. Exposed services include SIC codes 9, 19, 21, 27, 28, and 34, that is, retail trade, ambulatory health services, nursing and residential care facilities, food, accommodation, and social services and other services. Non-exposed services include SIC codes 8, 10, 11, 13, 14, 15, 16, 17, 20, 24 and 25, that is, wholesale trade, transport, information, real estate, professional services, management of businesses, administrative support, educational services, hospitals, arts, and recreation industries. In Panels (a) and (b), standard errors are clustered at the state level, and regressions are weighted by state-by-year average population. In Panel (c), standard errors are clustered at the state-by-industry level, and the regression is weighted by the average state-by-industry employment in the pre-period.\par}
\end{minipage}
\end{figure}

\paragraph{Capitalist-level fiscal effects} The fall in profits in exposed services implies a direct fiscal loss proportional to the corporate tax rate. However, the effect on capitalists' outcomes could generate additional fiscal externalities in other parts of the tax system. Figure \ref{es_fec} shows little support for that hypothesis. In Panels (a) and (b), data varies at the state-by-year level. In Panel (c), data varies at the state-by-industry-by-year level. Panel (a) shows no effect on business income per income tax return, Panel (b) shows no effect on dividend income per income tax return, and Panel (c) shows no effect on taxes on production and imports net of subsidies. These results suggest that the capitalist-level fiscal effects are mediated by the direct effect on profits and the corresponding loss in corporate tax revenue.

\subsection{Back to the optimal policy analysis} 

Results suggest that minimum wages benefit low-skill workers, hurt capitalists in exposed industries, and generate fiscal savings in transfers and fiscal costs in corporate tax revenue. I plug the estimates into the theoretical results to interpret the estimations through the lens of the optimal policy analysis. A modified version of equation \eqref{prop2} suggests that increasing the minimum wage increases social welfare if\footnote{The estimated elasticities correspond to the macro version of the sufficient statistics and, therefore, are relevant to the calibration of Proposition II.}
\begin{eqnarray}
\frac{d\log U^l}{d\overline{w}}\cdot U^l\cdot L_A^l\cdot g_1^l + \frac{d\log\Pi^S}{d\overline{w}}\cdot \Pi^S\cdot K_A^S\cdot g_K^S + \text{Fiscal effects} &>& 0, \label{prop2_v2}
\end{eqnarray}
where I omit the high-skill workers component -- because $dU^h/d\overline{w}$ is estimated to be zero -- and denote as $\Pi^S$ the average profit per establishment in exposed services and $g_K^S$ its corresponding marginal welfare weight. The fiscal effects component considers both worker- and capitalist-level fiscal externalities.

$U^l\cdot L_A^l$ equals the sum of total pre-tax income of low-skill workers plus total income maintenance transfers,\footnote{Note from equation \eqref{suffstat_2} that
\begin{eqnarray}
U^l\cdot L_A^l &=& \int E_m^l w_m^ldm - \int E_m^lT(w_m^l)dm + y_0\cdot L_A^l\cdot\rho^l, 
\end{eqnarray}
that is, $U^l\cdot L_A^l$ equals total pre-tax income plus the net tax liabilities which are composed by the taxes paid by employed workers and the transfers received by the unemployed workers. I use the total income maintenance benefits as a proxy for total net tax liabilities of low-skill workers.} so the first term of equation \eqref{prop2_v2} can be written as $(\epsilon_{U^l_{PT}}\cdot\mbox{PTW} + \epsilon_{IT}\cdot\mbox{IT})\cdot g_1^l$, where $\epsilon_{U^l_{PT}}$ is the pre-tax version of $d\log U^l/d\overline{w}$, $\epsilon_{IT}$ is the fiscal effect on income maintenance transfers, PTW accounts for total annual pre-tax wages, and IT accounts for total income maintenance benefits. Likewise, the second component of equation \eqref{prop2_v2} can be written as $\epsilon_{\Pi^S}\cdot \mbox{PTP}\cdot(1-t)\cdot g_K^S$, where $\epsilon_{\Pi^S}$ is the profit elasticity on exposed services, and PTP accounts for total annual pre-tax profits of exposed services. Finally, fiscal effects can be written as $- \epsilon_{IT}\cdot \mbox{IT} + \epsilon_{\Pi^S}\cdot t\cdot \mbox{PTP}$. Collecting terms, I can write equation \eqref{prop2_v2} as
\begin{eqnarray}
(\epsilon_{U^l_{PT}}\cdot\mbox{PTW} + \epsilon_{IT}\cdot\mbox{IT})\cdot g_1^l +\epsilon_{\Pi}\cdot\mbox{PTP}\cdot (1-t) \cdot g_K^S - \epsilon_{IT}\cdot \mbox{IT}+ \epsilon_{\Pi^S}\cdot t\cdot \mbox{PTP}&>& 0. \label{prop2_final}
\end{eqnarray}
Values for $\{\epsilon_{U^l_{PT}},\epsilon_{IT},\epsilon_{\Pi^S}\}$ can be taken from Tables \ref{T_w}, \ref{T_fiscal}, and \ref{T_cap}. I focus on the estimates using the stricter set of time fixed effects for making elasticities comparable and consider two values for $\epsilon_{\Pi^S}$ depending on the interpretation of the extensive margin response estimate.\footnote{I impute $\epsilon_{U^l_{PT}} = 0.017$ and $\epsilon_{IT} = -0.05$. Regarding $\epsilon_{\Pi^S}$, I first assume $\epsilon_{\Pi^S} = -0.047$. I also assume that the estimated decrease in the number of establishments reported in Table \ref{T_cap} is real, which yields $\epsilon_{\Pi^S} = -0.047 - 0.015 = -0.062$.} Likewise, values for $\{\mbox{PTW},\mbox{IT},\mbox{PTP}\}$ are directly observed in the data. I follow two approaches for their computation: the population-weighted average of treated states in the pre-event year -- to assess the welfare desirability of past minimum wage increases -- and the population-weighted average of all states in 2019 -- to predict the effects of small minimum wages today.\footnote{PTW is computed by multiplying the annualized average pre-tax sufficient statistic by state and year by the working-age population and the share of low-skill workers. IT and PTP are observed directly from the raw data.} Consequently, I impute values for $t$ using two assumptions. First, I consider statutory corporate tax rates, thus imputing $t=35\%$ for assessing past minimum wage increases and $t=21\%$ for assessing minimum wage increases today. Second, I consider the effective corporate tax rates estimated by \cite{zucman2014taxing}.\footnote{Effective corporate tax rates are computed by dividing all the corporate taxes
paid by US firms (to US and foreign governments) by total US corporate profits using national accounts data taken from the BEA NIPA tables.} For the first case, I set $t=20\%$, which is the average value for the period 1997-2017. For the second case, I consider $t=13\%$, which is the most recent available value of the series.

There are two unknowns left to quantitatively assess equation \eqref{prop2_final}: the SMWWs, $\{g_1^l,g_K^S\}$. I calibrate $g_K^S$ and then back up the welfare weight on low-skill workers that makes equation \eqref{prop2_final} hold with equality, $g_1^{l*}$, which can be interpreted as the minimum social value on redistribution toward low-skill workers such that increasing the minimum wage is welfare improving. $g_1^{l*}$ is a measure of the restrictions on social preferences that make the policy change desirable. The smaller $g_1^{l*}$, the weaker the required preferences for redistribution toward low-skill workers. For this purpose, I follow two approaches to calibrate $g_K^S$. First, I set $g_K^S = 1$, which emulates a scenario in which the social planner does not have a particular preference to redistribute from or to capitalists. Second, I assume that the social welfare function, $G$, is given by $G(V) = V^{1-\zeta}/(1-\zeta)$, with $\zeta>0$. I consider $\zeta\in\{1,1.5,2\}$. Under this functional form, higher $\zeta$ represents stronger preferences for redistribution, and $\{g_1^l,g_K^S\}$ are endogenous to final allocations. Therefore, relative welfare weights are proportional to average after-tax allocations. Formally, $g_1^l/g_K^S = (U^l/(1-t)\cdot \Pi^S)^{-\zeta}$, so $g_K^S$ can be jointly determined with $g_1^{l*}$.\footnote{For simplicity, I do not consider the participation and entry costs that ultimately matter for the computation of the welfare weights. To get an empirical estimate for the average post-tax sufficient statistic, I compute $\mbox{IT}/\mbox{PTW}=14\%$, to amplify the annualized average pre-tax sufficient statistic by 14\%. When $g_K^S = 1$, $g_1^{l*}$ is given by \begin{eqnarray}
g_1^{l*} = \frac{-\left(\epsilon_{\Pi}\cdot\mbox{PTP}\cdot (1-t)  - \epsilon_{IT}\cdot \mbox{IT}+ \epsilon_{\Pi^S}\cdot t\cdot \mbox{PTP}\right)}{\epsilon_{U^l_{PT}}\cdot\mbox{PTW} + \epsilon_{IT}\cdot\mbox{IT}}.\label{g_crit_1}
\end{eqnarray} 
When $g_K^S = g_1^l/\omega(\zeta)$, with $\omega(\zeta) = (U^l/(1-t)\cdot \Pi^S)^{-\zeta}$, $g_1^{l*}$ is given by 
\begin{eqnarray}
g_1^{l*} = \frac{-\left(  - \epsilon_{IT}\cdot \mbox{IT}+ \epsilon_{\Pi^S}\cdot t\cdot \mbox{PTP}\right)}{\epsilon_{U^l_{PT}}\cdot\mbox{PTW} + \epsilon_{IT}\cdot\mbox{IT} + \epsilon_{\Pi}\cdot\mbox{PTP}\cdot (1-t)\cdot \omega(\zeta)^{-1}}.\label{g_crit_2}
\end{eqnarray}}

\paragraph{Results} Table \ref{cb} summarizes the results. Each cell reports $g_1^{l*}$ for a different permutation of the 32 calibration choices discussed above. Table \ref{cb} suggests that past minimum wage increases have been welfare-improving, and that small minimum wage increases today are likely to be as well. When $g_K^S = 1$, the policy change is close to being welfare-neutral. Using a low value for $\epsilon_{\Pi^S}$ yields $g_1^{l*}$ close to 1, meaning that the policy breaks even. Using a higher value for $g_1^{l*}$ yields a minimum SMWW on low-skill workers that makes equation \eqref{prop2_final} true equal to 1.52 and 1.54. That is, if the planner does not care about inequality between low-skill workers and exposed services capitalists, a moderate preference for redistribution toward low-wage workers justifies the minimum wage reform. However, when preferences for redistribution are incorporated in the form of a concave social welfare function, the minimum wage becomes unambiguously welfare-improving. In all 24 cases, \eqref{prop2_final} is true even if $g_1^l = 0$. 

This exercise highlights the importance of including redistributive preferences in the analysis. Even if total output falls, the incorporation of distributional concerns makes the case for the minimum wage unambiguously favorable.\footnote{The degree of concavity of the social welfare function ($\zeta$) does not affect the analysis because average post-tax profits are several times larger than average post-tax incomes of active low-skill workers (between five and six times larger), so the redistributive forces in \eqref{prop2_final} manifest even when the concavity of the social welfare function is moderate.} Intuitively, the empirical analysis shows that minimum wages benefit low-skill workers, hurt firm owners in the exposed industries, and generate fiscal savings in transfers and fiscal costs in terms of corporate tax revenue. Total after-tax gains for low-skill workers are comparable to total after-tax losses for capitalists. Also, the net fiscal effect is positive in all cases (except when the corporate tax rate is calibrated at 35\%), however, it is small relative to baseline incomes: the net fiscal effect never represents more than 0.5\% of total pre-tax incomes of low-skill workers. Then, in the absence of preferences for redistribution, the policy is close to breaking even. When preferences for redistribution enter the analysis, the change in profits only affects the fiscal effect but plays a negligible role in the welfare assessment of the change in after-tax incomes. This fact makes a positive case for the minimum wage because the distinction between winners and losers is aligned with the social planner's preferences.

\section{Conclusion}
\label{sec6}

The desirability of the minimum wage has been a controversial policy debate for decades. The wide recent evidence on its effects on wages, employment, and other relevant labor market outcomes has encouraged economists to conceptually revisit its role as part of the available instruments for governments. Concerning inequality, a central question is whether there are rationales for governments to use the minimum wage to make tax-based redistribution more efficient. This paper aims to contribute to this discussion.

I propose a general theoretical framework based on an empirically grounded model of the labor market with positive profits that characterizes optimal redistribution for a social planner that can use the income tax system, the corporate tax rate, and a minimum wage to maximize social welfare. Conditions are derived that characterize the optimal minimum wage as a function of sufficient statistics for welfare, social preferences for redistribution, and fiscal externalities. Given a tax system, a minimum wage can increase social welfare when it increases the average post-tax wages of low-skill labor market participants and when corporate profit incidence is large. When chosen together with taxes, the minimum wage can help the government redistribute efficiently to low-skill workers by preventing firms from capturing low-wage income subsidies such as the EITC and from enjoying high profits that cannot be redistributed via corporate taxes due to capital mobility in unaffected industries. 

Event studies show that the average US state-level minimum wage reform over the last two decades increased average post-tax wages of low-skill labor market participants and reduced corporate profits in affected industries, namely low-skill labor-intensive services. They also show a substantial decrease in income maintenance transfers. A sufficient statistics analysis implies that US minimum wages typically remain below their optimum under the current tax and transfer system. Minimum wage changes in recent decades likely raised welfare, and small increases today would likely do so as well.

A general message of the paper is that there are rationales to complement tax-based redistribution with binding minimum wages. Governments should not make the tax system and the minimum wage compete for who is the most efficient redistributive policy. By contrast, social planners can benefit from using all instruments together to make redistribution more efficient. Optimal redistribution possibly consists of a binding minimum wage, a corporate tax rate, and a targeted EITC.

\newpage
\bibliographystyle{chicago}
\bibliography{referencias}

\newpage

\clearpage

\clearpage

\clearpage

\clearpage

\clearpage

\begin{table}[t!]
\centering
\def\sym#1{\ifmmode^{#1}\else\(^{#1}\)\fi}
\caption{Descriptive statistics}\label{ds}
\resizebox{\textwidth}{!}{
{\small\input{FT_SEPT2022/DS}}}
\floatfoot{\scriptsize Notes: This table shows descriptive statistics for the non-stacked panel. The unit of observation is a state-year pair. Nominal values are transformed to 2016 dollars using the R-CPI-U-RS index including all items. $U^l$ and $U^h$ are the average pre-tax wage including the unemployed annualized by computing Hourly Wage $\times$ Weekly Hours $\times$ Employment Rate $\times$ 52. Worker-level aggregates are computed using the CPS-MORG data and the Basic Monthly CPS files. Income maintenance benefits, medical benefits, and gross federal income taxes are taken from the BEA regional accounts. Profit per establishment corresponds to the gross operating surplus taken from the BEA regional accounts normalized by the number of private establishments reported in the QCEW data.}
\end{table}

\clearpage

\begin{table}[t!]
\centering
\def\sym#1{\ifmmode^{#1}\else\(^{#1}\)\fi}
\caption{Worker-level results: pre-tax wages including the unemployed}\label{T_w}

{\small\input{FT_SEPT2022/T1}}

\floatfoot{\scriptsize Notes: This table shows the estimated $\beta$ coefficient from equation \eqref{reg_did}. All columns represent different regressions using different dependent variables (all in logarithms) and fixed effects. Columns 1 to 3 use the average pre-tax wage of low-skill workers including the unemployed ($U^l$) as dependent variable. Columns 4 to 6 use the average pre-tax wage of high-skill workers including the unemployed ($U^h$) as dependent variable. Year FE means that the regression includes year-by-event fixed effects. Year x CR FE means that the regression includes year-by-census region-by-event fixed effects. Year x CD FE means that the regression includes year-by-census division-by-event fixed effects. $\Delta\log\mbox{MW}$ is the average change in the log of the real state-level minimum wage across events in the year of the event. The implied elasticity is computed dividing the point estimate by $\Delta\log\mbox{MW}$. Standard errors (in parentheses) are clustered at the state level and regressions are weighted by state-by-year population.}
\end{table}

\clearpage

\begin{table}[t!]
\centering
\def\sym#1{\ifmmode^{#1}\else\(^{#1}\)\fi}
\caption{Worker-level results: fiscal effects}\label{T_fiscal}
{\small\input{FT_SEPT2022/T2}}
\floatfoot{\scriptsize Notes: This table shows the estimated $\beta$ coefficient from equation \eqref{reg_did}. All columns represent different regressions using different dependent variables (all in logarithms) and fixed effects. Columns 1 to 3 use total income maintenance transfers per working-age individual as dependent variable. Columns 4 to 6 use total medical benefits per working-age individual as dependent variable. Columns 7 to 9 use total gross federal income taxes per working-age individual as dependent variable. Year x CR FE means that the regression includes year-by-census region-by-event fixed effects. Year x CD FE means that the regression includes year-by-census division-by-event fixed effects. $\Delta\log\mbox{MW}$ is the average change in the log of the real state-level minimum wage across events in the year of the event. The implied elasticity is computed dividing the point estimate by $\Delta\log\mbox{MW}$. Standard errors (in parentheses) are clustered at the state level and regressions are weighted by state-by-year population.}
\end{table}

\clearpage

\begin{table}[t!]
\centering
\def\sym#1{\ifmmode^{#1}\else\(^{#1}\)\fi}
\caption{Capitalist-level results: firm profits and number of establishments}\label{T_cap}

{\small\input{FT_SEPT2022/T3}}

\floatfoot{\scriptsize Notes: This table shows the estimated $\beta$ coefficient from equation \eqref{reg_did}. All columns represent different regressions using different dependent variables (all in logarithms) and fixed effects. Columns 1 to 4 use the average profit per establishment as dependent variable. Columns 5 to 8 use the total number of establishments as dependent variable. Year x CR FE means that the regression includes year-by-census region-by-event fixed effects. Year x CD FE means that the regression includes year-by-census division-by-event fixed effects. All regressions include state-by-industry-by-event fixed effects. $\Delta\log\mbox{MW}$ is the average change in the log of the real state-level minimum wage across events in the year of the event. The implied elasticity is computed dividing the point estimate by $\Delta\log\mbox{MW}$. Regressions consider a total of 25 industries that are grouped into three categories as follows. Manufacturing industries include SIC codes 41, 43, 44, 46, 50, 54, 56, and 57, that is, nonmetallic mineral products, fabricated metal products, machinery, electrical equipment, food and
beverages and tobacco, printing and related support activities, chemical manufacturing, plastics and rubber products. Exposed services include SIC codes 9, 19, 21, 27, 28, and 34, that is, retail trade, ambulatory health services, nursing and residential care facilities, food, accommodation, and social services and other services. Non-exposed services include SIC codes 8, 10, 11, 13, 14, 15, 16, 17, 20, 24 and 25, that is, wholesale trade, transport, information, real estate, professional services, management of businesses, administrative support, educational services, hospitals, arts, and recreation industries. Standard errors (in parentheses) are clustered at the state-by-industry level and regressions are weighted by the average state-by-industry employment in the pre-period.}
\end{table}

\clearpage

\begin{table}[t!]
\centering
\def\sym#1{\ifmmode^{#1}\else\(^{#1}\)\fi}
\caption{Welfare effects of minimum wage reforms under fixed taxes}\label{cb}

{\small
\begin{tabular}{ccccccccc}
\multicolumn{9}{c}{\textbf{Panel (a): Low $\epsilon_{\Pi^S}$}}\\
\toprule
 &\multicolumn{4}{c}{\textbf{Past minimum wage increases}} & \multicolumn{4}{c}{\textbf{Minimum wage increases today}} \\[5pt]
& $g_K^S = 1$ & $\zeta = 1$ & $\zeta = 1.5$ & $\zeta = 2$ & $g_K^S = 1$ & $\zeta = 1$ & $\zeta = 1.5$ & $\zeta = 2$  \\ [5pt]
Statutory $t$  & 0.98 & 0.00 & 0.00 & 0.00 & 0.99 & 0.00 & 0.00 & 0.00 \\[5pt]

Effective $t$  & 0.98 & 0.00 & 0.00 & 0.00 & 0.99 & 0.00 & 0.00 & 0.00 \\ [5pt]

\bottomrule
&&&&&&&&\\
\multicolumn{9}{c}{\textbf{Panel (b): High $\epsilon_{\Pi^S}$}}\\
\toprule
 &\multicolumn{4}{c}{\textbf{Past minimum wage increases}} & \multicolumn{4}{c}{\textbf{Minimum wage increases today}} \\[5pt]
& $g_K^S = 1$ & $\zeta = 1$ & $\zeta = 1.5$ & $\zeta = 2$ & $g_K^S = 1$ & $\zeta = 1$ & $\zeta = 1.5$ & $\zeta = 2$  \\ [5pt]
Statutory $t$  & 1.52 & 0.12 & 0.09 & 0.08 & 1.54 & 0.00 & 0.00 & 0.00 \\[5pt]

Effective $t$  & 1.52 & 0.00 & 0.00 & 0.00 & 1.54 & 0.00 & 0.00 & 0.00 \\ [5pt]

\bottomrule
\end{tabular}}

\floatfoot{\scriptsize Notes: This table shows estimates for $g_1^{l*}$ for different calibration choices. All cells consider $\epsilon_{U_{PT}^l} = 0.017$ and $\epsilon_{IT} = -0.05$. Panel (a) considers $\epsilon_{\Pi^S} = -0.047$ and Panel (b) considers $\epsilon_{\Pi^S} = -0.062$. Left-panels compute $\{\mbox{PTW},\mbox{IT},\mbox{PTP}\}$ using the population-weighted average of treated states in the pre-event year, while right-panels compute $\{\mbox{PTW},\mbox{IT},\mbox{PTP}\}$ using the population-weighted average of all states in 2019. Within each sub-panel, columns consider the different approaches for computing $g_1^{l*}$. The first column assumes $g_K^S = 1$, so $g_1^{l*}$ is computed using equation \eqref{g_crit_1}. Columns two to four assume $g_K^S = g_1^l/\omega(\zeta)$, with $\omega(\zeta) = (U^l/(1-t)\cdot\Pi^S)^{-\zeta}$, so $g_1^{l*}$ is computed using equation \eqref{g_crit_2} using $\zeta \in\{1,1.5,2\}$. Within each sub-panel, the rows consider either the statutory corporate tax rate or the effective corporate tax rate. The statutory and effective corporate tax rates are $(35\%,20\%)$ in the left sub-panel and $(21\%,13\%)$ in the right sub-panel, respectively. }
\end{table}

\clearpage

\appendix

\setcounter{footnote}{0} 
\begin{center}
\LARGE{Minimum Wages and Optimal Redistribution}\\
\vspace{.5cm}
\Large{\textbf{Online Appendix}}\\
\Large{Dami\'an Vergara - UC Berkeley\footnote{Email: \href{damianvergara@berkeley.edu}{damianvergara@berkeley.edu}. This version: \today.}}\\

\end{center}
\thispagestyle{empty}

\vspace{4cm}
\startcontents[sections]
\printcontents[sections]{l}{1}{\setcounter{tocdepth}{2}}

\newpage
\pagenumbering{roman} 

\section{Theory appendix}
\label{results}

\setcounter{equation}{0} \renewcommand{\theequation}{A.\Roman{equation}}

\paragraph{Firm's problem} The first-order conditions of firms are given by
\begin{eqnarray}
w^s:\quad&& \left(\phi_s - w^s\right)\cdot \tilde{q}^s_{w} = \tilde{q}^s, \label{foc_w}\\
v^s:\quad&& \left(\phi_s - w^s\right)\cdot \tilde{q}^s = \eta_v^s, \label{foc_v}
\end{eqnarray}
for $s\in\{l,h\}$, where $\phi_s = \partial \phi/\partial n^s$ and arguments are omitted from functions to simplify notation. Is direct from the FOCs that wages are below the marginal productivities, that is, that $\phi_s >w^s$. Moreover, defining the firm-specific labor supply elasticity as $\varepsilon^s = \left(\partial n^s/\partial w^s\right)\cdot \left(w^s / n^s\right) = \widetilde{q}_w^s\cdot w^s / \widetilde{q}^{s}$, we can rearrange \eqref{foc_w} and write $\phi_s/w_s = 1/\varepsilon^s + 1$, which is the standard markdown equation \citepapp{robinson1933economics}. In this model, $\varepsilon^s$ is endogenous and finite because of the matching frictions.

Also, combining both FOCs yields $\tilde{q}^{s2} = \eta_v^s\cdot \tilde{q}^s_w$. Differentiating and rearranging terms yields
\begin{eqnarray}
\frac{dw^s}{dv^s} = \frac{\eta_{vv}^s\cdot \tilde{q}_w^s}{2\tilde{q}^s\cdot \tilde{q}_w^s-\eta_v^s\cdot \tilde{q}_{ww}^s} > 0,\label{dwdv}
\end{eqnarray}
provided $\tilde{q}_{ww}^s < 0$.\footnote{Ignoring the superscripts, note that $\tilde{q}_w = q_{\theta}\cdot(\partial\theta/\partial w)$, which is positive in equilibrium since $U$ is fixed. Then
\begin{eqnarray*}
\tilde{q}_{ww} = q_{\theta\theta}\cdot \left(\frac{\partial\theta}{\partial w}\right)^2 + q_{\theta}\cdot \frac{\partial^2 \theta}{\partial w^2}.
\end{eqnarray*}
In principle the sign of $\tilde{q}_{ww}$ is ambiguous, since $q_{\theta\theta}>0$ and $\partial^2 \theta/\partial w^2>0$. I assume that the second term dominates so $\tilde{q}_{ww}<0$. If $\mathcal{M}(L,V) = L^{\delta}V^{1-\delta}$, $\sign\left[\tilde{q}_{ww}\right] = \sign\left[\frac{-(1-T'(w))^2}{1-\delta} - T''(w)\right]$, so the condition holds as long as the tax system is not too concave. For the result above, $\tilde{q}_{ww}<0$ is a sufficient but not necessary condition, that is, $\tilde{q}_{ww}$ is allowed to be \textit{moderately} positive, which is plausible since the opposite forces in $\tilde{q}_{ww}$ are interrelated. $q_{\theta\theta}>0$ follows the concavity and constant returns to scale of the matching function. To see why $\partial^2 \theta/\partial w^2>0$, recall that $dU = p_{\theta}\cdot d\theta_m\cdot (w_m - T(w_m) - y_0) + p_m\cdot(1-T'(w))\cdot dw_m$. Setting $dU=0$ and differentiating again yields
\begin{eqnarray*}
0&=& \left(y_m\cdot p_{\theta\theta}\cdot\frac{\partial \theta_m}{\partial w_m} + 2\cdot p_{\theta}\cdot (1-T'(w_m))\right)\cdot\frac{\partial\theta_m}{\partial w_m} + p_{\theta}\cdot y_m\cdot \frac{\partial^2\theta_m}{\partial w_m^{2}} - p_m\cdot T''(w_m),
\end{eqnarray*}
which implies that $\partial^2 \theta/\partial w^2>0$ as long as the tax system is not ``too concave''.} Moreover, differentiating \eqref{foc_v} yields
\begin{eqnarray}
\left(d\phi_s - dw^s\right)\cdot \tilde{q}^s + \left(\phi_s-w^s\right)\cdot\tilde{q}^s_w\cdot dw^s &=& \eta_{vv}^s\cdot dv^s. \label{aiv}
\end{eqnarray}
Note that 
\begin{eqnarray}
d\phi_s &=& \phi_{ss}\cdot\left(\tilde{q}^s_w\cdot dw^s\cdot v^s+\tilde{q}^s\cdot dv^s\right) + \phi_{sj}\cdot\left(\tilde{q}^j_w\cdot dw^j\cdot v^j+\tilde{q}^j\cdot dv^j\right),\label{dphi}
\end{eqnarray}
where $j$ is the other skill-type. Replacing \eqref{foc_w} and \eqref{dphi} in \eqref{aiv}, yields
\begin{eqnarray}
\left(\phi_{ss}\cdot\left[\tilde{q}^s_w\cdot dw^s\cdot v^s+\tilde{q}^s\cdot dv^s\right]+\phi_{sj}\cdot\left[\tilde{q}^j_w\cdot dw^j\cdot v^j+\tilde{q}^j\cdot dv^j\right]\right)\cdot\tilde{q}^s = \eta_{vv}^s\cdot dv^s. \label{tot_dif}
\end{eqnarray}
Rearranging terms gives
\begin{eqnarray}
\frac{dv^s}{dv^j} &=& \left[\phi_{sj}\cdot\left(\tilde{q}^j_w\cdot \frac{dw^j}{dv^j}\cdot v^j+\tilde{q}^j\right)\right]^{-1}\cdot\left[\frac{\eta_{vv}^s}{\psi\cdot \tilde{q}^s} - \phi_{ss}\cdot\left(\tilde{q}^s_w\cdot \frac{dw^s}{dv^s}\cdot v^s+\tilde{q}^s\right)\right],
\end{eqnarray}
which, given \eqref{dwdv}, implies that $\sign{(dv^s/dv^j)} = \sign{\phi_{sj}}$. 

\paragraph{Efficiency properties of the decentralized equilibrium} Without loss of generality, consider a case where there is a unique skill type. A social planner that only cares about efficiency decides on sequences of vacancies and applicants to maximize total output net of costs for firms and workers, internalizing the existence of matching frictions. The objective function is given by
\begin{eqnarray}
\mathcal{V} &=& K\cdot\int_{\psi^*}^{\overline{\psi}}\left[\phi\left(\psi,n\right) - \eta(v_{\psi}) - \xi\right]dO(\psi) - \alpha\cdot\int_0^{c^*}c \cdot dF(c), \label{obj_ef}
\end{eqnarray}
subject to
\begin{eqnarray}
n &=& q\left(\frac{K\cdot v_{\psi}\cdot o(\psi)}{L_{\psi}}\right)\cdot v_{\psi},\label{r1}\\
\int_{\psi^*}^{\overline{\psi}}L_{\psi}d\psi &=& \alpha \cdot F(c^*),\label{r2}
\end{eqnarray}
where $\{c^*,\psi^*\}$ are the thresholds for workers and firms to enter the labor market, and $\{v_{\psi},L_{\psi}\}$ are the sequences of vacancies and applicants, with $\theta_{\psi} = \left(K\cdot v_{\psi}\cdot o(\psi)\right)/L_{\psi}$. The planner chooses $\{c^*,\psi^*\}$ and $\{v_{\psi},L_{\psi}\}$ to maximize \eqref{obj_ef} subject to \eqref{r1} (matches are endogenous to the number of applicants and vacancies) and \eqref{r2} (the distribution of applicants across firms has to be consistent with the number of active workers). The Lagrangian is given by
\begin{eqnarray}
\mathcal{L} &=& K\cdot\int_{\psi^*}^{\overline{\psi}}\left[\phi\left(\psi,q\left(\frac{K\cdot v_{\psi}\cdot o(\psi)}{L_{\psi}}\right)\cdot v_{\psi}\right) - \eta(v_{\psi}) - \xi\right]dO(\psi)\nonumber\\&& - \alpha\cdot\int_0^{c^*}c \cdot dF(c) + \mu\cdot\left[\alpha \cdot F(c^*) - \int_{\psi^*}^{\overline{\psi}}L_{\psi}d\psi\right], \label{lag_ef}
\end{eqnarray}
where $\mu$ is the multiplier. The first order conditions with respect to $v_{\psi}$, $L_{\psi}$, and $c^*$ are given by
\begin{eqnarray}
v_{\psi}:\quad&& \phi_n\cdot\left(q_{\theta}\cdot \theta_{\psi} + q\right) = \eta_v, \label{foc_vpsi}\\
L_{\psi}:\quad&& -\theta_{\psi}^2 \cdot q_{\theta}\cdot \phi_n = \mu,\label{foc_Lpsi}\\
c^*:\quad&& -\alpha \cdot c^* \cdot f(c^*) + \mu\cdot\alpha\cdot f(c^*) = 0.\label{foc_cstar}
\end{eqnarray}
Equation \eqref{foc_cstar} implies that $\mu = c^*$. Using that and combining \eqref{foc_vpsi} and \eqref{foc_Lpsi} yields 
\begin{eqnarray}
q\cdot\phi_n - \frac{c^*}{\theta_{\psi}} &=& \eta_v.\label{opt_eff}
\end{eqnarray}
To assess the efficiency of vacancy posting decisions and applications decisions, I check whether \eqref{opt_eff} is consistent with the decentralized equilibrium. In the absence of taxes, the threshold for workers' entry is given by $U = p(\theta_{\psi})\cdot w_{\psi}$, that holds for any $\psi$. We also know, from the properties of the matching function, that $p(\theta_{\psi}) = \theta_{\psi}\cdot q(\theta_{\psi})$. Replacing in \eqref{opt_eff} yields $q\cdot\left(\phi_n - w_{\psi}\right) = \eta_v$, which coincides with the decentralized first order condition of the firms for vacancies (see equation \eqref{foc_v}). Then, the decentralized equilibrium is efficient in terms of vacancy posting and applications.

The first order condition with respect to $\psi^*$ is given by
\begin{eqnarray}
\psi^*:\quad&& -K\cdot\left(\phi\left(\psi^*,q\left(\theta_{\psi^*}\right)\cdot v_{\psi^*}\right) - \eta\left(v_{\psi^*}\right) - \xi\right)\cdot o\left(\psi^*\right) - \mu\cdot L_{\psi^*} = 0.\label{opt_psi}
\end{eqnarray}
Equation \eqref{opt_psi} can be written as
\begin{eqnarray}
\phi\left(\psi^*,q\left(\theta_{\psi^*}\right)\cdot v_{\psi^*}\right) - \eta\left(v_{\psi^*}\right) - \frac{\mu\cdot L_{\psi^*}}{K\cdot o\left(\psi^*\right)} &=& \xi
\end{eqnarray}
Note that $\left(\mu\cdot L_{\psi^*}\right)/\left(K\cdot o\left(\psi^*\right)\right) = \left(c^*\cdot v_{\psi^*}\right)/\left(\theta_{\psi^*}\right)$. Then, $c^* = p\left(\theta_{\psi^*}\right)\cdot w_{\psi^*}$ and  $p(\theta_{\psi^*}) = \theta_{\psi^*}\cdot q(\theta_{\psi^*})$ imply that $\left(c^*\cdot v_{\psi^*}\right)/\theta_{\psi^*} =  w_{\psi^*}\cdot q(\theta_{\psi^*})\cdot v_{\psi^*}$, which implies that equation \eqref{opt_psi} is equivalent to $\Pi\left(\psi^*\right) = \xi$, which coincides with the decentralized equilibrium. Therefore, the decentralized equilibrium is efficient. $\qedsymbol$

\paragraph{Microfounding $\phi(\psi,n^l,n^h,t)$} I provide two examples to microfound the production function as a function of $t$. I consider a capital-allocation problem, and an effort-allocation problem.

Regarding the capital-allocation problem, assume that the structural production function, $\widetilde{\phi}$ depends on capital, $k$, as well as the other inputs described in Section \ref{sec2} (except for $t$), with $\widetilde{\phi}_k>0$ and $\widetilde{\phi}_{kk}<0$. Capitalists have a fixed endowment of capital, $\overline{k}$, that can be invested domestically, $k_D$, or abroad, $k_A$, with $k_D + k_A = \overline{k}$. If capitalists invest $k_D$ domestically, they get after-tax profits $(1-t)\cdot \widetilde{\Pi}(k_D,\psi)$, where $\widetilde{\Pi}(k_D,\psi)$ is the value function that optimizes wages and vacancies given capital. If capitalists invest $k_A$ abroad, they get $k_A\cdot r^*\cdot (1-t^*)$, where $r^*\cdot (1-t^*)$ is the after-tax return of capital abroad. Capitalists choose $k_D$ to maximize $(1-t)\cdot \widetilde{\Pi}(k_D,\psi) + \left(\overline{k} - k_D\right)\cdot r^*\cdot (1-t^*)$. The first-order condition is given by $(1-t)\cdot \widetilde{\Pi}_k = r^*\cdot (1-t^*)$, which characterizes the optimal capital invested domestically, $k_D^*$, as a function of $\psi$ and $t$.\footnote{This is well-defined given decreasing returns to capital in $\widetilde{\phi}$. Since capitalists own the capital, we have that $\widetilde{\Pi}_k = \partial \widetilde{\Pi}(k_D,\psi)/\partial k_D > 0$ and $\widetilde{\Pi}_{kk} = \partial^2 \widetilde{\Pi}(k_D,\psi)/\partial k_D^2 <0$.} Then, $k_D^* = k_D(\psi,t)$, so $\widetilde{\phi}(k_D(\psi,t),n^l,n^h,\psi) = \phi(\psi,n^l,n^h,t)$ and $\widetilde{\Pi}(k_D(\psi,t),\psi) = \Pi(\psi,t)$. If capitalists have no investment opportunities abroad or transportation costs are large (which would be analogous to $r^*\cdot (1-t^*)\to0$), then $k_D = \overline{k}$ and the production function no longer depends on $t$.

A similar argument can be developed with respect to the minimum wage, $\overline{w}$. If $\overline{w}$ binds, then it also affects the allocation of capital to domestic investment. Then, $\phi$ can also be written as a function of $\overline{w}$.

Under this formulation, the behavioral response of profits to corporate taxes and minimum wages can be written as a formula of capital mobility. Define the elasticity of domestic capital to changes in corporate taxes by $\varepsilon_{k,t} = -(\partial k_D/\partial t)\cdot(t/k_D)$, and the elasticity of domestic capital to changes in the minimum wage (when it binds) by $\varepsilon_{k,\overline{w}} = -(\partial k_D/\partial \overline{w})\cdot(\overline{w}/k_D)$. $\varepsilon_{k,t}$ and $\varepsilon_{k,\overline{w}}$ are interpreted as the magnitude (absolute value) of the behavioral response. Both elasticities are related through the technological role of capital in the production function. Formally, $\varepsilon_{k,\overline{w}} = a\cdot\varepsilon_{k,t}$, with $a>0$.\footnote{Differentiating the first-order condition and setting $dr^* = dt^* = 0$ yields:
\begin{eqnarray*}
-dt\cdot \widetilde{\Pi}_k + (1-t)\cdot\left(\widetilde{\Pi}_{kk}\cdot dk_D + \widetilde{\Pi}_{k\overline{w}}\cdot d\overline{w}\right) = 0,
\end{eqnarray*}
where $\widetilde{\Pi}_{k\overline{w}}<0$. Then, $\varepsilon_{k,t} = -(\partial k_D/\partial t)\cdot(t/k_D) = -(\widetilde{\Pi}_k\cdot t)/((1-t)\cdot \widetilde{\Pi}_{kk}\cdot k_D)$, and $\varepsilon_{k,\overline{w}} = -(\partial k_D/\partial \overline{w})\cdot(\overline{w}/k_D) = (\widetilde{\Pi}_{k\overline{w}}\cdot \overline{w})/((1-t)\cdot \widetilde{\Pi}_{kk}\cdot k_D)= a\cdot \varepsilon_{k,t}$, 
where $a = -\widetilde{\Pi}_{k\overline{w}}\cdot \overline{w} / \widetilde{\Pi}_k\cdot t > 0$.} 

Then, pre-tax domestic profit effects to changes in the policy parameters are given by:
\begin{eqnarray}
\frac{\partial \Pi\left(\psi,t,\overline{w}\right)}{\partial t} = \frac{\widetilde{\Pi}\left(k_D(\psi,t,\overline{w}),\psi,\overline{w}\right)}{\partial t} = \widetilde{\Pi}_k\cdot \frac{\partial k_D}{\partial t} = -\gamma_t \cdot \varepsilon_{k,t} < 0,
\end{eqnarray}
where $\gamma_t =  \widetilde{\Pi}_k \cdot k_D / t >0$, and
\begin{eqnarray}
\frac{\partial \Pi\left(\psi,t,\overline{w}\right)}{\partial \overline{w}} =
\frac{\partial \widetilde{\Pi}\left(k_D(\psi,t,\overline{w}),\psi,\overline{w}\right)}{\partial \overline{w}} =  \widetilde{\Pi}_k\cdot \frac{\partial k_D}{\partial \overline{w}} +  \widetilde{\Pi}_{\overline{w}} = -\gamma_{\overline{w}}\cdot \varepsilon_{k,t} + \widetilde{\Pi}_{\overline{w}}<0,
\end{eqnarray}
where $\gamma_{\overline{w}} = \widetilde{\Pi}_k\cdot k_D\cdot a / \overline{w}>0$. Note that the envelope theorem does not hold for this object since pre-tax profits of domestic firms represent only a fraction of the value function of the capital allocation problem. The effect of corporate tax rates on pre-tax profits is driven by the reduction in capital. The effect of minimum wages on pre-tax profits is driven by both the reduction in capital and the direct effect on labor costs (see below). 

As an alternative microfoundation, assume that the structural production function, $\widetilde{\phi}$, depends on the managerial effort of the capitalist, $e$, as well as the other inputs described in Section \ref{sec2}, with $\widetilde{\phi}_e>0$ and $\widetilde{\phi}_{ee}<0$. If capitalists exert effort $e$, they get after-tax profits $(1-t)\cdot \widetilde{\Pi}(e,\psi)$, where $\widetilde{\Pi}(e,\psi)$ is the value function that optimizes wages and vacancies given effort. Exerting effort $e$ has a cost $c(e)$, with $c_e>0$ and $c_{ee}>0$. Optimal effort solves the first order condition $(1-t)\cdot \widetilde{\Pi}_e = c_e$, which characterizes the optimal effort, $e^*$, as a function of $\psi$ and $t$. Then, $e^* = e(\psi,t)$, so $\widetilde{\phi}(e(\psi,t),n^l,n^h,\psi) = \phi(\psi,n^l,n^h,t)$ and $\widetilde{\Pi}(e(\psi,t),\psi) = \Pi(\psi,t)$. If effort plays little role in revenue or the costs are negligible, then the production function no longer depends on $t$.

\paragraph{Firms' responses to changes in the minimum wage} To see the effect of the minimum wage on firms' decisions, note that the four first order conditions (equations \eqref{foc_w} and \eqref{foc_v} for $s=\{l,h\}$) hold for firms that are not constrained by the minimum wage, while \eqref{foc_w} no longer holds for firms that are constrained by the minimum wage. Then, for firms that operate in sub-markets with $w^l_m>\overline{w}$, it is sufficient to verify the reaction of one of the four endogenous variables to changes in the minimum wage and use the within-firm correlations to predict reactions in the other variables. For firms that operate in sub-markets where $w^l_m = \overline{w}$, it is necessary to first compute the change in low-skill vacancies and then infer the changes in high-skill vacancies and wages using the within-firm between-skill correlations that still hold for the firm. 

In both cases, it is easier to work with equation \eqref{foc_v} for $s=l$. For an unconstrained firm, totally differentiating the first order condition yields
\begin{eqnarray}
\left(\left[\phi_{ll}\cdot\left(q_{\theta}^l\cdot d\theta^l\cdot v^l+ q^l\cdot dv^l\right) + \phi_{lh}\cdot\left(q_{\theta}^h\cdot d\theta^h\cdot v^h + q^h\cdot dv^h\right)\right]- dw^l\right)\cdot q^l&&\nonumber\\ + (\phi_l -w^l)\cdot q_{\theta}^l\cdot d\theta^l &=& \eta_{vv}^l\cdot dv^l,\
\end{eqnarray}
where I omitted sub-market sub-indices to simplify notation. Rearranging terms gives
\begin{eqnarray}
dw^l\cdot\left[\frac{dv^l}{dw^l}\cdot\left(\eta_{vv}^l - \phi_{ll}\cdot q^{l2} - \phi_{lh}\cdot q^h\cdot q^l\cdot\frac{dv^h}{dv^l}\right) + q^l\right]\nonumber\\ = d\theta^l\cdot q_{\theta}^l\cdot\left[(\phi_l - w^l) + \phi_{ll}\cdot v^l\cdot q^l\right] + d\theta^h\cdot q_{\theta}^h\cdot \phi_{lh}\cdot q^l. \label{wage_sp}
\end{eqnarray}
Note that the sign and magnitude of $dw^l/d\overline{w}$ depends on $d\theta^l/d\overline{w}$. With the variation in wages it is possible to predict variation in vacancies (and, therefore, firm-size) and spillovers to high-skill workers.

On the other hand, for a constrained firm, totally differentiating the first order condition yields
\begin{eqnarray}
\left(\left[\phi_{ll}\cdot\left(q_{\theta}^l\cdot d\theta^l\cdot v^l+ q^l\cdot dv^l\right) + \phi_{lh}\cdot\left(q_{\theta}^h\cdot d\theta^h\cdot v^h + q^h\cdot dv^h\right)\right]- d\overline{w}\right)\cdot q^l&&\nonumber \\ + (\phi_l -\overline{w})\cdot q_{\theta}^l\cdot d\theta^l &=& \eta_{vv}^l\cdot dv^l,
\end{eqnarray}
where I omitted sub-market sub-indices to simplify notation. Rearranging terms gives
\begin{eqnarray}
\frac{dv^l}{d\overline{w}}\cdot\left(\eta_{vv}^l - \phi_{ll}\cdot q^{l2} - \phi_{lh}\cdot q^h\cdot q^l\cdot\frac{dv^h}{dv^l}\right) &=& \frac{d\theta^l}{d\overline{w}}\cdot q_{\theta}^l\cdot\left[(\phi_l - \overline{w}) + \phi_{ll}\cdot v^l\cdot q^l\right] \nonumber\\
&& + \frac{d\theta^h}{d\overline{w}}\cdot q_{\theta}^h\cdot \phi_{lh}\cdot q^l - q^l.
\end{eqnarray}
The sign and magnitude depends on the reaction on equilibrium sub-market tightness. However, note that the first-order effect is decreasing in productivity, since $\phi_l$ is decreasing in $\psi$ and $(\phi_l - \overline{w})\to 0$ as $\overline{w}$ increases. That is, among firms that pay the minimum wage, the least productive ones are more likely to decrease their vacancies, and therefore shrink and eventually exit the market.

Finally, to see the effect of the minimum wage on profits, we can use the envelope theorem and conclude that the total effect is equal to the partial effect ignoring general equilibrium changes on endogenous variables. This implies that for constrained firms
\begin{eqnarray}
\frac{d\Pi(\psi,t)}{d\overline{w}} = \frac{\partial \Pi(\psi,t)}{\partial \overline{w}} = q_{\theta}^l\cdot\frac{\partial \theta^l}{\partial \overline{w}}\cdot v^l\cdot (\phi_l - \overline{w}) - v^l\cdot q^l,\label{prof_ef_1}
\end{eqnarray}
where $\partial \theta^l/\partial \overline{w} = \partial \theta^l/\partial w + (\partial \theta^l/\partial U^l)\cdot(\partial U^l/\partial \overline{w})$. This effect is possibly negative given that the first-order condition with respect to low-skill wages holds with inequality and is stronger for less productive firms. When $\overline{w}=w^l$, the envelope theorem cancels out part of the effect, although profits still can be affected by general equilibrium effects through $U^l$. That is why unconstrained firms
\begin{eqnarray}
\frac{d\Pi(\psi,t)}{d\overline{w}}=\frac{\partial \Pi(\psi,t)}{\partial \overline{w}} = q_{\theta}^l\cdot\frac{\partial \theta^l}{\partial U^l}\cdot\frac{\partial U^l}{\partial \overline{w}}\cdot v^l\cdot (\phi_l - w^l),\label{prof_ef_2}
\end{eqnarray}
so the effect on profits is uniquely mediated by the effect on job-filling probabilities. 

\paragraph{Employment effects and workers' welfare} For simplicity, assume away taxes. In equilibrium, $U^s = p_m^s\cdot w_m^s$. Multiplying by $L_m^s$ at both sides and integrating over $m$ yields $L_A^s\cdot U^s = \int E_m^sw_m^sdm$, where $E_m^s = L_m^s\cdot p_m^s$ is the mass of employed workers of skill $s$ in sub-market $m$. Differentiating gives
\begin{eqnarray}
\frac{dU^s}{d\overline{w}} \cdot\left(L_A^s + U^s\cdot\alpha_s\cdot f_s(U^s)\right) &=& \int\left(\frac{dE_m^s}{d\overline{w}}\cdot w_m^s + E_m\cdot \frac{dw_m^s}{d\overline{w}}\right)dm,
\end{eqnarray}
where I used $L^s_A = \alpha_s\cdot F_s(U^s)$. The left-hand side is the welfare effect on workers times a positive constant. Then, the right-hand side can be used to calculate the wage-weighted disemployment effects, $\int (dE_m^s/d\overline{w})w_m^sdm$, that can be tolerated for the minimum wage to increase aggregate welfare for workers given employment-weighted wage effects. If both employment and wage effects are positive, the welfare effect on workers is unambiguously positive.

\paragraph{Additional discussion on the limitations of the model} I briefly discuss the implications of abstracting from dynamics, intensive margin responses, and informal labor markets.

\textit{Dynamics}: The model is static. The implications of this assumption for the optimal policy analysis are, in principle, ambiguous. \citeapp{dube2016minimum} and \citeapp{gittings2016getting} show that minimum wage shocks decrease employment flows -- separation, hires, and turnover rates -- while keeping the employment stock constant, thus increasing job stability. In the presence of labor market frictions, this induces a dynamic efficiency gain from minimum wage increases that is not captured by the model. On the other hand, \citeapp{sorkin2015there}, \citeapp{aaronson2018industry}, and \citeapp{hurst} argue that the long-run employment distortions of minimum wage shocks are larger than the short-run responses, because of long-run capital substitution through technological change.

\textit{Intensive margin responses}: The model assumes segmented labor markets. This assumption implies that the model abstracts from intensive margin responses \citepapp{saez2002optimal}. For example, increasing the minimum wage could induce high-skill workers to apply to low-skill vacancies. To the extent that these responses are empirically relevant, this is a caveat of the policy analysis. Note that this is different from changes in demand for skills, as suggested by \citeapp{butschek2019raising} and \citeapp{clemens2021dropouts}. The model can rationalize this by changes in the skill composition of posted vacancies mediated by $\phi$. Intensive margin responses could also affect incentives conditional on labor market segmentation. For example, workers may want to work more hours if the after-tax wage increases, or firms could offer jobs with shorter schedules \citepapp{jardim2022minimum}. This mechanism is muted in the model, mainly motivated by the fact that at the bottom of the wage distribution extensive margin responses tend to play a more important role to understand workers behavior. My empirical analysis finds no effect on hours worked conditional on employment, providing empirical support to the assumption.

\textit{Informality}: In some contexts, the interaction between the minimum wage and the degree of formality of the labor market may be a first order consideration. In the model, the costs of participating in the labor market, which are not taxed, may rationalize heterogeneity in outside options, including informal labor market opportunities. However, changing the characteristics of the formal sector may affect both the supply and demand for formal jobs. For detailed analyses, see \citeapp{bosch2010minimum}, \citeapp{meghir2015wages}, \citeapp{perez2020minimum}, and \citeapp{haanwinckel2020workforce}.

\paragraph{Additional discussion on the empirical effects of minimum wages} I briefly discuss the price and productivity effects documented in the empirical literature and its implications for the policy analysis.

\textit{Price effects}: The model assumes that output prices are fixed, ruling out price increases after minimum wage shocks. However, the empirical literature finds substantial passthrough to prices \citepapp{allegretto2016local,harasztosi2019pays,renkin2020pass,leung2021minimum,ashenfelter2021wages}. Modeling price increases after minimum wage shocks in the presence of limited employment effects is challenging: if employment does not fall and demand curves are downward sloping, prices should decrease rather than increase. \citeapp{bhaskar1999minimum} and \citeapp{sorkin2015there} reconcile limited employment effects with price increases in dynamic frameworks. Price effects matter for welfare since they can erode nominal minimum wage increases. Also, the unemployed and non-employed households can be made worse off given the absence of nominal improvements \citepapp{macurdy2015effective}. The distributional effect depends on which consumers buy the goods produced by firms that pay the minimum wage, and the relative importance of these goods in aggregate consumption. It also depends on the share of minimum wage workers since it affects the mapping from product-level prices to economy-level price indexes.

While more research is needed to assess the distributional impacts of the price effects, the available evidence suggests that they are unlikely to play a big role in the aggregate distributional analysis. Minimum wage workers represent a small share of the aggregate labor market, so it is unlikely that a small share of price increases can have first-order effects on aggregate price indexes. Also, \citeapp{harasztosi2019pays} show that the goods produced by firms that pay the minimum wage are evenly consumed across the income distribution, which neutralizes the potential unintended consequences through redistribution from high-income consumers to low-skill workers. \citeapp{ashenfelter2021wages} analyze McDonald's restaurants responses to local minimum wage shocks and show that the elasticity of the number of Big Mac's that can be purchased by minimum wage workers is around 80\% of the own-wage elasticity, meaning that even if workers spend all their money in Big Mac's, their real wage increases are still sizable. \citeapp{renkin2020pass} also suggest that the price effects do not neutralize the redistributive potential of the minimum wage, arguing that: ``the rise in grocery store prices following a \$1 minimum wage increase reduces real income by about \$19 a year for households earning less than \$10,000 a year. (...). The price increases in grocery stores offset only a relatively small part of the gains of minimum wage hikes. Minimum wage policies thus remain a redistributive tool even after accounting for price effects in
grocery stores.'' Based on these pieces of evidence, I conjecture that ignoring price effects is unlikely to dramatically affect the conclusions of the policy analysis and leave this extension to future research.

\textit{Productivity effects}: The model assumes that labor productivity is independent from the minimum wage. This abstracts from recent literature that finds that minimum wages can increase both workers' \citepapp{coviello2020minimum,ruffini,emanuel2022firm,ku2020does} and firms' \citepapp{riley2017raising,mayneris2018improving} productivities. Potential mechanisms include efficiency wages \citepapp{shapiro1984equilibrium} and effects on investment in training \citepapp{acemoglu1999structure}. \citeapp{harasztosi2019pays} argue that it is unlikely that productivity increases play a major role at the firm level as it would contradict the heterogeneous employment effects found between tradable and non-tradable sectors. If these effects are substantial, abstracting from these worker- and firm-specific increases in productivity after minimum wage hikes is likely to make the case for a positive minimum wage conservative. Note, however, that the model can accommodate aggregate increases in productivity through reallocation effects, as in 
\citeapp{dustmann2019reallocation}. Importantly, the main policy results depend on reduced-form profit elasticities that are robust to productivity increases.

\newpage
\section{Proofs}
\label{proofs}

\setcounter{equation}{0} \renewcommand{\theequation}{B.\Roman{equation}}

\paragraph{Proposition I} With no taxes, there is no budget constraint and the social welfare function is given by
\begin{eqnarray}
SW(\overline{w}) &=& \left(L_I^l + L_I^h + K_I\right)\cdot G(0) + \alpha_l\cdot\int_0^{U^l}G(U^l-c)dF_l(c)\nonumber\\&&+ \alpha_h\cdot\int_0^{U^h}G(U^h-c)dF_h(c)+ K\cdot\int_{\psi^*}^{\overline{\psi}}G\left(\Pi(\psi) - \xi\right)dO(\psi). \label{SW}
\end{eqnarray}
Replacing $L_I^l + L_I^h = 1 - L_A^l - L_A^h$, the total derivative with respect to the minimum wage is given by
\begin{eqnarray}
\frac{dSW}{d\overline{w}} &=& \left(\frac{dK_I}{d\overline{w}}-\frac{dL_A^l}{d\overline{w}}-\frac{dL_A^h}{d\overline{w}}\right)\cdot G(0)\nonumber\\
&&+ \alpha_l\cdot G(0)\cdot f_l(U^l)\cdot \frac{dU^l}{d\overline{w}}+\alpha_l\cdot \frac{dU^l}{d\overline{w}}\cdot\int_0^{U^l}G'(U^l-c)dF_l(c)\nonumber\\
&&+ \alpha_h\cdot G(0)\cdot f_h(U^h)\cdot \frac{dU^h}{d\overline{w}}+\alpha_h\cdot \frac{dU^h}{d\overline{w}}\cdot\int_0^{U^h}G'(U^h-c)dF_h(c)\nonumber\\
&& + K\cdot \left(\int_{\psi^*}^{\overline{\psi}}G'(\Pi(\psi)-\xi)\frac{d\Pi(\psi)}{d\overline{w}}dO(\psi) - \frac{d\psi^*}{d\overline{w}}\cdot G(0) \cdot o(\psi^*)\right).\label{p1eq1}
\end{eqnarray}
We have that $dL_A^s/d\overline{w} = d(\alpha_s\cdot F_s(U^s))/d\overline{w} = \alpha_s\cdot f_s(U^s)\cdot(dU^s/d\overline{w})$, for $s\in\{l,h\}$, and $dK_I/d\overline{w} = d(K\cdot O(\psi^*))/d\overline{w} = K\cdot o(\psi^*)\cdot (d\psi^*/d\overline{w})$. Then, equation \eqref{p1eq1} is reduced to
\begin{eqnarray}
\frac{dSW}{d\overline{w}} &=& \alpha_s\cdot \frac{dU^l}{d\overline{w}}\cdot\int_0^{U^l}G'(U^l-c)dF_l(c)+\alpha_h\cdot \frac{dU^h}{d\overline{w}}\cdot\int_0^{U^h}G'(U^h-c)dF_h(c)\nonumber\\&& + K\cdot \int_{\psi^*}^{\overline{\psi}}G'(\Pi(\psi)-\xi)\frac{d\Pi(\psi)}{d\overline{w}}dO(\psi).\label{p1eq2}
\end{eqnarray}
Using the marginal welfare weights definitions, equation \eqref{p1eq2} can be written as
\begin{eqnarray}
\frac{dSW}{d\overline{w}} &=& \gamma\cdot\left( \frac{dU^l}{d\overline{w}}\cdot L_A^l \cdot g_1^l +\frac{dU^h}{d\overline{w}}\cdot L_A^h \cdot g_1^h+ K\cdot\int_{\psi^*}^{\overline{\psi}}g_{\psi} \frac{d\Pi(\psi)}{d\overline{w}}dO(\psi)\right).\quad\qedsymbol  \label{p1eq3}
\end{eqnarray}


\paragraph{Proposition II}
With fixed taxes, the Lagrangian is given by
\begin{eqnarray}
\mathcal{L}(\overline{w},y_0) &=& \left(L_I^l + L_I^h+K_I\right)\cdot G(y_0) \nonumber\\&&+\alpha_l\cdot\int_0^{U^l-y_0}G(U^l-c)dF_l(c)+\alpha_h\cdot\int_0^{U^h-y_0}G(U^h-c)dF_h(c)\nonumber\\
&&+ K\cdot \int_{\psi^*}^{\overline{\psi}}G((1-t)\cdot\Pi(\psi,t) - \xi)dO(\psi)+ \gamma\cdot\left[\int\left(E_m^lT(w_m^l)+E_m^hT(w_m^h)\right)dm\right.\nonumber\\
&&\left.+ t\cdot K\cdot \int_{\psi^*}^{\overline{\psi}}\Pi(\psi,t)dO(\psi)- y_0\left(L_I^l +L_I^h +K_I + \rho^l\cdot L_A^l + \rho^h\cdot L_A^h\right)\right],\label{p2eq1}
\end{eqnarray}
where $\gamma$ is the budget constraint multiplier. Since $\rho^s\cdot L_A^s = L_A^s - \int E_m^sdm$, and using the fact that $L_I^l + L_I^h + L_A^l + L_A^h = 1$, equation \eqref{p2eq1} can be written as
\begin{eqnarray}
\mathcal{L}(\overline{w},y_0) &=& \left(L_I^l + L_I^h+K_I\right)\cdot G(y_0) \nonumber\\ &&+\alpha_l\cdot\int_0^{U^l-y_0}G(U^l-c)dF_l(c)+\alpha_h\cdot\int_0^{U^h-y_0}G(U^h-c)dF_h(c)\nonumber\\
&&+ K\cdot \int_{\psi^*}^{\overline{\psi}}G((1-t)\cdot\Pi(\psi,t) - \xi)dO(\psi)+ \gamma\cdot\left[\int\left(E_m^l(T(w_m^l)+y_0)\right.\right.\nonumber\\
&&\left.\left.+E_m^h(T(w_m^h)+y_0)\right)dm+ t\cdot K\cdot \int_{\psi^*}^{\overline{\psi}}\Pi(\psi,t)dO(\psi)- y_0\left(1 +K_I\right)\right].\label{p2eq2}
\end{eqnarray}
The total derivative with respect to $\overline{w}$, taking $y_0$, $t$, and $T(\cdot)$ as given, is given by
\begin{eqnarray}
\frac{d\mathcal{L}}{d\overline{w}} &=& \left(\frac{dK_I}{d\overline{w}}-\frac{dL_A^s}{d\overline{w}} - \frac{dL_A^h}{\overline{w}}\right)\cdot G(y_0)\nonumber\\ &&+ G(y_0)\cdot\alpha_l\cdot f_l(U^l-y_0)\cdot\frac{dU^l}{d\overline{w}} + \alpha_l\cdot\frac{dU^l}{d\overline{w}}\cdot\int_0^{U^l-y_0}G'(U^l-c)dF_l(c) \nonumber \\
&&+ G(y_0)\cdot\alpha_h\cdot f_h(U^h-y_0)\cdot\frac{dU^h}{d\overline{w}} + \alpha_h\cdot\frac{dU^h}{d\overline{w}}\cdot\int_0^{U^h-y_0}G'(U^h-c)dF_h(c)\nonumber \\
&& +K\cdot\left[\int_{\psi^*}^{\overline{\psi}}G'((1-t)\cdot\Pi(\psi,t)-\xi)(1-t)\frac{d\Pi(\psi)}{d\overline{w}}dO(\psi) - G(y_0)\cdot o(\psi^*)\cdot\frac{d\psi^*}{d\overline{w}}\right]\nonumber\\
&& \gamma\cdot\left[\int\left(\frac{dE_m^l}{d\overline{w}}\left(T(w_m^l) + y_0\right) + E_m^lT'(w_m^l)\frac{dw_m^l}{d\overline{w}}\right.\right.\nonumber\\&&\left.\left. + \frac{dE_m^h}{d\overline{w}}\left(T(w_m^h) + y_0\right) + E_m^hT'(w_m^h)\frac{dw_m^h}{d\overline{w}}\right)dm\right.\nonumber\\
&&\left. +t\cdot K\cdot\left(\int_{\psi^*}^{\overline{\psi}}\frac{d\Pi(\psi,t)}{d\overline{w}}dO(\psi) - \Pi(\psi^*,t)\cdot o(\psi^*)\cdot \frac{d\psi^*}{d\overline{w}}\right) - y_0\cdot\frac{dK_I}{d\overline{w}}\right].\label{p2eq3}
\end{eqnarray}
We have that $dK_I/d\overline{w} = K\cdot o(\psi^*)\cdot (d\psi^*/d\overline{w})$ and $dL_A^s/d\overline{w} = \alpha_s\cdot f_s(U^s-y_0)\cdot (dU^s/d\overline{w})$ for $s\in\{l,h\}$. Using the social marginal weights definitions, and grouping common terms, equation \eqref{p2eq3} can be written as
\begin{eqnarray}
\frac{d\mathcal{L}}{d\overline{w}}\cdot\frac{1}{\gamma} &=& \frac{dU^l}{d\overline{w}}\cdot L_A^l\cdot g_1^l+\frac{dU^h}{d\overline{w}}\cdot L_A^h\cdot g_1^h+ K\cdot (1-t)\cdot \int_{\psi^*}^{\overline{\psi}}g_{\psi}\frac{d\Pi(\psi,t)}{d\overline{w}}dO(\psi)\nonumber \\
&&+\int\left(\frac{dE_m^l}{d\overline{w}}\left(T(w_m^l) + y_0\right) + E_m^lT'(w_m^l)\frac{dw_m^l}{d\overline{w}}\right)dm\nonumber \\ 
&&+ \int\left(\frac{dE_m^h}{d\overline{w}}\left(T(w_m^h) + y_0\right) + E_m^hT'(w_m^h)\frac{dw_m^h}{d\overline{w}}\right)dm\nonumber\\
&& + t\cdot K\cdot \int_{\psi^*}^{\overline{\psi}}\frac{d\Pi(\psi,t)}{d\overline{w}}dO(\psi) - \frac{dK_I}{d\overline{w}}\cdot\left(t\cdot \Pi(\psi^*,t) + y_0\right).\label{p2eq4} \quad\qedsymbol
\end{eqnarray}


\paragraph{Proposition III} Assuming either $\max_i w_i^l < \min_j w_j^h$ or that the social planner can implement skill-specific income tax schedules, allows to solve the problem by pointwise maximization on final allocations. That is, the planner chooses $\Delta y^s_m = y^s_m - y_0$, for all $m$ and $s\in\{l,h\}$, and then recover taxes by $T(w_m^s) + y_0 = w_m^s - \Delta y_m^s$. The Lagrangian is given by
\begin{eqnarray}
\mathcal{L}\left(\overline{w},\{\Delta y_m^s\},y_0\right) &=& \left(L_I^l + L_I^h+K_I\right)\cdot G(y_0)\nonumber\\ &&+\alpha_l\cdot\int_0^{U^l-y_0}G(U^l-c)dF_l(c)+\alpha_h\cdot\int_0^{U^h-y_0}G(U^h-c)dF_h(c)\nonumber\\
&&+ K\cdot \int_{\psi^*}^{\overline{\psi}}G((1-t)\cdot\Pi(\psi,t) - \xi)dO(\psi)\nonumber\\
&&+ \gamma\cdot\left[\int\left(E_m^l(w_m^l - \Delta y_m^l)+E_m^h(w_m^h - \Delta y_m^h)\right)dm \right.\nonumber\\
&&\left.+ t\cdot K\cdot \int_{\psi^*}^{\overline{\psi}}\Pi(\psi,t)dO(\psi)- y_0\left(1 +K_I\right)\right].\label{p3eq1}
\end{eqnarray}
The main difference with respect to Proposition II is that the social planner leaves $\Delta y_m^s$ constant, for all $m$ and $s\in\{l,h\}$ when choosing $\overline{w}$. Then, the first order condition of the minimum wage is given by
\begin{eqnarray}
\frac{\partial\mathcal{L}}{\partial\overline{w}}\cdot\frac{1}{\gamma} &=& \frac{\partial U^l}{\partial\overline{w}}\cdot L_A^l\cdot g_1^l+\frac{\partial U^h}{\partial\overline{w}}\cdot L_A^h\cdot g_1^h+ K\cdot (1-t)\cdot \int_{\psi^*}^{\overline{\psi}}g_{\psi}\frac{\partial \Pi(\psi,t)}{\partial\overline{w}}dO(\psi)\nonumber \\
&&+\int\left(\frac{\partial E_m^l}{\partial\overline{w}}\left(T(w_m^l) + y_0\right) + E_m^l\frac{\partial w_m^l}{\partial\overline{w}}\right)dm\nonumber \\ 
&&+ \int\left(\frac{\partial E_m^h}{\partial\overline{w}}\left(T(w_m^h) + y_0\right) + E_m^h\frac{\partial w_m^h}{\partial \overline{w}}\right)dm\nonumber\\
&& + t\cdot K\cdot \int_{\psi^*}^{\overline{\psi}}\frac{\partial \Pi(\psi,t)}{\partial\overline{w}}dO(\psi) - \frac{\partial K_I}{\partial\overline{w}}\cdot\left(t\cdot \Pi(\psi^*,t) + y_0\right),\label{p3eq2} 
\end{eqnarray}
where, as in previous propositions, common terms are cancelled and the definition of the social marginal welfare weight is used. Since $\Delta y_m^s$ is fixed, we have that $\partial U^s/\partial \overline{w} = p_{\theta}\cdot(\partial \theta_m^s/\partial \overline{w})\cdot \Delta y_m^s$, $\partial E_m^s/\partial \overline{w} = p_{\theta}\cdot(\partial \theta_m^s/\partial \overline{w})\cdot L_A^s + p\cdot \alpha_s\cdot f_s(U^s-y_0)\cdot(\partial U^s/\partial \overline{w})$, profit effects are given by equations \eqref{prof_ef_1} and \eqref{prof_ef_2}, and wage spillovers are given by equation \eqref{wage_sp}, except for firms that are constrained by $\overline{w}$ for which $\partial w_m^l/\partial \overline{w} = 1$. When $\Delta y_m^s$ is fixed, changes in $\overline{w}$ do not affect $L_A^s$ and $L_m^s$ in partial equilibrium, so $\partial \theta_m^s/\partial \overline{w}$ is mediated by potential changes in vacancies, which in turn can generate a general equilibrium effect on applicants.

The first order condition with respect to $y_0$ yields
\begin{eqnarray}
\frac{\partial \mathcal{L}}{\partial y_0} = \left(L_I^l + L_I^h + K_I\right)\cdot G'(y_0) + \alpha_l\cdot \int_{0}^{U^l-y_0}G'(U^l-c)dF_l(c)&&\nonumber\\ +  \alpha_h\cdot \int_{0}^{U^l-y_0}G'(U^l-c)dF_l(c)  - \gamma\cdot (1 + K_I) &=& 0,
\end{eqnarray}
after noting that $\partial U^s/\partial y_0 = 1$ when $\Delta y_m^s$ is fixed, so $\partial L_I^s/\partial y_0 = \partial K_I / \partial y_0 = 0$. This implies that $\omega_0\cdot g_0 + \omega_1^l\cdot g_1^l + \omega_1^h\cdot g_1^h=1$, where $\omega_0 = \left(L_I^l + L_I^h + K_I\right)/(1 + K_I)$, $\omega_1^s = L_A^s/(1 + K_I)$, and $\omega_0 + \omega_1^l + \omega_1^h = 1$.

Finally, after simplifying terms, the first order condition with respect to $t$ yields
\begin{eqnarray}
\frac{\partial \mathcal{L}}{\partial t}\cdot\frac{1}{\gamma} &=& L_A^l\cdot g_1^l\cdot \frac{\partial U^l}{\partial t} + L_A^h\cdot g_1^h\cdot \frac{\partial U^h}{\partial t} + K\cdot \int_{\psi^*}^{\overline{\psi}}g_{\psi}\left[-\Pi(\psi,t) + (1-t)\cdot\frac{\partial \Pi(\psi,t)}{\partial t}\right]dO(\psi)\nonumber\\
&& + \int \left(\frac{\partial E_m^l}{\partial t}\left(w_m^l - \Delta y_m^l\right) + E_m^l\cdot\frac{\partial w_m^l}{\partial t}\right)dm +\int \left(\frac{\partial E_m^h}{\partial t}\left(w_m^h - \Delta y_m^h\right) + E_m^h\cdot\frac{\partial w_m^h}{\partial t}\right)dm\nonumber\\
&& + t\cdot K\cdot\int_{\psi^*}^{\overline{\psi}}\frac{\partial \Pi(\psi,t)}{\partial t}dO(\psi) + K\cdot\int_{\psi^*}^{\overline{\psi}}\Pi(\psi,t)dO(\psi)\nonumber\\&& - t\cdot K\cdot \Pi(\psi^*,t)\cdot o(\psi^*)\cdot\frac{\partial\psi^*}{\partial t} - y_0\cdot\frac{\partial K_I}{\partial t} = 0.
\end{eqnarray}
From equations \eqref{foc_w} and \eqref{foc_v}, it follows that wages and vacancies (and, therefore, employment and profits) decrease with $t$, because $\phi_{tn}\leq 0$. Since $\Delta y_m^s$ being fixed, this implies that tightness decreases and, therefore, $\partial U^s/\partial t = p_{\theta}\cdot(\partial \theta_m^s/\partial t)\cdot \Delta y_m^s < 0$ for $s\in\{l,h\}$. Reordering terms yields
\begin{eqnarray}
K\cdot \int_{\psi^*}^{\overline{\psi}}\left(1-g_{\psi}\right)\Pi(\psi,t)dO(\psi) &=& -\left(L_A^l\cdot g_1^l\cdot \frac{\partial U^l}{\partial t} + L_A^h\cdot g_1^h\cdot \frac{\partial U^h}{\partial t}\right)\nonumber\\
&& - K\cdot \int_{\psi^*}^{\overline{\psi}}\frac{\partial \Pi(\psi,t)}{\partial t}\left[g_{\psi}(1-t) + t\right]dO(\psi)\nonumber\\
&& - \int \left(\frac{\partial E_m^l}{\partial t}\left(w_m^l - \Delta y_m^l\right)+ E_m^l\cdot\frac{\partial w_m^l}{\partial t}\right)dm\nonumber\\ 
&& -\int \left(\frac{\partial E_m^h}{\partial t}\left(w_m^h - \Delta y_m^h\right) + E_m^h\cdot\frac{\partial w_m^h}{\partial t}\right)dm\nonumber \\
&& + \frac{\partial K_I}{\partial t}\left(\Pi(\psi^*,t) + y_0\right).
\end{eqnarray}
The right-hand side is positive,\footnote{Provided the optimal income tax system is not giving employment subsidies that are large enough to encourage increases in the corporate tax rate to decrease employment, to a degree that more than compensates for all the negative welfare effects and fiscal externalities. This would imply that the income tax system is not optimal in the first place.} which implies that
\begin{eqnarray}
\int_{\psi^*}^{\overline{\psi}}\omega_{\psi}(1- g_{\psi})dO(\psi) > 0,
\end{eqnarray}
with $\omega_{\psi} = \Pi(\psi,t)/\int_{\psi^*}^{\overline{\psi}}\Pi(\psi,t)dO(\psi)$ and $\int_{\psi^*}^{\overline{\psi}}\omega_{\psi}dO(\psi) = 1$, so the profit-weighted average welfare weight on active capitalists is smaller than one. \qedsymbol

\paragraph{Proposition IV} The objective function of the planner is given by
\begin{eqnarray}
\mathcal{L}(\overline{w},\Delta y^l,\Delta y^h,t,y_0) &=& (L_I^l+L_I^h)\cdot G(y_0) + \alpha_l\cdot\int_0^{U^l-y_0}G(U^l-c)dF_l(c) + \alpha_h\cdot\int_0^{U^h-y_0}G(U^h-c)dF_h(c)\nonumber\\
&& + K_S\cdot G\left((1-t)\cdot\Pi^S\right)+ K_M\cdot G\left((1-t)\cdot\Pi^M\right)\nonumber\\&&+\gamma\cdot\left[E^l\cdot(\overline{w} - \Delta y^l) + E^h\cdot(w^h - \Delta y^h) - y_0 + t\cdot\left(K_S\cdot\Pi^S + K_M\cdot\Pi^M\right)\right],
\end{eqnarray}
where $U^s = p^s(\theta^s)\cdot \Delta y^s - y_0$ and $E^s = p^s(\theta^s)\cdot L_A^s$, for $s\in\{l,h\}$. The first-order condition with respect to $\overline{w}$, after cancelling terms, is given by
\begin{eqnarray}
\frac{\partial \mathcal{L}}{\partial \overline{w}}\cdot\frac{1}{\gamma} &=& \frac{\partial U^l }{\partial \overline{w}}\cdot L_A^l\cdot g_1^l +\frac{\partial E^l}{\partial \overline{w}}\cdot(\overline{w}-\Delta y^l) + E^l + K_S\cdot\frac{\partial \Pi^S}{\partial\overline{w}}\cdot\left((1-t)\cdot g_K^S + t\right). \label{focmw}
\end{eqnarray}
Since $\Delta y^l$ is fixed, we have that \begin{eqnarray}
\frac{\partial U^l }{\partial \overline{w}} &=& p_{\theta}\cdot\frac{\partial \theta^l }{\partial \overline{w}}\cdot \Delta y^l,\\
\frac{\partial E^l }{\partial \overline{w}} &=& p_{\theta}\cdot \frac{\partial \theta^l }{\partial \overline{w}}\cdot L_A^l + p^l(\theta^l)\cdot \alpha_l\cdot f_l(U^l - y_0)\cdot\frac{\partial U^l }{\partial \overline{w}},\\
\frac{\partial \Pi^S}{\partial \overline{w}} &=& \left(\phi_n^S - \overline{w}\right)\cdot q_{\theta}\cdot \frac{\partial \theta^l}{\partial \overline{w}}\cdot v^l - q^l\cdot v^l. \label{profit_mw}
\end{eqnarray}

If follows that equation \eqref{focmw} can be written as
\begin{eqnarray}
\frac{\partial \mathcal{L}}{\partial \overline{w}}\cdot\frac{1}{\gamma} &=& p_{\theta}\cdot\frac{\partial \theta^l }{\partial \overline{w}}\cdot \Delta y^l\cdot L_A^l\cdot g_1^l +\left[p_{\theta}\cdot \frac{\partial \theta^l }{\partial \overline{w}}\cdot L_A^l + p^l(\theta^l)\cdot \alpha_l\cdot f_l(U^l - y_0)\cdot p_{\theta}\cdot\frac{\partial \theta^l }{\partial \overline{w}}\cdot \Delta y^l\right]\cdot(\overline{w}-\Delta y^l)\nonumber\\&&
+ E^l + K_S\cdot\left[\left(\phi_n^S - \overline{w}\right)\cdot q_{\theta}\cdot \frac{\partial \theta^l}{\partial \overline{w}}\cdot v^l - q^l\cdot v^l\right]\cdot\left((1-t)\cdot g_K^S + t\right),\nonumber\\
&=& \varepsilon_{\theta,\overline{w}}^l\cdot\frac{\theta^l}{\overline{w}}\cdot \left(p_{\theta}\cdot \Delta y^l\cdot L_A^l\cdot g_1^l+\left[p_{\theta}\cdot L_A^l + p^l(\theta^l)\cdot \alpha_l\cdot f_l(U^l - y_0)\cdot p_{\theta}\cdot \Delta y^l\right]\cdot(\overline{w}-\Delta y^l)\right.\nonumber\\
&& \left.+ K_S\cdot\left(\phi_n^S - \overline{w}\right)\cdot q_{\theta}\cdot v^l \cdot\left((1-t)\cdot g_K^S + t\right)\right) + E^l\cdot \left( 1 - \left((1-t)\cdot g_K^S + t\right)\right),\\
&=& \varepsilon_{\theta,\overline{w}}^l\cdot \theta^l\cdot\left( p_{\theta}\cdot L_A^l\cdot\left((1-\tau_l)\cdot g_1^l + \tau_l\right) +  K_S\cdot\frac{\left(\phi_n^S - \overline{w}\right)}{\overline{w}}\cdot q_{\theta}\cdot v^l \cdot\left((1-t)\cdot g_K^S + t\right)\right)\nonumber \\
&& + E^l\cdot\left(1 + \varepsilon_{L,\overline{w}}^l\cdot \tau_l - \left((1-t)\cdot g_K^S + t\right)\right),\nonumber\\
&\equiv& \varepsilon_{\theta,\overline{w}}^l\cdot a + E^l\cdot\left(1 + \varepsilon_{L,\overline{w}}^l\cdot \tau_l - \left((1-t)\cdot g_K^S + t\right)\right),\label{focmw_v2}
\end{eqnarray}
with $\varepsilon_{\theta,\overline{w}}^l = (\partial \theta^l/\partial \overline{w})/(\overline{w}/\theta^l)$, $\varepsilon_{L,\overline{w}}^l = (\partial L_A^l/\partial \overline{w})/(\overline{w}/L_A^l)$, both holding $\Delta y^l$ fixed, $\sign{\varepsilon_{\theta,\overline{w}}^l} = \sign{\varepsilon_{L,\overline{w}}^l}$, $a = $ is possibly positive, and $\Delta y^l = (1-\tau_l)\cdot \overline{w}$ and $\overline{w} - \Delta y^l = \tau_l\cdot \overline{w}$. Then, increasing $\overline{w}$ at the optimal tax allocation is desirable if equation \eqref{focmw_v2} is positive, which means that the fiscal externality to the social planner compensates the tightness distortions (given by potential negative vacancy distortions) and the (properly weighted) reduction in profits. When $\varepsilon_{\theta,\overline{w}}^l\to 0$, $\varepsilon_{L,\overline{w}}^l\to 0$, and the condition for equation \eqref{focmw_v2} being positive is reduced to $g_K^S < 1$.

Regarding low-skill workers after-tax allocations, the first order condition is given by
\begin{eqnarray}
\frac{\partial \mathcal{L}}{\partial \Delta y^l}\cdot\frac{1}{\gamma} &=& \frac{\partial U^l }{\partial \Delta y^l}\cdot L_A^l\cdot g_1^l +\frac{\partial E^l}{\partial \Delta y^l}\cdot(\overline{w}-\Delta y^l) - E^l\nonumber\\&&+ K_S\cdot\frac{\partial \Pi^S}{\partial\Delta y^l}\cdot\left((1-t)\cdot g_K^S + t\right). \label{focdeltal}
\end{eqnarray}
When $\overline{w}$ is fixed, we have that \begin{eqnarray}
\frac{\partial U^l }{\partial \Delta y^l} &=& p_{\theta}\cdot\frac{\partial \theta^l }{\partial \Delta y^l}\cdot \Delta y^l + p^l(\theta^l),\\
\frac{\partial E^l }{\partial \Delta y^l} &=& p_{\theta}\cdot \frac{\partial \theta^l }{\partial \Delta y^l}\cdot L_A^l + p^l(\theta^l)\cdot\frac{\partial L_A^l }{\partial \Delta y^l}\nonumber\\ 
&=& p_{\theta}\cdot \frac{\partial \theta^l }{\partial \Delta y^l}\cdot L_A^l + \frac{p^l(\theta^l)\cdot L_A^l}{\Delta y^l}\cdot \varepsilon_{L,\Delta}^l,
\end{eqnarray}
where $\varepsilon_{L,\Delta}^l$ is the participation elasticity with respect to changes in the after-tax allocations holding $\overline{w}$ fixed, given by $(\partial L_A^l/\partial \overline{w})/(\overline{w}/L_A^l)$. The first order condition can be written as
\begin{eqnarray}
\frac{\partial \mathcal{L}}{\partial \Delta y^l}\cdot\frac{1}{\gamma} &=& \left(p_{\theta}\cdot\frac{\partial \theta^l }{\partial \Delta y^l}\cdot \Delta y^l + p^l(\theta^l)\right)\cdot L_A^l\cdot g_1^l +\left(p_{\theta}\cdot \frac{\partial \theta^l }{\partial \Delta y^l}\cdot L_A^l + \frac{p^l(\theta^l)\cdot L_A^l}{\Delta y^l}\cdot \varepsilon_{L,\Delta}^l\right)\cdot(\overline{w}-\Delta y^l)\nonumber\\&& - E^l + K_S\cdot q_{\theta}\cdot \frac{\partial\theta^l}{\partial \Delta y^l}\cdot v^l\cdot\left(\phi^S_n-\overline{w}\right)\cdot\left((1-t)\cdot g_K^S + t\right),\nonumber \\
&=& p_{\theta}\cdot \frac{\partial \theta^l }{\partial \Delta y^l}\cdot L_A^l \cdot \left(\Delta y^l\cdot g_1 + \overline{w} - \Delta y^l\right) + E^l\cdot\left(g_1 + \varepsilon_{L,\Delta}^l\cdot \frac{\overline{w} - \Delta y^l}{\Delta y^l} - 1\right)\nonumber\\
&&+ K_S\cdot q_{\theta}\cdot \frac{\partial\theta^l}{\partial \Delta y^l}\cdot v^l\cdot\left(\phi^S_n-\overline{w}\right)\cdot\left((1-t)\cdot g_K^S + t\right),\nonumber\\
&=& -\frac{p_{\theta}\cdot\theta^l\cdot L_A^l\cdot\varepsilon_{\theta,\Delta}^l}{\Delta y^l}\cdot
\left(\Delta y^l\cdot g_1 + \overline{w} - \Delta y^l\right) + E^l\cdot\left(g_1 + \varepsilon_{L,\Delta}^l\cdot \frac{\overline{w} - \Delta y^l}{\Delta y^l} - 1\right)\nonumber\\
&&- \frac{K_S\cdot v^l\cdot q_{\theta}\cdot\theta^l\cdot\left(\phi^S_n-\overline{w}\right)\cdot\varepsilon_{\theta,\Delta}^l}{\Delta y^l}\cdot\left((1-t)\cdot g_K^S + t\right),
\label{simp_foc}
\end{eqnarray}
where $\varepsilon_{\theta,\Delta}^l = -\left(\partial \theta^l/\partial \Delta y^l\right)/\left(\Delta y^l/\theta^l\right)>0$ is the elasticity of tightness to changes in after-tax allocations holding the minimum wage fixed, with $\partial \theta^l/\partial \Delta y^l<0$ as shown below.

Noting that $\Delta y^l = (1-\tau_l)\cdot \overline{w}$ and $\overline{w} - \Delta y^l = \tau_l\cdot \overline{w}$ implies that 
\begin{eqnarray}
\frac{\partial \mathcal{L}}{\partial \Delta y^l}\cdot\frac{1}{\gamma} &=& -p_{\theta}\cdot\theta^l\cdot L_A^l\cdot\varepsilon_{\theta,\Delta}^l\cdot
\left( g_1 + \frac{\tau_l}{1-\tau_l}\right) + E^l\cdot\left(g_1 + \varepsilon_{L,\Delta}^l\cdot \frac{\tau_l}{1-\tau_l} - 1\right)\nonumber\\
&&- \frac{K_S\cdot v^l\cdot q_{\theta}\cdot\theta^l\cdot\left(\phi^S_n-\overline{w}\right)\cdot\varepsilon_{\theta,\Delta}^l}{(1-\tau_l)\cdot\overline{w}}\cdot\left((1-t)\cdot g_K^S + t\right).\label{simp_foc3}
\end{eqnarray}
To see whether a negative $\tau_l$ is optimal when $\overline{w}$ is fixed at its optimal value, equation \eqref{simp_foc3} is evaluated at $\tau_l = 0$. At $\tau_l=0$, we have that
\begin{eqnarray}
\frac{\partial \mathcal{L}}{\partial \Delta y^l} = E^l\cdot\gamma\cdot\left[-g_1\cdot\varepsilon_{\theta,\Delta}^l\cdot b + g_1 - 1 + c\cdot\varepsilon_{\theta,\Delta}^l\cdot\left((1-t)\cdot g_K^S + t\right)\right],\label{simp_foc4}
\end{eqnarray}
where $b = p_{\theta}\cdot \theta^l/p(\theta^l) \in (0,1)$ since $p(\theta^l) > p_{\theta}\cdot \theta^l$ given Euler's theorem since the matching function has constant returns to scale, and $c = -\left(q_{\theta}\cdot \theta^l\cdot \left(\phi^S_n-\overline{w}\right)\right)/\left(\overline{w}\cdot q(\theta^l)\right) \in(0,1)$ since $q(\theta^l)>-q_{\theta}\cdot \theta$ and $\overline{w} > \phi_n^S - \overline{w}$. Then, a sufficient condition for having negative marginal tax rates for low-skill workers (i.e., increasing $\Delta y^l$ is welfare improving when $\tau_l = 0$) when $\overline{w}$ is optimal
is given by 
\begin{eqnarray}
-g_1\cdot\varepsilon_{\theta,\Delta}^l\cdot b + g_1 - 1 + c\cdot\varepsilon_{\theta,\Delta}^l\cdot\left((1-t)\cdot g_K^S + t\right) &>& 0,
\end{eqnarray}
which happens when $g_1 > \left(1 - c\cdot\varepsilon_{\theta,\Delta}^l\cdot\left((1-t)\cdot g_K^S + t\right)\right)/\left(1 -b\cdot \varepsilon_{\theta,\Delta}^l\right)$, where I assumed $\varepsilon_{\theta,\Delta}^l<1$. 

To show that $\partial \theta^l/\partial \Delta y^l<0$, note that $\theta^l = (K_S\cdot v^l)/L_A^l$, so we have that
\begin{eqnarray}
\frac{\partial \theta^l }{\partial \Delta y^l} = \left(\frac{K_S}{L_A^l}\cdot \frac{\partial v^l }{\partial \Delta y^l} - \frac{K_S\cdot v^l}{L_A^{l2}}\cdot \frac{\partial L_A^l }{\partial \Delta y^l}\right) = \left(\frac{K_S}{L_A^l}\cdot \frac{\partial v^l }{\partial \Delta y^l} - \frac{K_S\cdot v^l}{L_A^{l}\cdot\Delta y^l}\cdot \varepsilon_{L,\Delta }^l\right).\label{dtheta_2}
\end{eqnarray}
Differentiating the first-order condition for vacancies in firms paying the minimum wage yields
\begin{eqnarray}
\left(\phi^S_n-\overline{w}\right)\cdot q_{\theta}\cdot\left(\frac{\partial \theta^l }{\partial w}\cdot\frac{\partial \overline{w}}{\partial \Delta y^l} + \frac{\partial \theta^l }{\partial U^l}\cdot \frac{\partial U^l }{\partial \Delta y^l}\right) = \eta_{vv}^l\cdot \frac{\partial v^l }{\partial \Delta y^l}.
\end{eqnarray}
Since the planner holds fixed $\overline{w}$ when varying $\Delta y^l$, the previous expression can be simplified to
\begin{eqnarray}
\frac{\left(\phi^S_n-\overline{w}\right)\cdot q_{\theta}}{\eta_{vv}^l}\cdot \left(\frac{\partial \theta^l }{\partial \Delta y^l} \cdot \Delta y^l + \frac{p^l(\theta^l)}{p_{\theta}}\right) =  \frac{\partial v^l }{\partial \Delta y^l}.
\end{eqnarray}
Replacing in equation \eqref{dtheta_2} yields
\begin{eqnarray}
\frac{\partial \theta^l }{\partial \Delta y^l}\cdot\left(1 - \frac{\left(\phi^S_n-\overline{w}\right)\cdot q_{\theta}\cdot \Delta y^l \cdot K_S}{\eta_{vv}^l\cdot L_A^l}\right)  &=& \frac{\left(\phi^S_n-\overline{w}\right))\cdot q_{\theta}\cdot p^l(\theta^l) \cdot K_S}{\eta_{vv}^l\cdot L_A^l\cdot p_{\theta}} - \frac{\theta^l}{\Delta y^l}\cdot \varepsilon_{L,\Delta }^l,\nonumber \\
\Longleftrightarrow \frac{\partial \theta^l }{\partial \Delta y^l}\cdot\left(1 - \frac{\left(\phi^S_n-\overline{w}\right)\cdot q_{\theta}\cdot \Delta y^l \cdot \theta^l}{\eta_{vv}^l\cdot v^l}\right)&=& \frac{\theta^l}{\Delta y^l}\cdot\left(\frac{\left(\phi^S_n-\overline{w}\right)\cdot q_{\theta}\cdot p^l(\theta^l) \cdot \Delta y^l}{\eta_{vv}^l\cdot v^l\cdot p_{\theta}} -  \varepsilon_{L,\Delta }^l\right), \label{dtheta_3}
\end{eqnarray}
which implies that $\partial \theta^l/\partial \Delta y^l <0$ provided $\varepsilon_{L,\Delta }^l\geq0$. $\qedsymbol$

\paragraph{Proposition V} Abstracting from the income tax system and the firm-level entry decisions implies that $T(w) = -y_0$, for all $w$, which is funded by the corporate tax revenue. Equation \eqref{p3eq2} implies that increasing the minimum wage when the corporate tax rate is optimal is welfare improving if
\begin{eqnarray}
\frac{\partial U^l}{\partial\overline{w}}\cdot L_A^l\cdot g_1^l +  K_S\cdot \frac{\partial\Pi^S}{\partial\overline{w}}\cdot\left(g_K^S + t\cdot (1-g_K^S)\right) >0. \label{prop}
\end{eqnarray}
where I used that $\partial U^h/\partial \overline{w} = 0$ and $\partial \Pi^M/\partial \overline{w} = 0$  because high-skill workers work in firms non-affected by the minimum wage. I omit arguments of the profit functions to simplify notation. 

With no income taxes, $U^l = p^l(\theta^l)\cdot \overline{w} + y_0$ and $U^h = p^h(\theta^h)\cdot w^h + y_0$. Then, it follows that $dU^l = p_{\theta}\cdot d\theta^l\cdot \overline{w} + p^l(\theta^l)\cdot d\overline{w}$ and $dU^h = p_{\theta}\cdot d\theta^h\cdot w^h + p^h(\theta^h)\cdot dw^h$, so
\begin{eqnarray}
\frac{\partial U^l}{\partial t} &=& p_{\theta}\cdot\frac{\partial \theta^l}{\partial t}\cdot \overline{w}, \label{ult}\\
\frac{\partial U^h}{\partial t} &=& p_{\theta}\cdot\frac{\partial \theta^h}{\partial t}\cdot w^h + p^h(\theta^h)\cdot\frac{\partial w^h}{\partial t}, \label{uht}\\
\frac{\partial U^l}{\partial \overline{w}} &=& p_{\theta}\cdot\frac{\partial \theta^l}{\partial \overline{w}}\cdot \overline{w} + p^l(\theta^l), \label{ulw}\\
\frac{\partial U^h}{\partial \overline{w}} &=& 0. \label{uhw}
\end{eqnarray}
Differentiating the first order conditions of the firms with respect to vacancies we have that
\begin{eqnarray}
\left(\widetilde{\phi}^S_{nk}\cdot dk_D - d\overline{w}\right)q^l + \left(\widetilde{\phi}^S_{n} - \overline{w}\right)\cdot q_{\theta}\cdot\left(\frac{\partial \theta^l}{\partial w}\cdot d\overline{w} +  \frac{\partial \theta^l}{\partial U^l}\cdot dU^l\right) &=& \eta_{vv}^l\cdot dv^l, \label{vl} \\
\left(\widetilde{\phi}^M_{nk}\cdot dk_D - dw^h\right)q^h + \left(\widetilde{\phi}^M_{n} - w^h\right)\cdot q_{\theta}\cdot\left(\frac{\partial \theta^h}{\partial w}\cdot dw^h +  \frac{\partial \theta^h}{\partial U^h}\cdot dU^h\right) &=& \eta_{vv}^h\cdot dv^h, \label{vac_m}
\end{eqnarray}
where, for analytical simplicity, I assumed that the differential effects driven by the curvature of $\widetilde{\phi}$ are second-order (i.e., $\widetilde{\phi}_{nn}$ is small relative to the first-order effects). Recall also from the first order condition of manufacturing firms with respect to wages that $\left(\widetilde{\phi}^M_n - w^h\right)\cdot q_{\theta}\cdot \frac{\partial \theta^h}{\partial w^h} = q^h$. Then, \eqref{vac_m} simplifies to
\begin{eqnarray}
\widetilde{\phi}^M_{nk}\cdot dk_D \cdot q^h + \left(\widetilde{\phi}^M_{n} - w^h\right)\cdot q_{\theta}\cdot \frac{\partial \theta^h}{\partial U^h}\cdot dU^h &=& \eta_{vv}^h\cdot dv^h. \label{vac_m2}
\end{eqnarray}
Finally, we also know that $\theta^l = K_S\cdot v^l/L_A^l$ and $\theta^h = K_M\cdot v^h/L_A^h$. Then:
\begin{eqnarray}
d\theta^l &=& \frac{K_S}{L_A^l}\cdot dv^l - \frac{K_S\cdot v^l}{L_A^{l2}}\cdot f_l(U^l-y_0)\cdot dU^l, \label{thetal}\\
d\theta^h &=& \frac{K_M}{L_A^h}\cdot dv^h - \frac{K_M\cdot v^h}{L_A^{h2}}\cdot f_h(U^h-y_0)\cdot dU^h. \label{thetah}
\end{eqnarray}

First, consider the comparative statics with respect to $t$, i.e., equations \eqref{ult} and \eqref{uht}. Setting $d\overline{w}=0$ and combining equations \eqref{ult}, \eqref{vl}, and \eqref{thetal} yields
\begin{eqnarray}
\frac{\widetilde{\phi}^S_{nk}\cdot q^l}{\eta_{vv}^l}\cdot\frac{\partial k_D}{\partial t} + \left(\widetilde{\phi}_n^S - \overline{w}\right)\cdot \frac{q_{\theta}}{\eta_{vv}^l}\cdot \frac{\partial \theta^l}{\partial U^l}\cdot p_{\theta}\cdot \overline{w}\cdot \frac{\partial \theta^l}{\partial t}\nonumber\\ = \frac{\partial \theta^l}{\partial t}\cdot\frac{L_A^l}{K_S}\cdot\left(1 + \frac{K_S\cdot v^l}{L_A^{l2}}\cdot f_l(U_l-y_0)\cdot p_{\theta}\cdot \overline{w}\right),
\end{eqnarray}
which implies that $\partial \theta^l/\partial t = -a_t^l\cdot \varepsilon_{k,t}^S$, with $a_t^l>0$ provided $\widetilde{\phi}_{nk}^S>0$. Then, $\partial U^l / \partial t = -a_t^l\cdot p_{\theta}\cdot\overline{w}\cdot \varepsilon_{k,t}^S$. Then, $\left(\partial U^l / \partial t\right)/\partial \varepsilon_{k,t}^S < 0$.

Also, from equation \eqref{dwdv}, we know that $dw/dv>0$ for firms that are not constrained by the minimum wage. Then, combining equations \eqref{uht}, \eqref{vac_m2}, and \eqref{thetah} yields
\begin{eqnarray}
\frac{\widetilde{\phi}^M_{nk}\cdot q^h}{\eta_{vv}^h}\cdot\frac{\partial k_D}{\partial t} + \left(\widetilde{\phi}_n^M - w^h\right)\cdot \frac{q_{\theta}}{\eta_{vv}^h}\cdot \frac{\partial \theta^h}{\partial U^h}\cdot p_{\theta}\cdot w^h\cdot \frac{\partial \theta^h}{\partial t}\nonumber \\ = \frac{\partial \theta^h}{\partial t}\cdot\frac{L_A^h}{K_M}\cdot\left( 1+ \frac{K_M\cdot v^h}{L_A^{h2}}\cdot f_h(U_h-y_0)\cdot p_{\theta}\cdot w^h\right)\cdot\frac{\left(1 - \left(\widetilde{\phi}_n^M - w^h\right)\cdot \frac{q_{\theta}}{\eta_{vv}^h}\cdot \frac{\partial \theta^h}{\partial U^h}\cdot p^h(\theta^h)\cdot\frac{\partial w^h}{\partial v^h} \right)}{\left(1 - \frac{v^h}{L_A^h}\cdot f_h(U^h-y_0)\cdot\frac{\partial w}{\partial v}\right)}.
\end{eqnarray}
The only term that has ambiguous sign is the last denominator. I assume it is positive, which economically implies that the increase in the corporate tax rate generates a decrease in posted vacancies that is attenuated by a change in the posted wage and is expected to happen when the density is negligible.\footnote{Assuming the contrary would imply that when capital mobility is larger the distortion of the corporate tax rate is smaller because of a huge labor participation effect if the density is large at $U^h - y_0$.} Then, this expression implies that $\partial \theta^h/\partial t = -a_t^h\cdot \varepsilon_{k,t}^M$, with $a_t^h>0$ provided $\widetilde{\phi}_{nk}^M>0$. Then, 
\begin{eqnarray}
\frac{\partial U^h}{\partial t} &=& -a_t^h\cdot\varepsilon_{k,t}^M\cdot\left(p_{\theta}\cdot w^h + p^h(\theta^h)\cdot\frac{\partial w^h}{\partial v^h}\cdot b\right),
\end{eqnarray}
where $b = \frac{L_A^h}{K_M}\cdot\left( 1+ \frac{K_M\cdot v^h}{L_A^{h2}}\cdot f_h(U_h-y_0)\cdot p_{\theta}\cdot w^h\right)\cdot \left(1 - \frac{v^h}{L_A^h}\cdot f_h(U^h-y_0)\cdot\frac{\partial w}{\partial v}\right) > 0$ under the assumption used above. Then, $\left(\partial U^h / \partial t\right)/\partial \varepsilon_{k,t}^M < 0$.

Finally, allowing $d\overline{w}$ to be non-zero, and combining equations \eqref{ulw}, \eqref{vl}, and \eqref{thetal} yields
\begin{eqnarray}
\widetilde{\phi}^S_{nk}\cdot \frac{\partial k_D}{\partial \overline{w}}\cdot q^l - q^l + \left(\widetilde{\phi}^S_{n} - \overline{w}\right)\cdot q_{\theta}\cdot\frac{\partial \theta^l}{\partial w}
+  \left(\widetilde{\phi}^S_{n} - \overline{w}\right)\cdot q_{\theta}\cdot\frac{\partial \theta^l}{\partial U^l}\left(p_{\theta}\cdot\frac{\partial \theta^l}{\partial\overline{w}}\cdot\overline{w} + p^l(\theta^l)\right) \nonumber \\
= \eta_{vv}^l\cdot\left( \frac{\partial\theta^l}{\partial \overline{w}}\cdot\frac{L_A^l}{K_S}\cdot\left(1+\frac{K_S\cdot v^l}{L_A^{l2}}\cdot f_l(U^l-y_0)\cdot p_{\theta}\cdot\overline{w}\right) + \frac{v^l}{L_A^l}\cdot f_l(U^l - y_0)\cdot p^l(\theta^l)\right).
\end{eqnarray}
Noting that $- q^l + \left(\widetilde{\phi}^S_{n} - \overline{w}\right)\cdot q_{\theta}\cdot\frac{\partial \theta^l}{\partial w} \leq 0$ because the firm is possibly deviating from the first order condition, then we have that $\partial \theta^l/\partial \overline{w} = -a_{\overline{w}}^l\cdot\varepsilon_{k,t}^S -b_{\overline{w}}^l$,  with $b_{\overline{w}}^l>0$ and $a_{\overline{w}}^l>0$ provided $\widetilde{\phi}_{nk}^S>0$. Then $\partial U^l/\partial \overline{w} = -\left(a_{\overline{w}}^l\cdot\varepsilon_{k,t}^S +b_{\overline{w}}^l\right)\cdot p_{\theta}\cdot \overline{w} + p^l(\theta)^l$. While the sign of $\partial U^l/\partial \overline{w}$ is ambiguous, it follows that $\left(\partial U^l / \partial \overline{w}\right)/\partial \varepsilon_{k,t}^S < 0$.

Now, consider the first order condition of the planner with respect to the corporate tax rate:
\begin{eqnarray}
\frac{\partial U^l}{\partial t}\cdot L_A^l\cdot g_1^l + \frac{\partial U^h}{\partial t}\cdot L_A^h\cdot g_1^h && \nonumber \\
+K_S\cdot g_K^S\cdot\left(-\Pi^S + (1-t)\cdot\frac{\partial \Pi^S}{\partial t}\right) + K_M\cdot g_K^M\cdot\left(-\Pi^M + (1-t)\cdot\frac{\partial \Pi^M}{\partial t}\right)&& \nonumber\\
+K_S\cdot \Pi^S + K_M\cdot\Pi^M + K_S\cdot t\cdot\frac{\partial \Pi^S}{\partial t} + K_M\cdot t\cdot\frac{\partial \Pi^M}{\partial t} &=& 0. \label{foct}
\end{eqnarray}
Grouping terms yields
\begin{eqnarray}
K_S\cdot \left(g_K^S + t\cdot (1-g_K^S)\right) &=& \frac{A + B - \left(g_K^M + t\cdot(1-g_K^M)\right)\cdot\gamma_t^M\cdot\varepsilon_{k,t}^M}{\gamma_t^S\cdot\varepsilon_{k,t}^S},\label{op_t}
\end{eqnarray}
where $A = K_S\cdot \Pi^S\cdot (1-g_K^S) + K_M\cdot \Pi^M\cdot (1-g_K^M)$ and $B = (\partial U^l/\partial t)\cdot L_A^l\cdot g_1^l + (\partial U^h/\partial t)\cdot L_A^h\cdot g_1^h$. Replacing \eqref{op_t} in \eqref{prop} gives
\begin{eqnarray}
\frac{\partial U^l}{\partial\overline{w}}\cdot L_A^l\cdot g_1^l  + \left(-\gamma_{\overline{w}}^S\cdot\varepsilon_{k,t}^S + \widetilde{\Pi}_{\overline{w}}^S\right)\cdot \frac{A + B - \left(g_K^M + t\cdot(1-g_K^M)\right)\cdot\gamma_t^M\cdot\varepsilon_{k,t}^M}{\gamma^S_t\cdot\varepsilon_{k,t}^S}>0. \label{prop3_mod}
\end{eqnarray}
Name the LHS $\mathcal{F}(\varepsilon_{k,t}^S,\varepsilon_{k,t}^M)$, so increasing the minimum wage is desirable if $\mathcal{F}(\varepsilon_{k,t}^S,\varepsilon_{k,t}^M) > 0$. Assuming the welfare weights are fixed (or that the effect of capital mobility on them are of second-order), note that:
\begin{eqnarray}
\frac{\partial \mathcal{F}(\varepsilon_{k,t}^S,\varepsilon_{k,t}^M)}{\partial \varepsilon_{k,t}^S} &=& \frac{\partial\left(\partial U^l/\partial\overline{w}\right)}{\partial \varepsilon_{k,t}^S}\cdot L_A^l\cdot g_1^l-\frac{\gamma_{\overline{w}}^S}{\gamma_t^S} \cdot\frac{\partial B}{\partial \varepsilon_{k,t}^S}  +\widetilde{\Pi}_{\overline{w}}^S\cdot\frac{\left(\frac{\partial B}{\partial \varepsilon_{k,t}^S}   - C\cdot \gamma_t^{S}\right)}{\gamma_t^S\cdot \varepsilon_{k,t}^S} ,
\end{eqnarray}
where $C = \left(A +B - \left(g_K^M + t\cdot(1-g_K^M)\right)\cdot\gamma_t^M\cdot\varepsilon_{k,t}^M\right)/\gamma^S_t\cdot\varepsilon_{k,t}^S>0$ following \eqref{op_t}, provided the optimal corporate tax rate has an interior solution. Then, the sign of $\partial \mathcal{F}(\varepsilon_{k,t}^S,\varepsilon_{k,t}^M)/\partial \varepsilon_{k,t}^S$ is ambiguous: the first term is negative and the second and third a positive. On the other hand:
\begin{eqnarray}
\frac{\partial \mathcal{F}(\varepsilon_{k,t}^S,\varepsilon_{k,t}^M)}{\partial \varepsilon_{k,t}^M} = \frac{\left(-\gamma_{\overline{w}}^S\cdot\varepsilon_{k,t}^S + \widetilde{\Pi}^S_{\overline{w}}\right)}{\gamma_t^S\cdot \varepsilon_{k,t}^S}\cdot\left(\frac{\partial B}{\partial \varepsilon_{k,t}^M} - \left(g_K^M + t\cdot(1-g_K^M)\right)\cdot \gamma_t^M\right) > 0,
\end{eqnarray}
which is unambiguously positive. \qedsymbol


\newpage

\section{Additional figures and tables}
\label{robust}

\setcounter{figure}{0} \renewcommand{\thefigure}{C.\Roman{figure}}

\setcounter{table}{0} \renewcommand{\thetable}{C.\Roman{table}}

\begin{figure}[h!]
\centering
\caption{State-level events by year}
\label{events}
\includegraphics[scale = 0.75]{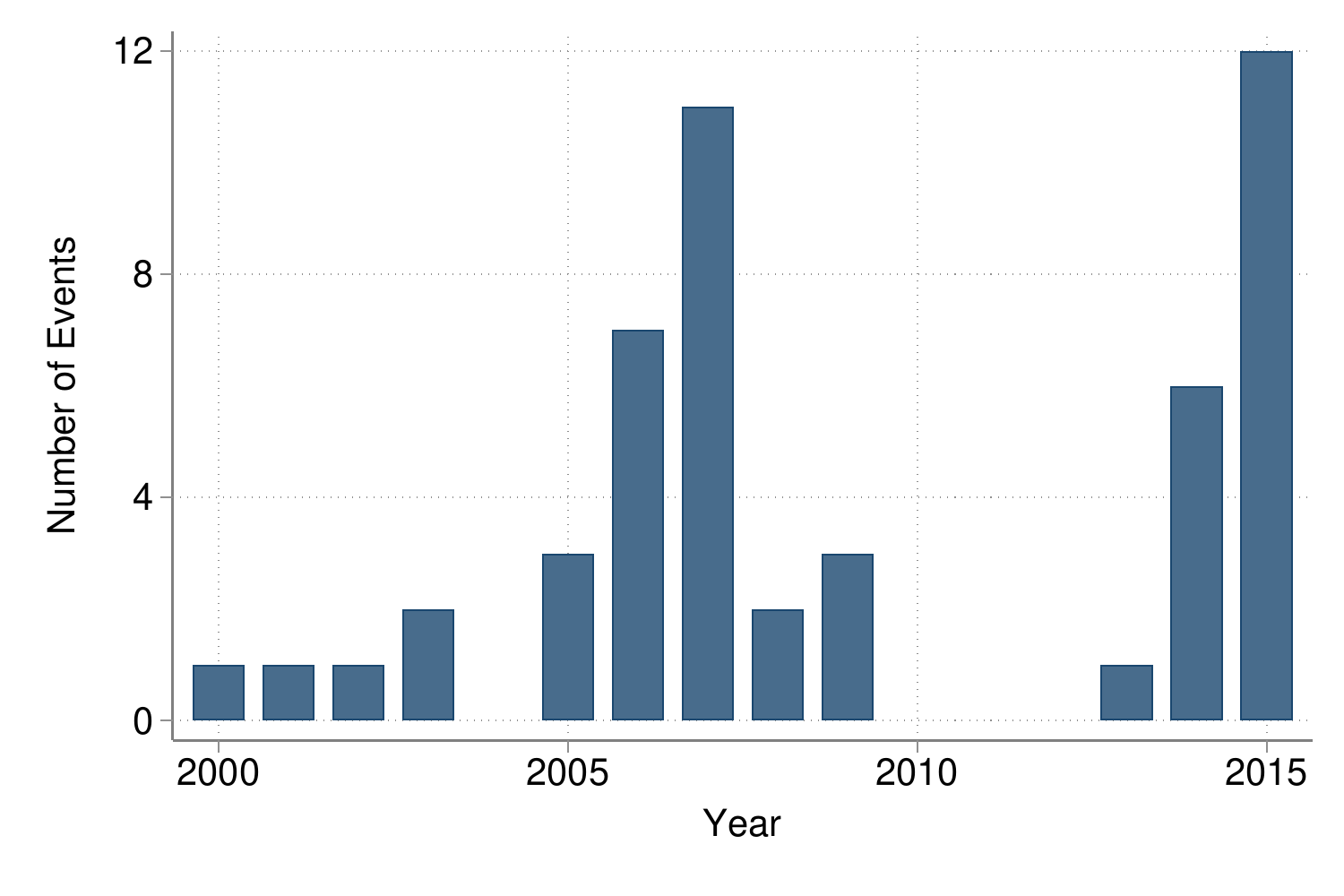}
\begin{minipage}{\textwidth} 
{\scriptsize \vspace{0.5cm}Notes: This figure plots the annual frequency of state-level minimum wage increases classified as events following \citeapp{cengiz2019effect,cengiz2021ml}. Data on minimum wages is taken from \citeapp{vaghul2016historical}. A state-level hourly minimum wage increase above the federal level is classified as an event if the increase is of at least \$0.25 (in 2016 dollars) in a state with at least 2\% of the working population affected, where the affected population is computed using the NBER Merged Outgoing Rotation Group of the CPS, treated states do not experience other events in the three years previous to the event, and the event-timing allows to observe the outcomes from three years before to four years after.\par}
\end{minipage}
\end{figure}

\clearpage

\begin{figure}[t!]
\centering
\caption{Minimum wage effects on low-skill workers' welfare: change in percentile considered}
\label{wage_perc}
\includegraphics[width=0.8\textwidth]{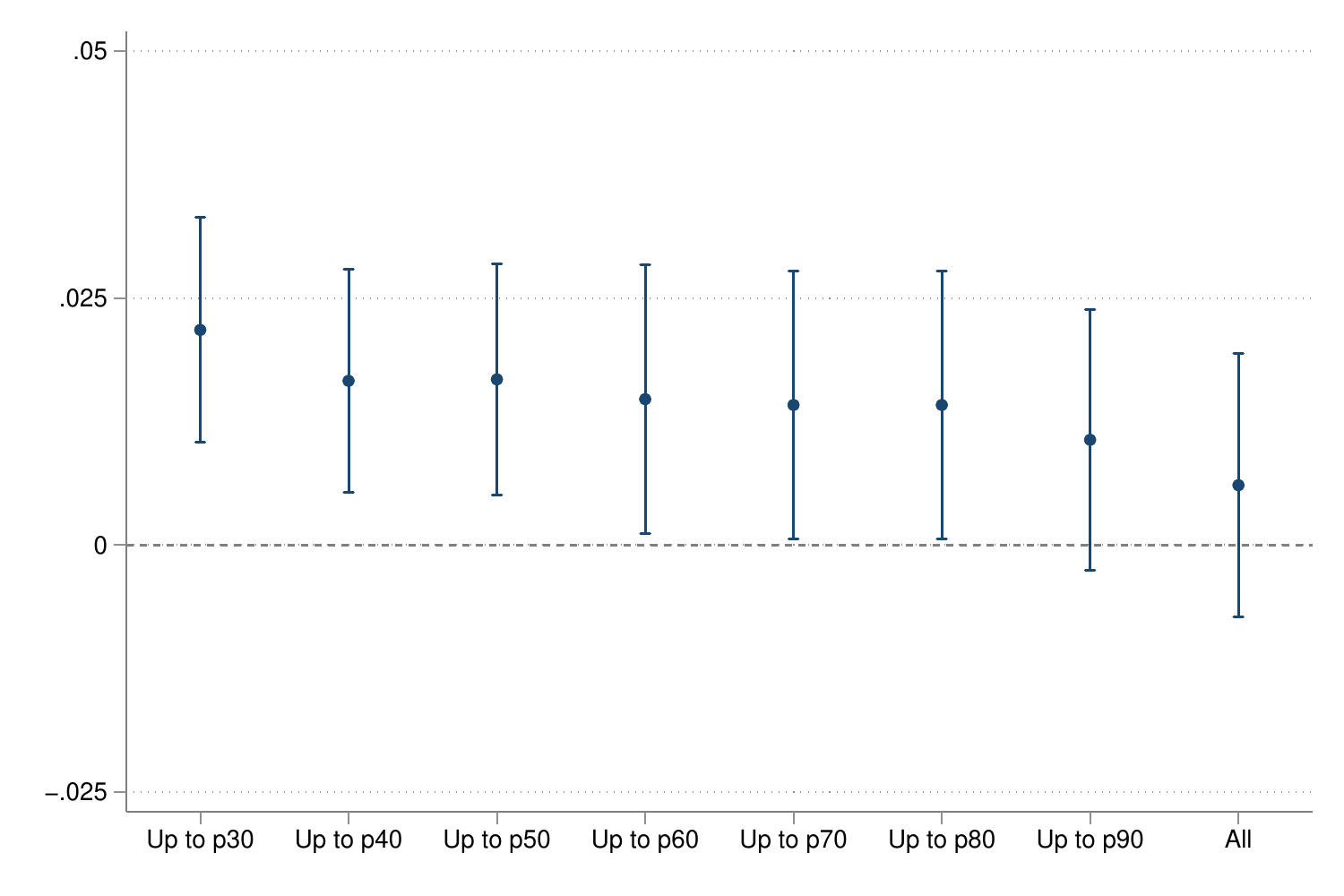}
\begin{minipage}{\textwidth} 
{\scriptsize \vspace{0.5cm} Notes: This figure plots the estimated $\beta$ coefficient with its corresponding 95\% confidence intervals from equation \eqref{reg_did} using the average pre-tax wage of active low-skill workers including the unemployed as dependent variable. Each coefficient comes from a different regression where the sufficient statistic is computed using different percentiles to truncate the sample of employed low-skill workers when computing the average wage. Low-skill workers are defined as not having a college degree. All regressions include year-by-event fixed effects. Standard errors are clustered at the state level, and regressions are weighted by state-by-year average population.\par}
\end{minipage}
\end{figure}

\clearpage

\begin{table}[h!]
    \centering
    \caption{List of Events}
    \label{list_events}
    {\footnotesize
    \begin{tabular}{c c c c c c c}
    \toprule
    State & Events (year) & Total &$\quad$& State & Events (year) & Total\\
    \midrule
    Alabama     &  -  & 0 &&Montana & 2007 & 1 \\
    Alaska     & 2003, 2015  & 2&& Nebraska & 2015 & 1 \\
    Arizona & 2007 & 1 &&Nevada & 2006 & 1\\
    Arkansas & 2006, 2015 & 2&& New Hampshire & - & 0\\
    California & 2007, 2014 & 2 && New Jersey & 2006, 2014 & 2\\
    Colorado & 2007, 2015 & 2&&  New Mexico & 2008 & 1\\
    Connecticut & 2009, 2015 & 2 &&New York & 2005, 2013 & 2 \\
    Delaware & 2000, 2007, 2014 & 3&&North Carolina & 2007 & 1 \\
    District of Columbia & 2014 & 1 &&North Dakota & - & 0 \\
    Florida & 2005, 2009 & 2 &&Ohio & 2007 & 1 \\
    Georgia & - & 0&&Oklahoma & - & 0\\
    Hawaii & 2002, 2015 & 2&&Oregon & 2003 & 1 \\
    Idaho & - & 0&&Pennsylvania & 2007 & 1 \\
    Illinois & 2005 & 1&&Rhode Island & 2006, 2015 & 2 \\
    Indiana & - & 0&&South Carolina & - & 0\\
    Iowa & 2008 & 1 &&South Dakota & 2015 & 1\\
    Kansas & - & 0&& Tennessee & - & 0\\
    Kentucky & - & 0&&Texas & - & 0 \\
    Louisiana & - & 0 &&  Utah & -  & 0 \\
    Maine & - & 0&&Vermont & 2009, 2015 & 2 \\
    Maryland & 2015 & 1&&Virginia & -  & 0 \\
    Massachusetts & 2001, 2007, 2015 & 3 &&Washington & 2007 & 1 \\
    Michigan & 2006, 2014  & 2&& West Virginia & 2006, 2015 & 2\\
    Minnesota & 2014 & 1&&Wisconsin & 2006 & 1 \\
    Mississippi & - & 0&&Wyoming & - & 0 \\
    Missouri & 2007 & 1&&&&\\

    \bottomrule
    \end{tabular}
    }
\floatfoot{\scriptsize Notes: This table details the list of events considered in the event-studies. Data on minimum wages is taken from \citeapp{vaghul2016historical}. A state-level hourly minimum wage increase above the federal level is classified as an event if the increase is of at least \$0.25 (in 2016 dollars) in a state with at least 2\% of the working population affected, where the affected population is computed using the NBER Merged Outgoing Rotation Group of the CPS, treated states do not experience other events in the three years previous to the event, and the event-timing allows to observe the outcomes from three years before to four years after.\par}    
    
\end{table}

\newpage

\section{Simulation appendix}
\label{simulation}

\setcounter{figure}{0} \renewcommand{\thefigure}{D.\Roman{figure}}

\setcounter{table}{0} \renewcommand{\thetable}{D.\Roman{table}}

\setcounter{equation}{0} \renewcommand{\theequation}{D.\Roman{equation}}

\paragraph{Functional forms} To simulate the model, I impose the following structure. Matching functions are given by $\mathcal{M}^s(L^s,V^s) = \delta_{0s}L^{s\delta_{1s}}V^{s1-\delta_{1s}}$, for $s\in\{l,h\}$. Revenue functions are given by $\widetilde{\phi}^I(k,n) = \psi^I\cdot k^{\beta_k^I}\cdot n^{\beta_n^I}$, for $I\in\{S,M\}$. The vacancy cost functions are given by $\eta^s(v) = \frac{\kappa_{0s}v^{1+\kappa_{1s}}}{1+\kappa_{1s}}$, for $s\in\{l,h\}$. The outside option is uniformly distributed with upper bound $\lambda_s$, for $s\in\{l,h\}$.

\paragraph{Calibration} To simulate the model, I need to impute parameter values. For a subset of parameters, I take values from the related literature (\textit{calibrated parameters}). The rest are chosen to match empirical moments (\textit{estimated parameters}). Whenever relevant and possible, I use values for 2019 (last year of my sample) to better approximate current policy analysis. Monetary values are in 2022 dollars.

\begin{table}[h!]
\centering
\def\sym#1{\ifmmode^{#1}\else\(^{#1}\)\fi}
\caption{Calibrated parameters}\label{fixed_parameters}
{\small
\begin{tabular}{ccc} \toprule
Parameters & Value & Source \\ \midrule
$\{\alpha_l,\alpha_h\}$  & $\{0.68,0.32\}$ & CPS \\
$\{\beta_n^S,\beta_k^S\}$ & $\{0.65,0.14\}$ & BEA, \citeapp{lamadon2019imperfect} \\
$\{\beta_n^M,\beta_k^M\}$ & $\{0.44,0.35\}$ & BEA, \citeapp{lamadon2019imperfect}\\
$ r^*(1-t^*)$ ($I = S$) & 0.032 & \citeapp{piketty2014capital}, \citeapp{bachas2022globalization} \\
$ r^*(1-t^*)$ ($I = M$) & 0.052 & \citeapp{piketty2014capital}, \citeapp{bachas2022globalization} \\
$t$ & 0.2 & US statutory corporate tax rate\\
$\{y_0,\tau\}$ & $\{15.92, 0.276\}$ & \citeapp{piketty2018distributional}\\
$\zeta$ & 1 & -
\\\bottomrule
\end{tabular}
}
\floatfoot{\scriptsize Notes: All monetary values are in thousands of dollars of 2019.}
\end{table}

\textit{Calibrated parameters:} Table \ref{fixed_parameters} summarizes the calibrated parameters. I set $\{\alpha_l,\alpha_h\} = \{0.68,0.32\}$, based on the distribution of skill within the working age population in the CPS Basic files. To compute factor shares, I use data from the BEA tables on the Composition of Gross Output by Industry and define the labor share ($LS$) as compensation of employees over the sum of compensation of employees and gross operating surplus. I do this for each of the industries used in the empirical analysis of the groups ``exposed services'' and ``manufacturing''. Then, I define $\beta_n^I = b\cdot LS$ and $\beta_k^I = b\cdot(1- LS)$, for $I\in\{S,M\}$, where $b$ is a returns to scale parameter, which I set equal to 0.79 based on \citeapp{lamadon2019imperfect}. This yields $\{\beta_n^S,\beta_k^S\} = \{0.65,0.14\}$ and $\{\beta_n^M,\beta_k^M\} = \{0.44,0.35\}$. To calibrate the foreign return to capital, I use the fact that the ratio of global capital to global output is around 500\%, and the global capital share of output is around 30\%, so the global pre-tax return is around 30\%/500\% = 6\% \citepapp{piketty2014capital}. Since the global capital tax rate is around 30\% \citepapp{bachas2022globalization}, this implies that the global after-tax return is around 4.2\%. To accommodate differential capital mobility based on differential transportation costs paid in units of investment returns, I assume that the global after-tax return is 3.2\% for $I=S$ and 5.2\% for $I=M$. The estimation below is done assuming a fixed tax system, which I define as follows. I use $t=20\%$, which is the statutory corporate tax rate. For the income tax system, I use \citeapp{piketty2018distributional} files and estimate linear regressions of taxes paid (post-tax incomes minus pre-tax incomes, including all taxes and transfers apportioned) over pre-tax incomes, restricting to working-age units whose total income is almost exclusively composed by labor income and whose annual incomes are lower than \$250,000. The relationship is surprisingly linear, being the current tax system reasonably approximated by a universal lump-sum of almost \$16,000 and a flat income tax rate of 27.6\%. Finally, I set $\zeta = 1$ so the social welfare function is logarithmic.

\begin{table}[h!]
\centering
\def\sym#1{\ifmmode^{#1}\else\(^{#1}\)\fi}
\caption{Estimated parameters}\label{calibrated_parameters}
{\small
\begin{tabular}{cccc}
\multicolumn{4}{c}{\textbf{Panel (a): Moments}} \\
\toprule
Moment & Source & Data & Model \\ \midrule
Unemployment rates ($s = \{l,h\}$) & CPS & $\{0.049,0.024\}$ & $\{0.046,0.054\}$\\
Job-filling rates ($I = \{S,M\}$) & JOLTS & $\{0.825,0.774\}$ & $\{0.752,0.831\}$\\
Ratio employment to establishments ($I = \{S,M\}$) & QCEW & $\{9.90,29.89\}$& $\{9.94,25.86\}$\\
Annual pre-tax earnings ($s = \{l,h\}$) & CPS & $\{13.20,82.42\}$ & $\{13.20,76.85\}$ \\
Labor force participation ($s = \{l,h\}$) & CPS & $\{0.570,0.737\}$ & $\{0.583,0.675\}$\\
Profit per establishment ($I = \{S,M\}$) &BEA, QCEW & $\{198.08,314.64\}$ & $\{199.94,345.96\}$\\
Average markdown ($s = \{l,h\}$) & \citeapp{bergermw} & $\{0.72,0.72\}$ & $\{0.516,0.794\}$\\
\bottomrule
\end{tabular}
\begin{tabular}{cccc}
&&&\\
\multicolumn{4}{c}{\textbf{Panel (b): Parameters}} \\
\toprule
Parameter & Value & Parameter & Value \\ \midrule
$\delta_{0l}$  & 0.85 &   $\delta_{0h}$  & 0.92\\
$\delta_{1l}$  & 0.51 & $\delta_{1h}$  & 0.79 \\
$\lambda_l$  & 15.62 & $\lambda_h$  & 77.97 \\
$K_S$  & 0.038 & $K_M$  &  0.008\\
$\psi^S$  & 31.46 & $\psi^M$  & 36.78 \\
$\kappa_{0l}$  & 0.727 &   $\kappa_{0h}$  & 0.239\\
$\kappa_{1l}$  & 0.987 & $\kappa_{1h}$  & 1.233 \\
\bottomrule
\end{tabular}
}
\floatfoot{\scriptsize Notes: All monetary values are in thousands of dollars of 2019.}
\end{table}

\textit{Estimated parameters:} Table \ref{calibrated_parameters} summarizes the moments matched and the estimated parameters. I solve the model to calibrate the following parameters by matching the following moments, separately for low-skill workers in exposed services and high-skill workers in manufacturing. I match skill-specific unemployment rates computed using the CPS and industry-specific job-filling rates (hires over postings) computed using JOLTS data to discipline the matching function parameters, $\{\delta_{0l},\delta_{01},\delta_{1l},\delta_{1h}\}$. I match industry-specific ratios of establishments to employment computed using the QCEW files to discipline the mass of capitalists, $\{K_S,K_M\}$. I match skill-specific labor force participation rates computed using the CPS to discipline the upper bounds of the opportunity cost distribution, $\{\lambda_l,\lambda_h\}$. Finally, I match the average profit per establishment computed using BEA data, the skill-specific average pre-tax annual earnings computed using the CPS, and the average wage markdown estimated by \citeapp{bergermw} to discipline the productivity and vacancy creation functions, $\{\psi^S,\psi^M,\kappa_{0l},\kappa_{0h},\kappa_{1l},\kappa_{1h}\}$.

\begin{figure}[t!]
\centering
\caption{Simulation results}
\label{sim_1}
\subfigure[Social welfare versus minimum wage]{\includegraphics[width=0.49\textwidth]{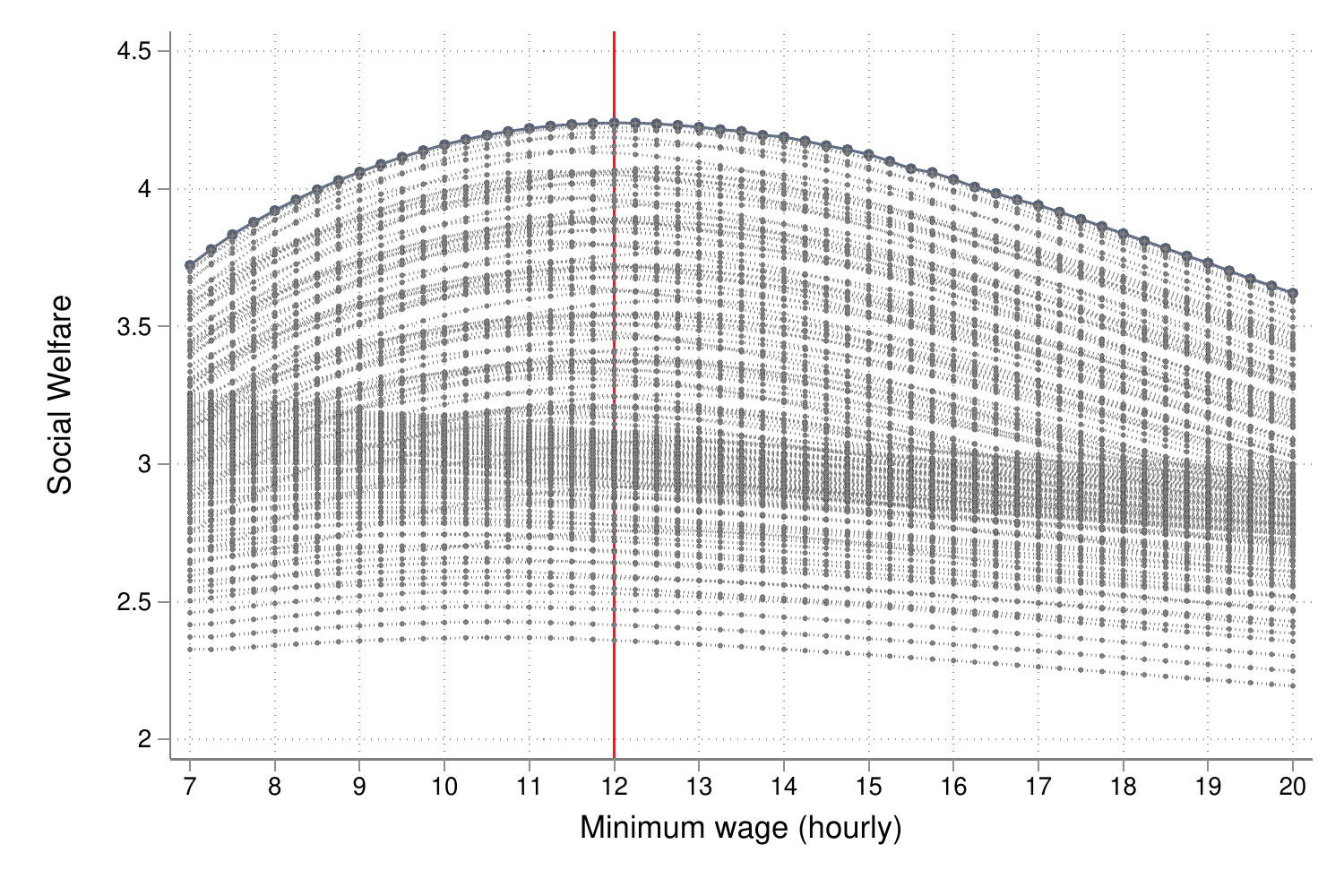}}
\subfigure[Optimal hourly minimum wage given taxes]{\includegraphics[width=0.49\textwidth]{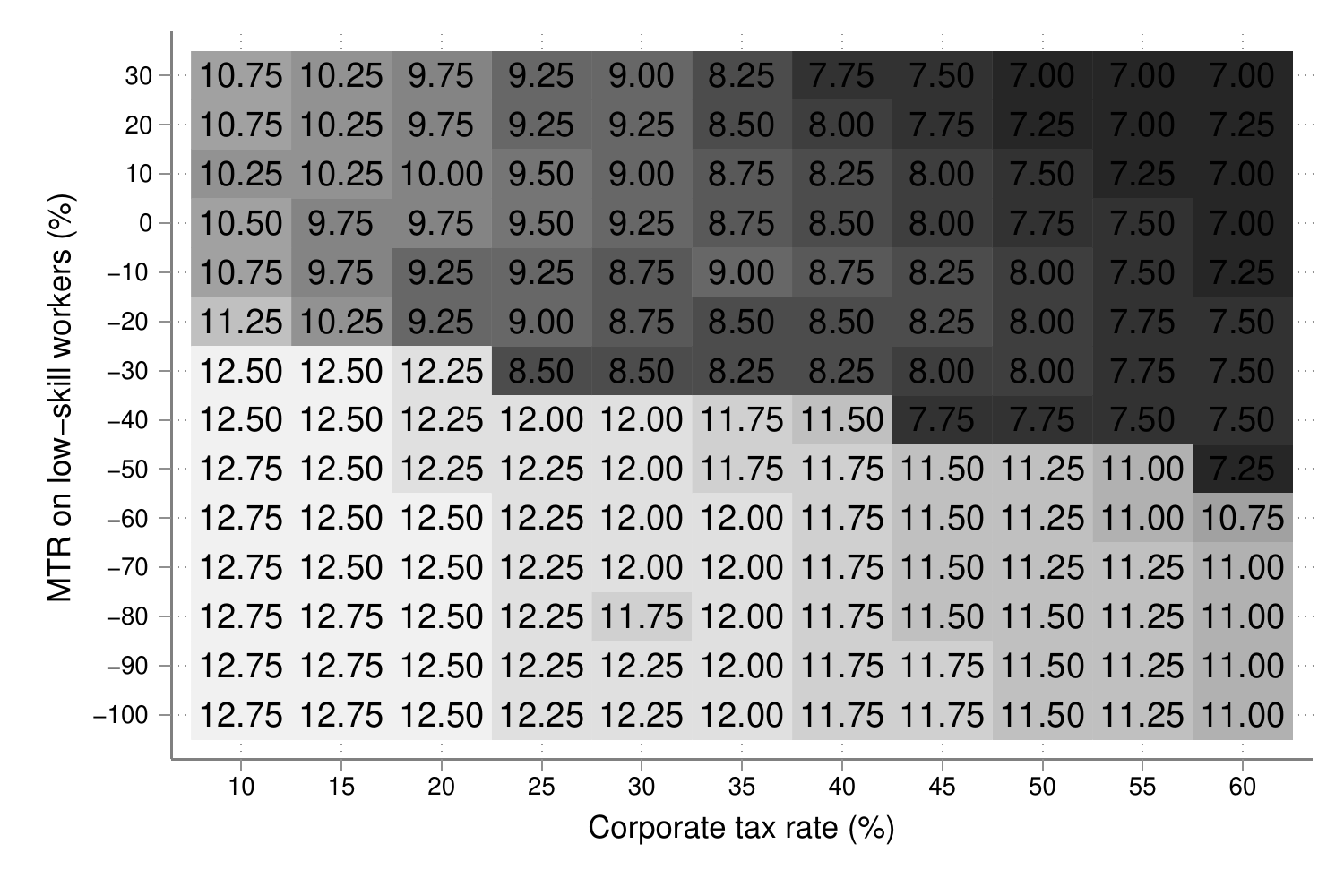}}
\begin{minipage}{\textwidth} 
{\scriptsize \vspace{0.5cm} Notes: This figure presents the results of the simulation exercises using the calibration procedures described in this appendix.\par}
\end{minipage}
\end{figure}

\paragraph{Exercise and results} I solve the model for different combinations of tax parameters and minimum wages. The policy parameters are reduced to $\{\tau_l,\tau_h,t,\overline{w}\}$, with $y_0$ recovered using the budget constraint. To simplify the analysis, I fix $\tau_h = 0.3$ and focus the attention on $\tau_l$, $t$, and $\overline{w}$. I solve the model for 154 permutations of tax parameters and 53 hourly minimum wage values and compute social welfare for each combination. Figure \ref{sim_1} summarizes the results. Panel (a) plots social welfare against hourly minimum wages given tax parameters. Each gray line represents one of the 154 tax combinations. The blue line represents the social welfare envelope (i.e., the tax system that maximizes social welfare given a minimum wage). Panel (b) shows the optimal hourly minimum wage (i.e., the minimum wage that maximizes social welfare) for each of the 154 tax combinations.

There are three messages from Panel (a). First, given taxes, social welfare generally follows a concave trajectory against minimum wages. That is, social welfare increases with the minimum wage until a point that starts decreasing. This is explained by the fact that wage effects tend to dominate employment effects at low minimum wages, but the effect is reverted at some point. Second, different tax systems yield very different levels of social welfare. Third, the envelope suggests that the optimal minimum wage (\$12 dollars the hour) is far from the market wage (\$7 dollars the hour). The exact number should be not taken literally given the simplicity of the exercise and, if anything, should be considered a lower bound given the efficiency properties of the model. Interestingly, at the optimal minimum wage of \$12 dollars the hour, the tax system consists of a substantial EITC (with $\tau_l = -100\%$) and a corporate tax rate of 35\%, suggesting that the joint optimum uses all instruments in tandem.

Similarly, there are three messages from Panel (b). First, the optimal minimum wage varies with the tax system. That is, there is vast heterogeneity in the turning points of each of the gray lines plotted in Panel (a). Second, optimal minimum wages seem to be larger when the EITC is larger, and when the corporate tax rate is lower. This supports the intuition developed throughout the paper, which suggests that minimum wages complement tax-based transfers to low-skill workers and substitute profit redistribution based on corporate tax rates. Third, together with Panel (a), it is suggested that social welfare is maximized when both the minimum wage and the corporate tax rate are set at ``intermediate'' values. Since the distortions of each policy are increasing in their values, the planner benefits from redistributing profits using both instruments, rather than just using larger corporate tax rates or large minimum wages.

\newpage
\bibliographystyleapp{chicago}
\bibliographyapp{referencias}

\end{document}

%% file: FT_SEPT2022/DS.tex
\begin{tabular}{lccccc}
\toprule
 & Obs. & Mean & Std. Dev. & Min & Max \\ \midrule  
 \textbf{Low-skill workers}: & & & & & \\  
 $U^l$ (annualized) &         1,173 &     19,396.69 &      1,225.82 &     16,176.45 &     24,002.46\\  
 Hourly wage &         1,173 &        11.55 &         0.62 &         9.74 &        13.99\\  
 Weekly hours worked &         1,173 &        34.83 &         1.57 &        29.84 &        38.50\\   
 Employment rate &         1,173 &         0.93 &         0.03 &         0.79 &         0.97\\   
 Participation rate &         1,173 &         0.61 &         0.05 &         0.47 &         0.72\\ \midrule 
 \textbf{High-skill workers}: & & & & & \\  
 $U^h$ (annualized) &         1,173 &     61,401.02 &      7,771.19 &     42,370.24 &     89,741.55\\  
 Hourly wage &         1,173 &        29.73 &         3.87 &        20.70 &        43.24\\  
 Weekly hours worked &         1,173 &        40.85 &         0.89 &        37.56 &        44.02\\   
 Employment rate &         1,173 &         0.97 &         0.01 &         0.92 &         1.00\\   
 Participation rate &         1,173 &         0.78 &         0.04 &         0.65 &         0.88\\ \midrule 
 \textbf{Fiscal variables (per working-age individual)}: & & & & & \\ 
 Income maintenance benefits &         1,173 &      1,056.56 &       328.81 &       402.09 &      2,194.19\\
 Medical benefits &         1,173 &      4,540.73 &      1,388.37 &      1,691.10 &      9,536.34\\
 Gross federal income taxes &         1,173 &      7,179.38 &      2,091.98 &      3,780.21 &     16,346.43\\ \midrule
 \textbf{Capitalists}: & & & & & \\ 
 Profit per establishment (exposed services) &         1,173 &    170,217.33 &     50,459.38 &     95,477.16 &    539,061.13\\ 
 Establishments (exposed services) &         1,173 &     70,313.94 &    103,291.48 &      5,397 &    914,454\\ 
 Profit per establishment (non-exposed services) &         1,173 &    943,530.83 &    261,578.15 &    441,352.63 &   1,765,250.88\\ 
 Establishments (non-exposed services) &         1,173 &     59,394.48 &     64,998.93 &      5,436 &    438,230\\ 
 Profit per establishment (manufacturing) &         1,173 &   1,957,057.43 &   1,359,081.14 &   -237,418.27 &   7,436,421.00\\
 Establishments (manufacturing) &         1,173 &      4,314.69 &      4,723.98 &        92 &     30,725\\ 
\bottomrule
\end{tabular}

%% file: FT_SEPT2022/T1.tex
\begin{tabular}{lcccccc}
& \multicolumn{3}{c}{\textbf{Low-skill Workers}} & \multicolumn{3}{c}{\textbf{High-skill Workers}} \\ \toprule
 $\hat{\beta}$ &        0.017  &        0.013 &        0.015 &        0.000 &       -0.003 &        0.002\\  
  & (0.006)  & (0.006) & (0.005) & (0.007) & (0.006) & (0.008) \\  \midrule
 Year FE & Y & N & N & Y & N & N \\ 
 Year x CR FE & N & Y & N & N & Y & N \\ 
 Year x CD FE & N & N & Y & N & N & Y \\ 
 Obs. &        10,300  &        10,300 &         9,653 &        10,300 &        10,300 &         9,653 \\  
 Events &           50  &           50 &           50 &           50 &           50 &           50 \\  
 $\Delta \log \mbox{MW}$ &        0.131  &        0.131 &        0.127 &        0.131 &        0.131 &        0.127 \\  
 Elasticity &        0.128  &        0.099 &        0.120 &        0.002 &       -0.023 &        0.013 \\  
\bottomrule
\end{tabular}

%% file: FT_SEPT2022/T2.tex
\begin{tabular}{lccccccccc}
& \multicolumn{3}{c}{\textbf{Income maintenance}} & \multicolumn{3}{c}{\textbf{Medical benefits}} & \multicolumn{3}{c}{\textbf{Gross federal}} \\
& \multicolumn{3}{c}{\textbf{transfers}} &  && & \multicolumn{3}{c}{\textbf{income taxes}} \\ \toprule
 $\hat{\beta}$ &       -0.040  &       -0.049 &       -0.050 &        0.004 &        0.001 &        0.006 &       -0.000 &       -0.006 &        0.005\\  
  & (0.015)  & (0.012) & (0.015) & (0.009) & (0.009) & (0.009)  & (0.009) & (0.009) & (0.008) \\  \midrule
 Year FE & Y & N & N & Y & N & N & Y & N & N\\ 
 Year x CR FE & N & Y & N & N & Y & N & N & Y & N \\ 
 Year x CD FE & N & N & Y & N & N & Y & N & N & Y \\ 
 Obs. &        10,300  &        10,300 &         9,653 &        10,300 &        10,300 &         9,653 &        10,300 &        10,300 &         9,653\\  
 Events &           50  &           50 &           50 &           50 &           50 &           50  &           50 &           50 &           50 \\  
 $\Delta\log\mbox{MW}$ &        0.131  &        0.131 &        0.127 &        0.131 &        0.131 &        0.127  &        0.131 &        0.131 &        0.127 \\  
 Elast. &       -0.306  &       -0.371 &       -0.389 &        0.029 &        0.008 &        0.044  &       -0.002 &       -0.049 &        0.042 \\  
\bottomrule
\end{tabular}

%% file: FT_SEPT2022/T3.tex
\begin{tabular}{lcccccccc}
& \multicolumn{4}{c}{\textbf{Profits per establishment}} & \multicolumn{4}{c}{\textbf{Establishments}} \\ 
& \multicolumn{3}{c}{\textbf{All industries}} & \textbf{Exp. Serv.} & \multicolumn{3}{c}{\textbf{All industries}} & \textbf{Exp. Serv.}  \\\toprule
 $\hat{\beta}$ &       -0.005  &       -0.007 &       -0.007 & -0.047 &        0.012 &        0.005 &       -0.000  & -0.015 \\  
  & (0.011)  & (0.009) & (0.011) & (0.014) & (0.013) & (0.007) & (0.006)  & (0.010)\\   \midrule
 Year FE & Y & N & N & N & Y & N & N & N\\ 
 Year x CR FE & N & Y & N & N & N & Y &N & N\\ 
 Year x CD FE & N & N & Y & Y & N & N &Y &Y\\ 
 Obs. &       519,311  &       519,311 &       519,311 &    519,311 &     552,792 &       552,792 &       552,792&            552,792 \\  
 Events &           50  &           50 &           50 &           50 &           50 &           50 &           50 &           50 \\  
 $\Delta\log \mbox{MW}$ &        0.132  &        0.132 &        0.132 &        0.132 &        0.131 &        0.131 &        0.131 &        0.131 \\  
 Elast. &       -0.034  &       -0.053 &       -0.051 &       -0.357&        0.094 &        0.041 &       -0.003  &       -0.116 \\  
\bottomrule
\end{tabular}